\documentclass[preprint,aps,tightenlines,draft]{revtex4}
\tighten
\begin{document}
\preprint{Bicocca-FT-01-27  October 2001}

\title
 {A library of extended high-temperature expansions of basic \\
 observables for the  spin-$S$ Ising   models  on  two-  and  
  three-dimensional lattices  \\}
\author{P. Butera\cite{pb} and M. Comi\cite{mc}}
\address
{Istituto Nazionale di Fisica Nucleare\\
Dipartimento di Fisica, Universit\`a di Milano-Bicocca\\
Via Celoria 16, 20133 Milano, Italy}
\date{\today}
\begin{abstract}
We present an on-line library of unprecedented extension for
 high-temperature expansions of basic  observables in the 
Ising models of general spin $S$, with nearest-neighbor interactions.
 We have tabulated through  order $\beta^{25}$ the series 
for the nearest-neighbor correlation function, the  
susceptibility and  the second correlation moment 
 in two dimensions on the square lattice, and, in three
dimensions,  on the simple-cubic and the 
 body-centered cubic lattices.
The expansion of the second field derivative of the 
susceptibility  is  also  tabulated through $\beta^{23}$ 
for the same lattices. We have thus added several terms (from four up to
thirteen) to the series already published for spin 
 $S=1/2,1,3/2,2,5/2,3,7/2,4,5,\infty$.   
\end{abstract}

\maketitle

 \widetext

\section{Introduction}

With the modern developments
 of the renormalization group theory the understanding of 
critical phenomena for  ordered systems has reached a stage of maturity:
the conceptual and computational paradigm has become widely accepted 
and essentially only more or less important ``details''  still remain to be 
worked out\cite{fisherrmp}.
The question of a complete a priori definition of the universality class 
of critical behavior, the actual computation of the corrections to the scaling 
behavior, and the accurate determination of the universal and nonuniversal 
critical parameters are some  strictly intertwined examples of these
 details which must always be faced in the study of specific models.
Indeed the identification of the universality class in terms of the spatial
dimensionality of the system, the number of components and symmetry 
properties of the order parameter and the range of the interactions is 
not always unambiguous and ought to be checked in each case.
On the other hand the size of the corrections to scaling (and, as a 
consequence, the size of the asymptotic critical region) may vary
strongly among models 
in the same universality class and unless these corrections can be somehow 
brought under control, the estimates of the critical parameters and therefore
the identification of the universality class may be not very reliable.
High-temperature(HT) expansions are  a source of valuable numerical 
benchmarks in the study of spin models and as such are also a good tool 
to investigate these questions. Therefore  it is important to provide
 easy access to a 
body of information in the form of exact high-temperature(HT) expansions
as extensive as presently possible for a family of models expected to belong
 to the same universality class  such as the Ising models of general spin $S$.
These models  have a great historical and 
conceptual interest: in the past 
their study\cite{dosy} contributed to shape   the modern formulation
 of the universality hypothesis\cite{univ} and, still today, they remain
 the simplest and phenomenologically most  
useful  discrete-state lattice models\cite{dg,dc,fisherising}.  
It is worth remarking that in spite of this, for many years,  
high-temperature  series expansions  both  in two and in three 
dimensions,  have been available only for  few observables 
 and  have been generally  too short to make  numerically accurate  
discussions of these models possible\cite{camp,camp2,saul}.
 For instance, in three dimensions 
on the simple-cubic (sc) lattice,  the expansions  of
  $\chi(\beta;S)$ and of the second moment of the correlation 
function $\mu_2(\beta;S) $,   for spin $S > 1/2$, can be found explicitly
 in the literature\cite{camp2} 
only up to the order $\beta^{12}$.   
The subsequent work by R.Roskies and P. Sackett\cite{sack} who  made 
an extension through $\beta^{15}$ possible, 
 improved the situation only slightly. 
 On the fcc lattice, the HT series initially derived to order
 $\beta^{12}$ in  Ref.\cite{camp2}, 
 were later extended  in Ref.\cite{mckest} 
 to order $\beta^{14}$. Only in the case of the 
body-centered-cubic (bcc) lattice, 
 remarkable progress  occurred  already two decades ago,
 with the computation by B.G.Nickel\cite{nick21} of 
 expansions for  $\chi(\beta;S)$ and  $\mu_2(\beta;S)$ through $\beta^{21}$. 
(Exclusively the series for $S=1/2,1,2,\infty$ 
were published\cite{nr90} at the time.) 

 Also   for the second field-derivative of the 
susceptibility $\chi_4(\beta;S)$ 
and for the nearest-neighbor correlation function $G(\beta;S)$,  
 the published data  are scarce. 
 On  the sc lattice, series for $\chi_4(\beta;S)$ 
 could be  derived  from the  data  of Ref.\cite{lw}
  up to order $\beta^{14}$, and, on the bcc lattice,  
 from the data of Ref.\cite{bakin} up to $\beta^{10}$.
 On the fcc lattice, 
  series  for $\chi_4(\beta;S)$  are available\cite{mck}  through $\beta^{13}$.
  For general spin, only expansions\cite{camp2,mckest} of 
 $G(\beta;S)$ up  to order $\beta^{14}$ on the fcc lattice 
have been published. 

In two dimensions the Ising model is (partially) solved only for spin $S=1/2$  
 and,  by taking advantage of this property, very long series\cite{nick21}  
have been computed
 for $\chi(\beta;1/2)$ and $\mu_2(\beta;1/2)$ on the square (sq) lattice. 
For $S>1/2$ the models are not solvable and they even lack any 
simple duality property.
Series up to $\beta^{21}$ for 
 $\chi(\beta;S)$ and  $\mu_2(\beta;S)$  on the 
sq lattice  have 
 been tabulated in Ref.\cite{nr90} for $S=1,2$ and $\infty$.
On the square and the triangular
 lattices   series
for $\chi(\beta;S)$ , $\mu_2(\beta;S)$  and $\chi_4(\beta;S)$ 
 up to $\beta^{10}$ 
can be computed from the data of  Ref. \cite{bakin}.  

 A summary of the HT expansions published until now  for the  
 Ising models of  general spin appears in Table 1.

 We have long undertaken   
 a systematic project 
 to realize more flexible and efficient algorithms and 
 codes for graphical HT expansions
 in  two-dimensional\cite{bc2d,bc2d1,bc2d2} 
and   in three-dimensional\cite{bc,bc23,bcesse} lattice spin models.
 By  taking advantage also  of the 
steady increase of computer performances in the last decade,  
we have been able
 to extend,  in successive steps, 
 the HT series for some of the most widely studied 
spin models,  as well as
 to update their numerical analyses whenever it was possible to improve 
 significantly the determination of the critical parameters.  
   We have now added several terms (from four up to thirteen) to  
the HT expansions of various observables for the general spin-$S$ 
Ising models on  the sq, the sc and the bcc lattices.
In Ref.\cite{bcesse} we have already 
presented an analysis of these series in the three-dimensional case, 
 which  adds  further support to the validity 
of hyperscaling and of universality with respect to the lattice 
 structure and to the the value of the spin. 
Moreover,  our extension of the existing computations has confirmed the 
remarkably fast convergence properties of the HT expansions on the bcc lattice 
and has made possible to understand quantitatively how the approach to 
scaling depends on the spin $S$. In particular, the spread of the critical 
exponents estimates as a function of $S$, already observed long ago in the 
study of shorter expansions
has been simply explained in terms of the pattern of signs and sizes 
of the leading corrections to scaling and this insight has been used 
also to 
improve the determination of the universal critical parameters
in the spirit of the suggestions of Refs.\cite{zinn81,fishnick}

 In deriving these data  by an appropriately  renormalized
 linked-cluster method\cite{lw,bakin,lcm}, 
we have essentially used  the  same thoroughly tested 
code which recently produced \cite{bc23}  series  
through $\beta^{23}$ 
for $\chi(\beta;1/2)$ and $\mu_2(\beta;1/2)$ on 
 the sc and the bcc lattices.  
 The correctness of our procedures is 
 ensured  by numerous internal consistency checks, as well as by their
 ability to reproduce established results in simpler particular cases,
 such as the two-dimensional spin 1/2 Ising model or 
the one-dimensional spin-$S$ Ising models. 
 Of course, we have made sure that  our code also  reproduces, through
 $\beta^{21}$, 
the existing computation  of Ref.\cite{nick21} 
for $S=1,2,\infty$,   on the bcc lattice and,
 as far as there is overlap, also the recent computation
 of Ref.\cite{cam} for $S=1/2$ on the same lattice. 

  Our expansions of $G(\beta;S)$, 
 $\chi(\beta;S)$ and $\mu_2(\beta;S)$  
to order $\beta^{25}$  and   of  $\chi_4(\beta;S)$ up to  $\beta^{23}$ 
for spin $S=1/2,1,3/2,2,5/2,3,7/2,4,5,\infty$, in two-dimensions for 
the square lattice case and, in three dimensions, 
for the simple cubic and the body-centered cubic lattice cases are
now made available in electronic form in order to provide
 a convenient reference for possible future extensions and 
 further work of analysis and phenomenological 
applications.

In the next section we shall  set our notations and definitions.  In  
 the following sections we shall report our data first on the 
 square lattice and then on the three-dimensional lattices.

\section{ The basic observables of the spin-$S$ Ising model}

  The spin-$S$ Ising model with nearest-neighbor interactions
 is defined by  the Hamiltonian:

\begin{equation}
H \{ s \} = -{\frac {J} {2}} \sum_{\langle \vec x,{\vec x}' \rangle } 
s({\vec x})  s({\vec x}') 
\label{hamilt} \end{equation}

where $s(\vec x)=s^z(\vec x)/S$  and  $s^z(\vec x)$ denotes a  
classical spin  variable at the
lattice site $\vec x$, taking the $2S+1$ values 
$-S, -S+1,\ldots ;,S-1, S$.    
 The sum runs over  all nearest-neighbor pairs of  sites. 
We have computed the expansions in terms of 
the usual HT variable $\beta=J/k_BT$
 called  "inverse temperature" for brevity.

 We shall first tabulate the expansion coefficients of 
the nearest-neighbor correlation function   

\begin{equation}
G(\beta;S)= \langle s(\vec 0)  s(\vec \delta) \rangle_c=
 \sum_{r=0}^\infty h_r(S) \beta^r
\end{equation}
  where $\vec \delta$ is a nearest-neighbor lattice vector.

We shall then report the expansions of 
the zero-field reduced susceptibility  defined by

\begin{equation}
\chi(\beta;S) = \sum_{\vec x} \langle s(0)  s(\vec x) \rangle_c = 
\sum_{r=0}^\infty c_r(S) \beta^r 
\label{chi} \end{equation}

of the second  moment of the correlation function  defined by

\begin{equation}
 \mu_{2}(\beta;S)=\sum_{\vec x} \vec x^2 \langle s(0)  s(\vec x) 
\rangle_c = 
 \sum_{r=1}^\infty d_r(S) \beta^r 
\end{equation}

and finally the expansion of the second field derivative of the
susceptibility  defined by

\begin{equation}
 \chi_{4}(\beta;S)=  \sum_{x,y,z}
\langle  s(0)  s(x) s(y)  s(z)\rangle_{c}=
 \sum_{r=0}^\infty e_r(S) \beta^r. 
\label{chi4}\end{equation}

 Notice that our 
definitions ensure both that a non-trivial limit as 
$S \rightarrow \infty$ exists and  that the universal
 combinations\cite{aha} of the critical amplitudes 
have an $S$-independent  numerical value.

\normalsize

\section{The nearest-neighbor correlation 
 function  on the sq lattice}

The HT  expansion coefficients of the nearest-neighbor 
correlation  function  
$G^{sq}(\beta;S) =  \sum_{r=1}^\infty h^{sq}_r(S) \beta^r $ 
 for the spin $S=1/2$ Ising model on the sq lattice are

\scriptsize

\[ G^{sq}(\beta;1/2 )=\beta^{ } +   5/3\beta^{3 } +   32/15\beta^{5 } 
+   425/63\beta^{7 } +   57272/2835\beta^{9 } +   2233388/31185\beta^{11 } +   
 1610847424/6081075\beta^{13 } + \]\[  131895762457/127702575\beta^{15 } +   
 45158782261352/10854718875\beta^{17 } 
+   6394645028904284/371231385525\beta^{19 } + \]\[  
 14200564385208064384/194896477400625\beta^{21 } 
+   181984936857098828252/
  580103632733625\beta^{23 } +  \]\[ 
5069199284121682414346464/3698160658676859375\beta^{25 }+ \ldots ;\]

\normalsize
  for  spin $S=3/2$ we have
\scriptsize

\[ G^{sq}(\beta;3/2 )=25/81\beta^{ } +   7253/19683\beta^{3 } +  
 679840/1594323\beta^{5 } +   6851317741/13559717115\beta^{7 } + 
 1251974959000/1977006755367\beta^{9 } \]\[ 
+   5331888182912068/6291117925114275\beta^{11 } +   
 6746481788866169536/5564619627122078523\beta^{13 } 
+ \]\[  2182601527790409015012797/
  1183177248216831945952875\beta^{15 } +  \]\[
  27527769875866309256121176/9309914690254729083297765\beta^{17 } 
+ \]\[  111444149709460874134231431957196/
  22566534965575694193232535027625\beta^{19 } 
+ \]\[   65349691500118048675470701211090304/
  7677135195288851164537708416398025\beta^{21 } +  \]\[ 
 8444772359659820696832198670870769210516/
  561884041275194381838825923490159736875\beta^{23 } +  \]\[  
 25838995777708349903292324409667101343006176/
  955764754209105643507842895856761712424375\beta^{25 }+ \ldots ;\]

\normalsize
  for  spin $S=5/2$ we have
\scriptsize

\[ G^{sq}(\beta;5/2 )=49/225\beta^{ } +   233093/1171875\beta^{3 } 
+   6483223904/35595703125\beta^{5 } + \]\[  
 855104825086547/5005645751953125\beta^{7 } 
+   2500940769232024/15016937255859375\beta^{9 } + \]\[  
 383123818941655862501588/2258406579494476318359375\beta^{11 } + \]\[   
 1537747218046159096559533888/8493221886456012725830078125\beta^{13 } +\]\[   
 1209747610743746978965198908326279/5971796638914383947849273681640625
\beta^{15 } + \]\[  
 795740460737593343930709000570424271912/
  3357791955145585234276950359344482421875\beta^{17 } +  \]\[ 
 52798557204520305860139314810193100397116364/
  183093885309277766282320953905582427978515625\beta^{19 } + \]\[    
 10254588072751165409121916374050257904618220928/
  28271849937462008028887794353067874908447265625\beta^{21 } +  \]\[    
 \frac{1178139652080331063144759568919240418737530117381479315204}
{  2513642668229607495502531122610889724455773830413818359375}\beta^{23 }
 + \]\[   
 \frac{178579817009039110776854210230801209607291419223370338300384}
{  288555918546766166575545664585433769389055669307708740234375}\beta^{25 }
+ \ldots ;\]

\normalsize
 for  spin $S=7/2$ we have
\scriptsize

\[ G^{sq}(\beta;7/2 )=9/49\beta^{ } +   17079/117649\beta^{3 } 
+   8045630304/69206436005\beta^{5 } +   
 2266864432981/23737807549715\beta^{7 } 
+  \]\[  41677311445977272/512950283341791435\beta^{9 } +   
 238456535650856217955324/3319144833668313129497325\beta^{11 } +   \]\[ 
 59961452714349749109428032/906126539591449484352769725\beta^{13 } +   \]\[ 
 101648942978587880831493976910117/
1599073218845916605769285080647875\beta^{15 } +  \]\[  
 203430414537407786169580316252865558248/
  3198199207108055126786560547703411379875\beta^{17 } +    \]\[
 1770503808520331658366369498591297008751916/
  26797711156358393907344590829206883951972625\beta^{19 } +   \]\[ 
 36841527716991001934503706951700415743468112768/
  519679767005671761231623697769015498362466048125\beta^{21 } +    \]\[
\frac{ 15719394509080615926994542095301655865604593803567699628}
  {201088818343892961175151145921521688462555731267277373125}\beta^{23 } + 
 \]\[  
\frac{ 65139386914524301943373798081580887937411391163315449412064}
{739001407413806632318680461261592205099892312407244346234375}\beta^{25 }+
 \ldots ;\]

\normalsize
  for  spin $S=1$ we have
\scriptsize

\[ G^{sq}(\beta;1 )=4/9\beta^{ } +   2/3\beta^{3 } +   6667/7290\beta^{5 }
 +   1203689/918540\beta^{7 } +    46204667/22044960\beta^{9 } +  \]\[  
 9078206483/2424945600\beta^{11 } +   74910520638409/10213870867200\beta^{13 }
 +    \]\[
 99151316308624243/6434738646336000\beta^{15 } +
   1606930917692748413929/
  47256720618691584000\beta^{17 } +  \]\[  
1255367644480518227446001/16161798451592521728000\beta^{19 } +  \]\[
   1235252363986014705453635129/6787955349668859125760000\beta^{21 } +  \]\[  
40316042891302386155107919977307/92737045987175953376133120000\beta^{23 } +
 \]\[   
 176103666056855156648908280252073049/
166926682776916716077039616000000\beta^{25 }+ \ldots ;\]

\normalsize
  for  spin $S=2$ we have
\scriptsize

\[ G^{sq}(\beta;2 )=1/4\beta^{ } +   2449/9600\beta^{3 }
 +   158549/614400\beta^{5 } +   2757501893/10321920000\beta^{7 } +   \]\[ 
 686898352153/2378170368000\beta^{9 }
 +   274925562580157/837115969536000\beta^{11 } +  \]\[  
 102860665321816423009/261180182495232000000\beta^{13 } +   
 436799902265689091451577/877565413183979520000000\beta^{15 } +    \]\[
 12601867668042988291932285077/19095823390883394355200000000\beta^{17 } +
   \]\[ 
 2380739033401154920978550610184709/
2612308639872848347791360000000000\beta^{19 } +   \]\[ 
 351344127300423485018239380557971837/
270072523999162167648583680000000000\beta^{21 } +  \]\[
   1354683151035248192129414800144202499532681/
  710614825146595495516953378816000000000000\beta^{23 } +    \]\[
 778169719143474473494410060469857176023407817/
  272876092856292670278510097465344000000000000\beta^{25 }+ \ldots ;\]

\normalsize
  for  spin $S=3$ we have
\scriptsize

\[ G^{sq}(\beta;3 )=16/81\beta^{ } +   3272/19683\beta^{3 }
 +   55142018/390609135\beta^{5 } 
+   11664708943/94918019805\beta^{7 } +    \]\[
 61417397550727/553561891502760\beta^{9 } +   25208558053755418927/
  241679586211189988400\beta^{11 } +  \]\[  
911467676611299752881/8903391403395325636800\beta^{13 } +   
 779192054623011781506550921/7420887700815969965016432000\beta^{15 } +  \]\[  
 900071722553226867972685615531201/
8011352893394495063192859461376000\beta^{17 } +    \]\[
 27718218134604228943480241737301971369/
  221930497852814302240568592799037952000\beta^{19 } +    \]\[
 22107242125000401957828466973541074747329/
  154083174223525358412737623000474920960000\beta^{21 } +    \]\[
 4031547346486912973138626505196058860443932093/
  23803645842888085127332211276813522339450880000\beta^{23 } +    \]\[
 \frac{21517412450907114856223112197872662972795241297373287}
 {105274004104756845284139437592835483898455461888000000}\beta^{25 }+ \ldots ;
\]

\normalsize
  for  spin $S=4$ we have
\scriptsize

\[ G^{sq}(\beta;4 )=25/144\beta^{ } +   3209/24576\beta^{3 }
 +   610193015/6115295232\beta^{5 } +   
 77173653478721/986274815016960\beta^{7 } +  \]\[  
76966644305986375/1211934492692840448\beta^{9 } +   
 9129004991161435541627/170640376571151935078400\beta^{11 } +  \]\[  
 345022287361099204644769801/7359869605045165573634260992\beta^{13 } + \]\[   
 1269598636005133846915774036643707/
29674994247542107592893340319744000\beta^{15 } +   \]\[ 
 90670463081344606074551418584391104329/
  2231635535400440198781016998995967344640\beta^{17 } +   \]\[ 
 195510179528760340700971856205445073340905129/
  4884603859884483507091890007402373323948032000\beta^{19 } +  \]\[  
 855479765211125092056795820534553381014177632569/
  21007704280591186667300800543836127191635696025600\beta^{21 } +   \]\[ 
\frac{ 78205243808960288938134345361958571887027610472490804963}
{1836846437641195470391446636991290686029612026251182080000}\beta^{23 } 
+  \]\[  
\frac{ 1541377499643922367827546156319043361983835916725658961481081}
{33856753538602514910255144413023469924897808867861788098560000}\beta^{25 }+ 
\ldots ;\]

\normalsize
  for  spin $S=5$ we have
\scriptsize

\[ G^{sq}(\beta;5 )=4/25\beta^{ } +   131042/1171875\beta^{3 } +   
23341969/292968750\beta^{5 } +   
 11206938159841/192260742187500\beta^{7 } +  \]\[  
73864636161005089/1674975585937500000\beta^{9 } +    
 32318430505151947246253/934696197509765625000000\beta^{11 } +   \]\[ 
 718352640618017304457700699/25517206192016601562500000000\beta^{13 } +  \]\[  
 1995489547673686958264400139481291/
83728332817554473876953125000000000\beta^{15 } +  \]\[   
 1705403640523931467150473252543034493/
  81336094737052917480468750000000000000\beta^{17 } +  \]\[ 
 7038534726821791589579004478921004229151104131/
  368139873544704169034957885742187500000000000000\beta^{19 } +   \]\[
 372025903186092331039340189188879376168821717433/
  20670955466413870453834533691406250000000000000000\beta^{21 } + \]\[  
\frac{ 141011928278517684498811943551667443581498425585788687}
{  8100016350313962902873754501342773437500000000000000000}\beta^{23 } + \]\[  
\frac{ 717846184697087160194012195983695606989379300548453191241209}
{  41674584122365339135285466909408569335937500000000000000000000}\beta^{25 }+
 \ldots ;\]

\normalsize
  for  spin $S=\infty$ we have
\scriptsize

\[ G^{sq}(\beta; \infty)=1/9\beta^{ } +   37/675\beta^{3 }
 +   24784/893025\beta^{5 } +   2902951/200930625\beta^{7 } +   
 113284504/14587563375\beta^{9 } + \]\[   
39089225413804/9059970923128125\beta^{11 } +   
 61932079507744/24961144380046875\beta^{13 } + \]\[  23523494724734209913/
  15906364450462970859375\beta^{15 } +   34701383932521722281609816/
  37984636903172331356750390625\beta^{17 }
 + \]\[  603767152738540662108159101404/
  1034131739688866721187529384765625\beta^{19 }
 + \]\[   57409874526515597913036541772128/
  149197006444202862411328103056640625\beta^{21 } + \]\[  
 48773386285997925513086743266213915964/
  187216133611346856825294786918049072265625\beta^{23 } + \]\[  
 49827201787066781774584268457305739008/
  276128143553859410902859535220266357421875\beta^{25 }+ \ldots ;\]

\normalsize

\section{The susceptibility on the square lattice}

\normalsize

On the sq lattice, the HT expansion coefficients of the susceptibility 
$\chi^{sq}(\beta;S) = \sum_{r=0}^\infty c_r^{sq}(S) \beta^r $ 
  for the spin $S=1/2$ Ising model are

\scriptsize

\[ \chi^{sq}(\beta;1/2 )=1 + 4 \beta^{ }+ 12 \beta^{2 }+ 104/3 \beta^{3 }+ 
92 \beta^{4 }
+ 3608/15 \beta^{5 }+ 3056/5 \beta^{6 }+   484528/315 \beta^{7 }+ 400012/105
\beta^{8 }
+ 26548808/2835 \beta^{9 }+  \]\[ 
  107828128/4725 \beta^{10 }+ 
 8625391648/155925 \beta^{11 }+ 6926560288/51975 \beta^{12 }
+ 1945106545216/6081075 \beta^{13 }+ \]\[  
  10844734860976/14189175 \beta^{14 }+  8130987171952/4465125 \beta^{15 }
+ 920355608477852/212837625 \beta^{16 }+  \]\[ 
  10109628948182728/986792625 \beta^{17 }+ 
 29198178356609824/1206079875 \beta^{18 }+ 
  9635362499516230208/168741538875 \beta^{19 }+ \]\[ 
 415782048556042942544/3093594879375 \beta^{20 }+ 
  4735391065845611373232/14992036723125 \beta^{21 }+ \]\[ 
  529562920319138348552816/714620417135625 \beta^{22 }+ 
  85616154520095267692857616/49308808782358125 \beta^{23 }+  \]\[
  66773068948180944546678128/16436269594119375 \beta^{24 }+  
  3192145249472459217984684656/336196423516078125\beta^{25 } +      \ldots ;\]

\normalsize
  for  spin $S=3/2$ we have
\scriptsize

\[ \chi^{sq}(\beta;3/2 )=5/9 + 100/81 \beta^{ }+ 1660/729 \beta^{2 }+ 
81512/19683 \beta^{3 }
+ 1246660/177147 \beta^{4 }+ 18903160/1594323 \beta^{5 }+  \]\[
  832962352/43046721 \beta^{6 }+ 
  425532036784/13559717115 \beta^{7 }
+  1220802674948/24407490807 \beta^{8 }+ \]\[
  156804173480680/1977006755367 \beta^{9 }+
 2214390946267424/17793060798303 \beta^{10 }+\]\[ 
  8564391025518912544/44037825475799925 \beta^{11 }+ 
  71771456562578890528/237804257569319595 \beta^{12 }+ \]\[  
  2597775308518453343296/5564619627122078523 \beta^{13 }+  
  1259594660861192999294096/1752855182543454734745 \beta^{14 }+\]\[   
  1305294470581288253711166608/1183177248216831945952875 \beta^{15 }+ \]\[  
  718813751627813555580129316/425943809358059500543035 \beta^{16 }+ \]\[ 
  167874391518573832702113231128/65169402831783103583084355 \beta^{17 }+ \]\[ 
  103491884594913656246941138771936/26393608146872156951149163775 \beta^{18 }
+ \]\[ 
  12223137477235550630605671904629824/2051503178688699472112048638875
\beta^{19 }+ \]\[ 
  1834673160539690469858340871232977392/203098814690181247739092815248625
\beta^{20 }+ \]\[ 
  104991553074734012041413486716735410672/7677135195288851164537708416398025
\beta^{21 }+  \]\[
  78537933084999510402979180027801273673168/
   3800181921667981326446165666117022375 \beta^{22 }+ \]\[ 
  122687843372242930181812633343547747696101648/
   3933188288926360672871781464431118158125 \beta^{23 }+  \]\[
  998448638893877113233403310358995680977635824/
   21239216760202347633507619907928038053875 \beta^{24 }+  \]\[
  2940902156480564049561839939481179798123342768/
   41554989313439375804688821558989639670625\beta^{25 }+   \ldots ;\]

\normalsize
  for  spin $S=5/2$ we have
\scriptsize

\[ \chi^{sq}(\beta;5/2 )=7/15 + 196/225 \beta^{ }+ 115444/84375 \beta^{2 }
+ 67191544/31640625 \beta^{3 }+ 1467970756/474609375 \beta^{4 }+ 
 \]\[ 159159295496/35595703125 \beta^{5 }+ 83670619605776/13348388671875 
\beta^{6 }+ \]\[ 
  43738653398946128/5005645751953125 \beta^{7 }+ 
 898952840019390212/75084686279296875 \beta^{8 }+\]\[ 
  18390544873067954552/1126270294189453125 \beta^{9 }+  \]\[
  9306398597377304096864/422351360321044921875 \beta^{10 }+  \]\[
  1806243329502940793356250144/60976977646350860595703125 \beta^{11 }+ \]\[
  5166675260889697220402511968/130664952099323272705078125 \beta^{12 }+ \]\[ 
  1340734028723741702128385701184/25479665659368038177490234375 \beta^{13 }+
  \]\[
  665748312453898446630489586492496/9554874622263014316558837890625
\beta^{14 }+  \]\[
  1648738594433777028722362262500465168/17915389916743151843547821044921875
\beta^{15 }+ \]\[ 
  227814108361779586207533783583901072884/
   1881115941258030943572521209716796875 \beta^{16 }+  \]\[
  534021144150889173342876764955361476987608/
   3357791955145585234276950359344482421875 \beta^{17 }+  \]\[
  37462702931909585099584420205288618835477536/
   179881711882799208979122340679168701171875 \beta^{18 }+  \]\[
  49847871974871410495034119429734220793540214976/
   183093885309277766282320953905582427978515625 \beta^{19 }+  \]\[
  51329542513024532424691469521628041936421685104/
   144547804191535078643937595188617706298828125 \beta^{20 }+ \]\[ 
  666863355719953177463728536266957689176916020779344/
   1441864346810562409473277512006461620330810546875 \beta^{21 }+ \]\[ 
\frac{25033446554042550218612963546371575545460375428097368816}
   {41633833014154989573540888159186579287052154541015625} \beta^{22 }+  \]\[
\frac{1962492488318410478406889384860257908932513620239086775159216}
{2513642668229607495502531122610889724455773830413818359375} \beta^{23 }
+  \]\[
\frac{5452207218298190252797200447882850351016484614803999335225712}
{5386377146206301776076852405594763695262372493743896484375} \beta^{24 }
+  \]\[
\frac{378286527712599384181809139578084785964397079215201170473348176}
{288555918546766166575545664585433769389055669307708740234375}\beta^{25 }+
   \ldots ;\]
\normalsize
  for  spin $S=7/2$ we have
\scriptsize

\[ \chi^{sq}(\beta;7/2 )=3/7 + 36/49 \beta^{ }+ 2556/2401 \beta^{2 }
+ 179448/117649 \beta^{3 }+ 82882932/40353607 \beta^{4 }+\]\[ 
  190046638536/69206436005 \beta^{5 }+
 1725611880976/484445052035 \beta^{6 }+  
  109086602712112/23737807549715 \beta^{7 }+ 
\]\[   6779055367754612/1163152569936035 \beta^{8 }+ 
  3774543024934879688/512950283341791435 \beta^{9 }+ \]\[ 
  8086776445809091876576/879709735931172311025 \beta^{10 }+\]\[    
  37975521176898263039787808/3319144833668313129497325 \beta^{11 }+  \]\[
  985578365101400903358065248/69702041507034575719443825 \beta^{12 }+ \]\[ 
  2255230349421196875105730624/129446648513064212050395675 \beta^{13 }+  \]\[
603491178296568493651471623568/28254673007260652103000001425 \beta^{14 }+  \]\[
41768937116245274217112598155395344/1599073218845916605769285080647875
\beta^{15 }+  \]\[
  17455822912696365646848174840890751164/
   548482114064149395778864782662221125 \beta^{16 }+  \]\[
  123762314101985756924844123918603109666712/
   3198199207108055126786560547703411379875 \beta^{17 }+  \]\[
  1350503799046684667595063415448461973961568/
   28783792863972496141079044929330702418875 \beta^{18 }+  \]\[
  1521847577408850479492127177833643828382274752/
   26797711156358393907344590829206883951972625 \beta^{19 }+  \]\[ 
  450276781496340931659733124827843408670976426544/
   6565439233307806507299424753155686568233293125 \beta^{20 }+  \]\[
\frac{558747229815967918272509104679476816149457147745776}
  {6755836971073732896011108070997201478712058625625}\beta^{21 }+  \]\[
\frac{2537603226536820979889437109114979959231556813779309552}
   {5489772891861194216649910751872441179180597194483125}\beta^{22 }+  \]\[
\frac{24067923170469933836194853111678957625431650380294204757136}
{201088818343892961175151145921521688462555731267277373125} \beta^{23 }+ \]\[ 
\frac{55156416568188926153952440052170620862795213303279446554608}
{383896835020159289516197642213814132519424577873893166875} \beta^{24 }+  \]\[
  \frac{127322083795991568869056558984095239508092099477210579394413776}
 {739001407413806632318680461261592205099892312407244346234375}\beta^{25 }+ 
  \ldots ;\]

\normalsize
  for  spin $S=1$ we have
\scriptsize

\[ \chi^{sq}(\beta;1 )=2/3 + 16/9 \beta^{ }+ 104/27 \beta^{2 }
+ 664/81 \beta^{3 }
+ 3938/243 \beta^{4 }+ 115574/3645 \beta^{5 }+ 656587/10935 \beta^{6 }+  
  25923041/229635 \beta^{7 }+ \]\[ 1149698483/5511240 \beta^{8 }+ 
6337163281/16533720 \beta^{9 }+ 
  345754537763/496011600 \beta^{10 }+ 20657190676169/16368382800 \beta^{11 }
+  \]\[
  445804543457209/196420593600 \beta^{12 }
+ 31158644648901227/7660403150400 \beta^{13 }+ 
  16323535856666527/2249908617600 \beta^{14 }+ \]\[
 62282994349127515121/4826053984752000 \beta^{15 }+ 
  5299219029060247425763/231650591268096000 \beta^{16 }+  \]\[
  477999765992285118676009/11814180154672896000 \beta^{17 }+ 
  5058727374582200667543763/70885080928037376000 \beta^{18 }+  \]\[
  507690889276999207152864089/4040449612898130432000 \beta^{19 }+ \]\[ 
  53516700222258234369076992347/242426976773887825920000 \beta^{20 }+  \]\[
  1971519171391119362379195677467/5090966512251644344320000 \beta^{21 }+ \]\[ 
  227858729669257025618425144401761/
336003789808608526725120000 \beta^{22 }+ \]\[ 
  27497362898232739962632331076025603/
23184261496793988344033280000 \beta^{23 }+  \]\[
  9524691852860876575809034199871083/
4598531205975667109560320000 \beta^{24 }+  \]\[
  150788006235698923753947566239789820089/
41731670694229179019259904000000\beta^{25 }+   \ldots ;\]

\normalsize
  for  spin $S=2$ we have
\scriptsize

\[ \chi^{sq}(\beta;2 )=1/2 + 1 \beta^{ }+ 67/40 \beta^{2 }
+ 6649/2400 \beta^{3 }
+ 164873/38400 \beta^{4 }+ 1014053/153600 \beta^{5 }
+ 181370369/18432000 \beta^{6 }+ \]\[ 
  37624622237/2580480000 \beta^{7 }+ 1753399317637/82575360000 \beta^{8 }+ 
  91486793499677/2972712960000 \beta^{9 }+ \]\[
 5247947624188069/118908518400000 \beta^{10 }+ 
  1649234510879444113/26159874048000000 \beta^{11 }+  \]\[
  112292162542674887513/1255673954304000000 \beta^{12 }+ 
  8256642480783153764113/65295045623808000000 \beta^{13 }+  \]\[
  650573640394045709455481/3656522554933248000000 \beta^{14 }+  \]\[
  273910367897394464882415701/1096956766479974400000000 \beta^{15 }+ \]\[ 
  24512608354235862979525434109/70205233054718361600000000 \beta^{16 }+  \]\[
  2325886270755089765565842182613/4773955847720848588800000000 \beta^{17 }
+ \]\[ 
  105916636445131618707103111683283/156238555016318681088000000000 \beta^{18 }
+  \]\[
  614968503441623770992082049391656909/653077159968212086947840000000000
\beta^{19 }+ \]\[ 
  13636827597257675908753300674238672433/10449234559491393391165440000000000
 \beta^{20 }+  \]\[
  1585267062889503930552858180158939511721/
   877735702997277044857896960000000000 \beta^{21 }+ \]\[ 
  963549797072811827244431626197030280811413/
   386203709318801899737474662400000000000 \beta^{22 }+  \]\[
  24461147637517072043019132625435851516004009/
   7106148251465954955169533788160000000000 \beta^{23 }+  \]\[
  6221188738099381935618722274154049788740761837/
   1311904292578330145569760083968000000000000 \beta^{24 }+  \]\[
  445192462083565055391188189433243562204355250217/
   68219023214073167569627524366336000000000000\beta^{25 }+   \ldots ;\]

\normalsize
  for  spin $S=3$ we have
\scriptsize

\[ \chi^{sq}(\beta;3 )=4/9 + 64/81 \beta^{ }+ 32/27 \beta^{2 }+ 34592/19683 
\beta^{3 }
+ 1011784/413343 \beta^{4 }+ 1322056712/390609135 \beta^{5 }+  \]\[ 
  15958699324/3515482215 \beta^{6 }+  574745995708/94918019805 \beta^{7 }+ 
  1507192862971/189836039610 \beta^{8 }+\]\[ 
 204875089436161/19770067553670 \beta^{9 }+  
  388943349703560671/29061999303894900 \beta^{10 }+ \]\[ 
  1040390659576943784463/60419896552797497100 \beta^{11 }+  
  47850400902776295315697/2175116275900709895600 \beta^{12 }+ \]\[ 
  436573317203108078107927/15580934955941819864400 \beta^{13 }+ \]\[ 
  2581996364449332897421297/72711029794395159367200 \beta^{14 }+ \]\[ 
  83283484723239415184409169129/1855221925203992491254108000 \beta^{15 }+\]\[  
  5035009556856503954153574940151/89050652409791639580197184000 \beta^{16 }+
  \]\[
  142336601165096837008610672196826561/
2002838223348623765798214865344000 \beta^{17 }+  \]\[ 
  9633427518357863923658359873294499599/
108153264060825683353103602728576000 \beta^{18 }+ \]\[ 
  883340951328218413151078834984416445743/
   7926089209029082222877449742822784000 \beta^{19 }+  \]\[  
  22054920639151265976867396685907816174939/
   158521784180581644457548994856455680000 \beta^{20 }+  \]\[
  46766317332185369766453662034575450745743623/
   269645554891169377222290840250831111680000 \beta^{21 }+  \]\[
  26884455339558609632975911614739902880096460159/
   124576246359720252276698368195883973596160000 \beta^{22 }+  \]\[
  145240307262369115199919546243981446199535412249167/
   541532942925703936646807806547507633222507520000 \beta^{23 }+  \]\[
  12976423859036344681633067372252757656981025960889873/
   38990371890650683438570162071420549592020541440000 \beta^{24 }+  \]\[
\frac{ 67439387606722087762085803172777199219689945606262407}
 {163468950473224915037483598746638950152881152000000}\beta^{25 }+   \ldots ;\]
\normalsize
  for  spin $S=4$ we have
\scriptsize

\[ \chi^{sq}(\beta;4 )=5/12+ 25/36 \beta^{ }+ 3385/3456 \beta^{2 }
+ 226643/165888 \beta^{3 }
+ 57070525/31850496 \beta^{4 }+ 
  3567477815/1528823808 \beta^{5 }
+ \]\[ 432763246139/146767085568 \beta^{6 }+  
  913798397355041/246568703754240 \beta^{7 }+
 \]\[ 86713655787899419/18936476448325632 \beta^{8 }+ 
  5120094300386606165/908950869519630336 \beta^{9 }+ \]\[ 
  598260509519857768967/87259283473884512256 \beta^{10 }+ 
  9576634223917961316715289/1151822541855275561779200 \beta^{11 }+ \]\[ 
  63258473583359885895445847/6318569372463225938903040 \beta^{12 }+ \]\[ 
  66343569048972638222231822939/5519902203783874180225695744 \beta^{13 }+
 \]\[  
  266273832826692278273147098557313/18546871404713817245558337699840
\beta^{14 }+  \]\[
  34612806179135461758917891601668371/
2023295062332416426788182294528000 \beta^{15 }+  \]\[
  69426625634806042612873623113499250667/
   3418559337316850794701312804834508800 \beta^{16 }+  \]\[
  1220322468865715466631367653637562265979/
   50718989440919095426841295431726530560 \beta^{17 }+  \]\[
  1087399852749665991270845192384938458544949/
   38256609178293260550531719982788011622400 \beta^{18 }+  \]\[
  40934915317587848893053282194212622986253867849/
   1221150964971120876772972501850593330987008000 \beta^{19 }+  \]\[
  9248087708195995243517427259467654296221866693619/
   234460985274455208340410720355313919549505536000 \beta^{20 }+  \]\[
  146045285713711300005910665818129417556648718087343/
   3151155642088678000095120081575419078745354403840 \beta^{21 }+  \]\[
\frac{4521970521731046876773755898752476664939545285894505969}
   {83190508951141099202511170153591063678877356261376000} \beta^{22 }+  \]\[
\frac{29239864698324206104502270765093245977936672130571229829763}
{459211609410298867597861659247822671507403006562795520000} \beta^{23 }+ \]\[ 
\frac{75043678730400079645349794189141550807208914379336446837513}
{1007641474363170086614736440863793747764815740114934169600} \beta^{24 }+ 
 \]\[
 \frac{66956543482695033468966490917316595291874786123440781000961931}
 {769471671331875338869435100295987952838586565178677002240000}\beta^{25 }+
   \ldots ;\]

\normalsize
  for  spin $S=5$ we have
\scriptsize

\[ \chi^{sq}(\beta;5 )=2/5 + 16/25 \beta^{ }+ 2712/3125 \beta^{2 }
+ 1364168/1171875 \beta^{3 }+ 2868846/1953125 \beta^{4 }+  
  269618498/146484375 \beta^{5 }+\]\[ 40984657759/18310546875 \beta^{6 }+  
  130141778113561/48065185546875 \beta^{7 }
+ 22699700846250229/7049560546875000 \beta^{8 }+  \]\[
  39913033966137584473/10468597412109375000 \beta^{9 }+ 
  1298898154869049305481/290794372558593750000 \beta^{10 }+  \]\[
  775338563329361335085401/148701667785644531250000 \beta^{11 }+ \]\[ 
  39541366817118422395549873/6542873382568359375000000 \beta^{12 }+ \]\[ 
  44552238334924755115118526827/6379301548004150390625000000 \beta^{13 }+  \]\[
  29882618309677956285183855702901/3721259236335754394531250000000 \beta^{14 }
+  \]\[
  14830864619337874514578876363837727/1610160246491432189941406250000000 
\beta^{15 }+ \]\[ 
  652810202870499074094920311793929229/62020987272262573242187500000000000
 \beta^{16 }+  \]\[
  1708601037253456599421548637928816816171/
   142338165789842605590820312500000000000 \beta^{17 }+  \]\[
  16029527954687774747093620973041951641265927/
   1174289867766201496124267578125000000000000 \beta^{18 }+  \]\[
  1426231973348797814787490419667524918912843409211/
   92034968386176042258739471435546875000000000000 \beta^{19 }+  \]\[
  53848536509195646693073249377526693442631610849357/
   3067832279539201408624649047851562500000000000000 \beta^{20 }+ \]\[ 
  26832720376572572114452376146918908291097827101657/
   1351562472803983837366104125976562500000000000000 \beta^{21 }+ 
 \]\[
\frac{601662594892674083266485018959191442632291061417973421}
 {26843532445968012325465679168701171875000000000000000} \beta^{22 }+
  \]\[
\frac{17554625986420454083120726552438068308407227707019751212241}
{694576402039422318921424448490142822265625000000000000000} \beta^{23 }+
  \]\[
\frac{790579850061843094274876784454355309972689413129065982579463}
 {27783056081576892756856977939605712890625000000000000000000} \beta^{24 }+
  \]\[
\frac{333440742764073647737957744546498971274909600185246883674767449}
{10418646030591334783821366727352142333984375000000000000000000}\beta^{25}
+ \ldots ;\]
\normalsize
  for  spin $S=\infty$ we have
\scriptsize

\[ \chi^{sq}(\beta;\infty )=1/3 + 4/9 \beta^{ }+ 68/135 \beta^{2 }
+ 1144/2025 \beta^{3 }+ 25372/42525 \beta^{4 }+ 558952/893025 \beta^{5 }
+ \]\[ 8538752/13395375 \beta^{6 }+  
  129772336/200930625 \beta^{7 }+ 609461588/947244375 \beta^{8 }
+ 139652513752/218813450625 \beta^{9 }+ \]\[ 
  187100500124624/298680360103125 \beta^{10 }+ 
 22707256491351296/37063517412796875 \beta^{11 }+ \]\[ 
  727688359670109184/1223096074622296875 \beta^{12 }+  
  301969527280546432/524184031980984375 \beta^{13 }+ \]\[  
  103857542295122705168/187133699417211421875 \beta^{14 }+ 
  127264034209920737628784/238595466756944562890625 \beta^{15 }+  \]\[
  287690260266202572024676/563311190745685683984375 \beta^{16 }+ \]\[ 
  18537343856347832243549645896/37984636903172331356750390625 \beta^{17 }+
 \]\[ 
  2914365965366182018219800490384/6267465089023434673863814453125 \beta^{18 }
+  \]\[
  65345300052223877461553834064928/
147733105669838103026789912109375 \beta^{19 }+ \]\[ 
  4279144615315792327975513332471056/
10193584291218829108848503935546875 \beta^{20 }+ \]\[ 
  16188155599111527872631129341708656/
40690092666600780657634937197265625 \beta^{21 }+  \]\[
  2529237184729700018724544702552498575248/
   6720579155279117937318274402186376953125 \beta^{22 }+  \]\[
  652382502806003152776824755424575089659408/ 
   1834718109391199196887888911796880908203125 \beta^{23 }+  \]\[
  543035840542536669599528870539392049496624/
   1618868920051058114901078451585483154296875 \beta^{24 }+  \]\[
  3728495487078091052200511853170450534497744/
   11794616417514851979993571575837091552734375\beta^{25 }+ 
  \ldots ;\]
\normalsize

\section{The second correlation moment on the   square lattice}

\normalsize

On the sq lattice, the HT expansion coefficients of the second 
 correlation moment 
$\mu_{2}^{sq}(\beta;S) = \sum_{r=0}^\infty c_r^{sq}(S) \beta^r $ 
  for the spin $S=1/2$ Ising model are

\scriptsize

\[\mu^{sq}_2(\beta;1/2 )= 4\beta^{ }+ 32\beta^{2 }+ 488/3\beta^{3 }
+ 2048/3\beta^{4 }+ 38168/15\beta^{5 }+   394624/45\beta^{6 }
+ 8994736/315\beta^{7 }+\]\[  28064768/315\beta^{8 }+  
 761977928/2835\beta^{9 }+ 11186723584/14175\beta^{10 }
+ 353321963296/155925\beta^{11 }+  \]\[
 2986983755776/467775\beta^{12 }+ 107692763046976/6081075\beta^{13 }+  \]\[
2062173184868288/42567525\beta^{14 }+ 
 83609340468743312/638512875\beta^{15 }+  \]\[
223627234152593408/638512875\beta^{16 }+ 
 774906722333778296/834978375\beta^{17 }+  \]\[
238255570273199826304/97692469875\beta^{18 }+ 
 11807573369344881466816/1856156927625\beta^{19 }+  \]\[
152956234782459128486656/9280784638125\beta^{20 }+ 
8271153229057812207708016/194896477400625\beta^{21 }+  \]\[
 232979217196127169315240128/2143861251406875\beta^{22 }+ 
 13651702825360397567324321552/49308808782358125\beta^{23 }+  \]\[
 103851766888965746884178039296/147926426347074375\beta^{24 }+  \]\[
 6555143295186718737306258649936/3698160658676859375\beta^{25}+\ldots ; \]

\normalsize
  for  spin $S=3/2$ we have
\scriptsize

\[\mu^{sq}_2(\beta;3/2 )= 100/81\beta^{ }+ 4000/729\beta^{2 }+ 
321512/19683\beta^{3 }+  7336960/177147\beta^{4 }+ 
151171960/1594323\beta^{5 }+ \]\[
 8747144320/43046721\beta^{6 }+  5619988253104/13559717115\beta^{7 }+ 
19883967976448/24407490807\beta^{8 }+  \]\[
 3073001952304360/1977006755367\beta^{9 }+ 
51522201620963840/17793060798303\beta^{10 }+  \]\[
 232838218145832157984/44037825475799925\beta^{11 }+ 
2257495114191013931008/237804257569319595\beta^{12 }+  \]\[
93482514309943802617408/5564619627122078523\beta^{13 }+ 
 10292604711339908068518592/350571036508690946949\beta^{14 }+  \]\[
 60045664643376523073772829328/1183177248216831945952875\beta^{15 }+  \]\[
 37005137061792711053356306432/425943809358059500543035\beta^{16 }+  \]\[
 9609086914497404940087523789208/65169402831783103583084355\beta^{17 }+ \]\[ 
 1310818269945601381474555713360512/
5278721629374431390229832755\beta^{18}+  \]\[
 9371905193095267018570915766994142144
/22566534965575694193232535027625\beta^{19 }+  \]\[
 140178513162804236221138156780588172032/
203098814690181247739092815248625\beta^{20 }+  \]\[
 8755668232532131740423476743258309725808/
7677135195288851164537708416398025\beta^{21 }+  \]\[
 109601946761940123143544035596100706772672/
  58464337256430481945325625632569575\beta^{22 }+  \]\[
 12061906928074151223255417000628476671338968848/
  3933188288926360672871781464431118158125\beta^{23 }+  \]\[
 106078432106065313026373535950506304534195321344/
  21239216760202347633507619907928038053875\beta^{24 }+  \]\[
 7742397908093881256218164615301749066288434419024/
  955764754209105643507842895856761712424375\beta^{25}+ \ldots ;\]

\normalsize
  for  spin $S=5/2$ we have
\scriptsize

\[\mu^{sq}_2(\beta;5/2 )= 196/225\beta^{ }+ 10976/3375\beta^{2 }+ 
86423848/10546875\beta^{3}+  8378288128/474609375\beta^{4 }+ \]\[
 1225344666376/35595703125\beta^{5 }
+ 56030208192896/889892578125\beta^{6 }+ \]\[ 
 549385449492223568/5005645751953125\beta^{7 }+ 
13858697176580365312/75084686279296875\beta^{8}+  \]\[
 113207700193006626088/375423431396484375\beta^{9 }+ 
40658773108256913105664/84470272064208984375\beta^{10 }+  \]\[
45944615275033211742722577184/60976977646350860595703125\beta^{11 }+\]\[
 50537611100584193898256125952/43554984033107757568359375\beta^{12 }+  \]\[
44889278829720679134915516667712/25479665659368038177490234375\beta^{13 }+\]\[ 
5049836801602077524258869518163904/
1910974924452602863311767578125\beta^{14}+ \]\[
 23417426414534742681718300680167109296/
5971796638914383947849273681640625\beta^{15 }+ \]\[
61960516081710243397555892482435237888/
10749233950045891106128692626953125\beta^{16 }+  \]\[
 28211496624531632260552170622963320144203608/
  3357791955145585234276950359344482421875\beta^{17 }+ \]\[
 145754012663595180534108991152504974512774784/
  11992114125519947265274822711944580078125\beta^{18 }+ \]\[ 
 3197315676053299443513270610069047612274607132096/
  183093885309277766282320953905582427978515625\beta^{19 }+ \]\[
 342412052320914137684815200031449569036951896407296/
  13732041398195832471174071542918682098388671875\beta^{20 }+ \]\[ 
 17016520315518773765791671665752238219790672176341232/
  480621448936854136491092504002153873443603515625\beta^{21 }+ \]\[ 
\frac{ 2379699095556908997386916662750456516166429642177057472}
{  47581523444748559512618157896213233470916748046875}\beta^{22 }+ \]\[ 
\frac{176744716878166853779232505254657354053020935198995747805725616}
{2513642668229607495502531122610889724455773830413818359375}\beta^{23 }+  \]\[
\frac{176719081671437696871960488847942094351423950058967093746652672}
{1795459048735433925358950801864921231754124164581298828125}\beta^{24 }+ \]\[ 
\frac{39595901924275626309985550010068201679270795844065714982608273616}
{288555918546766166575545664585433769389055669307708740234375}\beta^{25}
+\dots\]

\normalsize
  for  spin $S=7/2$ we have
\scriptsize

\[\mu^{sq}_2(\beta;7/2 )=36/49\beta^{ }+ 864/343\beta^{2 }+ 
687480/117649\beta^{3 }+ 9556992/823543\beta^{4 }+ 
1444931401416/69206436005\beta^{5 }+  \]\[
 17079820514688/484445052035\beta^{6 }+ 
269498469292400/4747561509943\beta^{7 }+\]\[ 
 42733336962048/484445052035\beta^{8 }+ 
68326132131546502088/512950283341791435\beta^{9 }+ \]\[
 8234531405026730142208/41890939806246300525\beta^{10 }+ 
188880031737141308849085344/663828966733662625899465\beta^{11 }+\]\[ 
9415115210400538241089767424/23234013835678191906481275\beta^{12 }+ \]\[ 
515682142352365657924679124928/906126539591449484352769725\beta^{13 }+\]\[ 
81808352278734772420352001887168/
103600467693289057711000005225\beta^{14 }+ \]\[
346701645272557688921790365407117264/
319814643769183321153857016129575\beta^{15 }+ \]\[
38507888332358182953332855657797308416/
  26118195907816637894231656317248625\beta^{16 }+ \]\[
 6359494106675527178071952213414642404179992/
  3198199207108055126786560547703411379875\beta^{17 }+ \]\[
 178770821633044002498716060313171028499637376/
  67162183349269157662517771501771638977375\beta^{18 }+ \]\[
 159391310506664141168708982957969167358459712/
  45038170010686376314864858536482157902475\beta^{19 }+ \]\[
 602039883218443425922428601857210935907004042496/
  128734102613878558966655387316778168004574375\beta^{20 }+ \]\[
 41517158153577220291150797731111134687226646284481136/
  6755836971073732896011108070997201478712058625625\beta^{21 }+\]\[ 
\frac{ 9751920721803193656217320350737945860178424707093426496}
{1213798709136247343649995750089163865675266533070625}\beta^{22 }+ \]\[
\frac{ 420449105798071857928554986892307051332222925299620842495312}
{40217763668778592235030229184304337692511146253455474625}\beta^{23}+ \]\[
\frac{19065660073422542509827973833382504104648959509904382545857024}
{1407621728407250728226058021450651819237890118870941611875}\beta^{24 }+\]\[ 
\frac{12916352226053316808795694195248561646213550523977976212670213456}
{39001407413806632318680461261592205099892312407244346234375}
\beta^{25}\ldots ;\]

\normalsize
  for  spin $S=1$ we have
\scriptsize

\[\mu^{sq}_2(\beta;1)= 16/9\beta^{ }+ 256/27\beta^{2}+ 904/27\beta^{3}
+ 24320/243\beta^{4}+ 980854/3645\beta^{5}+ 2461376/3645\beta^{6}+ \]\[ 
 369648353/229635\beta^{7}+ 508702816/137781\beta^{8}+ 
45252925787/5511240\beta^{9}+ 
 552210632732/31000725\beta^{10}+ \]\[
88505425902511/2338340400\beta^{11}+ 16171349009218/204604785\beta^{12 }+ \]\[
 1245544003290163307/7660403150400\beta^{13 }+ 
1659972584243816071/5027139567450\beta^{14 }+ \]\[
 1066627622966649044731/1608684661584000\beta^{15 }+ 119281470867675431443/
  90488512214100\beta^{16 }+ \]\[
4384034168685468403383887/1687740022096128000\beta^{17 }+ 
 1874869412655938868022001/369193129833528000\beta^{18 }+ \]\[
 39828744775882343893543285337/4040449612898130432000\beta^{19 }+ 
 130933908242378115716013673/6887130021985449600\beta^{20 }+ \]\[
 61857060978357561290680909233929/1696988837417214781440000\beta^{21 }+ \]\[
 19208320588141289146888519876169/276318906092605696320000\beta^{22 }+ \]\[
 436924719364729901716379848165336357/
3312037356684855477719040000\beta^{23 }+ \]\[
 2777852460091521203424450932795711/
11146279565766340550016000\beta^{24 }+ \]\[
 1150929936359075978690696623077859785257/
2454804158484069354074112000000\beta^{25}+\ldots ;\]

\normalsize
  for  spin $S=2$ we have
\scriptsize

\[\mu^{sq}_2(\beta;2 )= \beta^{ }+ 4\beta^{2 }+ 25849/2400\beta^{3 }+ 
7429/300\beta^{4 }+ 7909157/153600\beta^{5 }+ 1156127/11520\beta^{6 }+  \]\[
 480809315741/2580480000\beta^{7 }+ 107610282541/322560000\beta^{8 }+ 
 344907886084921/594542592000\beta^{9 }+  \]\[
1830781874227837/1857945600000\beta^{10 }+ 
 8586070515696255773/5231974809600000\beta^{11 }+ \]\[
753481611301721849/280284364800000\beta^{12 }+ 
 283470132967498149228817/65295045623808000000\beta^{13 }+ \]\[
 1978297025943725674711039/285665824604160000000\beta^{14 }+ \]\[ 
 11987467366885172752753990997/1096956766479974400000000\beta^{15 }+ \]\[
 2342064312286041962029477013/137119595809996800000000\beta^{16 }+  \]\[
 126369200820404870784544946956757/
4773955847720848588800000000\beta^{17 }+ \]\[
 22311202841206997532585552658781/548030645784281088000000000\beta^{18 }+  \]\[
 40611620928599173319284433660332163789/
653077159968212086947840000000000\beta^{19 }+ \]\[
 90657730666342554269283537023455249/
960407588188547186688000000000\beta^{20 }+  \]\[
 125057820926169722485700149323351551025577/
  877735702997277044857896960000000000\beta^{21 }+\]\[ 
 67947742047876633132569038129325960897017/
  317601734637172614915686400000000000\beta^{22 }+ \]\[
 56799077903116789645487395441207941867024094081/
  177653706286648873879238344704000000000000\beta^{23 }+ \]\[
 4527341302044898518442476422248911205387237013/
  9517162836784761100673482752000000000000\beta^{24 }+ \]\[
 48089096890223440252344251761314431972160501718057/
  68219023214073167569627524366336000000000000\beta^{25}+\ldots ;\]
\normalsize
  for  spin $S=3$ we have
\scriptsize

\[\mu^{sq}_2(\beta;3 )= 64/81\beta^{ }+ 2048/729\beta^{2 }+ 
132896/19683\beta^{3 }+ 2459648/177147\beta^{4 }+ 
10100849672/390609135\beta^{5 }+ \]\[
 476603121664/10546446645\beta^{6 }+ 7146029210044/94918019805\beta^{7 }+\]\[ 
 14770605406208/122037454035\beta^{8}+ 
26164027625710087/138390472875690\beta^{9 }+ \]\[
 128368536721643776/444826519957575\beta^{10 }+ 26102923847098637161999/
  60419896552797497100\beta^{11 }+ \]\[
259517541325777872667648/407834301731383105425\beta^{12 }+ 
 705576875167326779526908263/763465812841149173355600\beta^{13 }+ \]\[
 569468737170049041814590368/429449519723146410012525\beta^{14 }+ \]\[
 3492363940164669486166029615337/1855221925203992491254108000\beta^{15 }+\]\[ 
 9669166965137437142337075132704/3652468165245360217156525125\beta^{16 }+ \]\[
 7393959844176819964495871891614242241/
2002838223348623765798214865344000\beta^{17 }+\]\[ 
 1619028086975275830492996151238392681/
316855265803200244198545711118875\beta^{18 }+ \]\[
 389727371016370736214570451655647652597833/
  55482624463203575560142148199759488000\beta^{19 }+ \]\[
 23399765179303484006653791286006725185026/
  2438201270355625879107809247059743125\beta^{20 }+\]\[ 
 3515889904392724511541181038716775437362073863/
  269645554891169377222290840250831111680000\beta^{21 }+ \]\[
 6432362632112069975058850838612974642512064861/
  364969471756992926591889750573878828895000\beta^{22 }+ \]\[
 1834274171048857671666329925577834347631428536216953/
  77361848989386276663829686649643947603215360000\beta^{23 }+ \]\[
 453484948114714027833387686996441697269028224318201/
  14278700643548834267054502711701861422859085000\beta^{24 }+ \]\[
\frac{ 1115131406895017286748663596417904154097640148534714323687}
{26318501026189211321034859398208870974613865472000000}\beta^{25 }+\ldots ;\]

\normalsize
  for  spin $S=4$ we have
\scriptsize

\[\mu^{sq}_2(\beta;4 )= 25/36\beta^{ }+ 125/54\beta^{2 }+ 
288881/55296\beta^{3 }+
 1256215/124416\beta^{4 }+  27033922615/1528823808\beta^{5 }+ \]\[
 2777808905/95551488\beta^{6 }+ 11237736659683361/246568703754240\beta^{7 }
+  
 637706602097023/9246326390784\beta^{8 }+ \]\[
30730356894974824135/302983623173210112\beta^{9 }+ 
 24844682623478614195/170428288034930688\beta^{10 }+ \]\[ 
33818048397999603287357407/164546077407896508825600\beta^{11 }+ 
4104928299267831518257003/14397781773190944522240\beta^{12 }+  \]\[
2152910828101259291195905268315/5519902203783874180225695744\beta^{13 }+ \]\[
3819217571901141130426522575431/7244871642466334861546225664\beta^{14 }+ \]\[ 
5230141901029903059246954146656789147/
7418748561885526898223335079936000\beta^{15 }+ \]\[
155909395729838428198406760858838609/
166921842642424355210025039298560\beta^{16 }+  \]\[
 97807124807276013777952022308137023473967/
  79701269121444292813607749964141690880\beta^{17 }+\]\[ 
 55801552690201549701430726155793324886557/
  34869305240631878105953390609311989760\beta^{18 }+  \]\[
 148896594487897544003651754583709926469646486457/
  71832409704183580986645441285329019469824000\beta^{19 }+ \]\[
 35940877310285092854424963408756863664486913149/
  13468576819534421434996020240999191150592000\beta^{20 }+  \]\[
\frac{ 17941640701630445913899151765380318159594010702584441}
{  5251926070147796666825200135959031797908924006400}\beta^{21 }+ \]\[
\frac{ 10876601452058926558707883889557137749933119405138111}
{  2499714812233807067383148141634346865350882099200}\beta^{22 }+  \]\[
\frac{ 361857535114298263458356057152592812597678881927788810489509}
{  65601658487185552513980237035403238786771858080399360000}\beta^{23 }+  \]\[
\frac{39965711198823372110613810026367365849872105318691597440189}
{5740145117628735844973270740597783393842537582034944000}\beta^{24}+ \]\[ 
\frac{74088929218366440544251194850535401261835943257271600246293207801}
{8464188384650628727563786103255867481224452216965447024640000}\beta^{25}
+\ldots ; \]

\normalsize
  for  spin $S=5$ we have
\scriptsize

\[\mu^{sq}_2(\beta;5 )= 16/25\beta^{ }+ 256/125\beta^{2 }+ 
5204168/1171875\beta^{3 }+ 48338176/5859375\beta^{4 }+ 
2034805058/146484375\beta^{5 }+  \]\[ 
 241280039872/10986328125\beta^{6 }+ 
1591718582528281/48065185546875\beta^{7 }+ 
 11584945581233824/240325927734375\beta^{8 }+ \]\[  713948604527910954841/
  10468597412109375000\beta^{9 }+ 
3085206848920157772116/32714366912841796875\beta^{10 }+  \]\[ 
 209254631040413445230852051/1635718345642089843750000\beta^{11 }+  \]\[
 261861081871019357148431194/1533485949039459228515625\beta^{12 }+ \]\[  
 286796489296904612503446641839/1275860309600830078125000000\beta^{13 }
+ \]\[   
 927256956094774595841861503059/3171527758240699768066406250\beta^{14 }+ \]\[  
 7876058633795911811844039266882910611/
20932083204388618469238281250000000\beta^{15 }+ \]\[   
 313814785508210643089805152539050763/
654127600137144327163696289062500\beta^{16 }+  \]\[ 
 6641766785411052004742939272264707158807/
  10949089676141738891601562500000000000\beta^{17 }+ \]\[   
 1676245871442533713159763595332123370379529/
  2201793502061627805233001708984375000000\beta^{18 }+  \]\[ 
 87345606965629938995595043418722063823007502943291/
  92034968386176042258739471435546875000000000000\beta^{19 }+  \]\[ 
 84548623916207453480078477475208733803749425657603/
  71902319051700033014640212059020996093750000000\beta^{20 }+  \]\[ 
 \frac{ 1400107459426560635640948807626903033873322748925282579}
{  966367168054848443716764450073242187500000000000000}\beta^{21 }+ \]\[  
 \frac{ 1340896031038431455783352327413352352150568975534390143}
{754974350042850346653722226619720458984375000000000}\beta^{22 }+  \]\[ 
 \frac{136835779829923175224156095157902909869547864207069619616531}
{ 63143309276311119901947677135467529296875000000000000000}\beta^{23 }+ \]\[  
 \frac{ 857174566443645713477002548675163352151717032192968383116903}
  {325582688455979211994417710229754447937011718750000000000}\beta^{24 }+ 
\]\[  
 \frac{ 33190993136061424862800928458658368747611145935196404885463577689}
{10418646030591334783821366727352142333984375000000000000000000}\beta^{25}
+\ldots ;\]

\normalsize
  for  spin $S=\infty$ we have
\scriptsize

\[\mu^{sq}_2(\beta;\infty)= 4/9\beta^{ }+ 32/27\beta^{2 }+ 1448/675\beta^{3 }+
 20224/6075\beta^{4 }+   4184168/893025\beta^{5 }+ 
1106368/178605\beta^{6 }+ \]\[
 1569917488/200930625\beta^{7 }+   5736232448/602791875\beta^{8 }+ 
54784888888/4862521125\beta^{9 }+ \]\[
 299610578338496/22975412315625\beta^{10 }+  
 1207444655769827584/81539738308153125\beta^{11 }+ \]\[
 1349348772110324224/81539738308153125\beta^{12 }+   66924940787412532096/
  3669288223866890625\beta^{13 }+ \]\[
1093318166543934948544/55039323358003359375\beta^{14 }+
 1702606769489789868894416/79531822252314854296875\beta^{15 }+ \]\[ 
 114532768278734103635275264/5010504801895835820703125\beta^{16 }+ \]\[  
 919459793247570137608951329224/37984636903172331356750390625\beta^{17 }
+\]\[    
 53155942283866615518891835176896/
2089155029674478224621271484375\beta^{18 }+
 \]\[   
 27472697484495976429107067149389536/
1034131739688866721187529384765625\beta^{19 }+ \]\[   
 1006261682495915580106986936770816/
36498767283136472512501037109375\beta^{20 }+\]\[   
 15565548691653887806849516436710333168/
547055690295410495508203044541015625\beta^{21 }+ \]\[   
 939491965116188391534804097416015076288/
  32155881125737406398652030632470703125\beta^{22 }+ \]\[  
 39133576063333116300268729065255000280903472/
  1310512935279427997777063508426343505859375\beta^{23 }+\]\[   
 278777984087166103934001241856931322430806016/
  9173590546955995984439444558984404541015625\beta^{24 }+\]\[   
 110575140481557845204099752974796626857362256/
  3589665866200172341737173957863462646484375\beta^{25 }+\ldots ;\]

\normalsize
\section{The second field derivative of the susceptibility on the sq lattice}

The HT  expansion coefficients of the second field derivative of the 
susceptibility 
$\chi_4^{sq}(\beta;S) =    \sum_{r=0}^\infty e^{sq}_r(S) \beta^r $ 
 for the spin $S=1/2$ Ising model on the sq lattice are

\scriptsize

\[ \chi^{sq}_4(\beta;1/2 )=-2 -32\beta^{ } -264\beta^{2 } -4864/3\beta^{3 }
 -8232\beta^{4 } -553024/15\beta^{5 } 
-2259616/15\beta^{6 } -180969728/315\beta^{7 } 
  -217858792/105\beta^{8 }\]\[  -20330135104/2835\beta^{9 }
 -5377792736/225\beta^{10 } -12048694416128/155925\beta^{11 } 
  -3450381618464/14175\beta^{12 }\]\[ -4559524221383168/6081075\beta^{13 } 
  -32137492094329792/14189175\beta^{14 } 
-4294238083842489856/638512875\beta^{15 }\]\[ 
  -66447301472480024/3378375\beta^{16 }
 -615947855084824982464/10854718875\beta^{17 }\]\[ 
  -5266523150993579779936/32564156625\beta^{18 } 
  -846203417984529146640128/1856156927625\beta^{19 } \]\[
  -3936026186821838336997344/3093594879375\beta^{20 } 
  -36092892170656596997895552/10257709336875\beta^{21 }\]\[ 
  -29853003233148167969372992/3093594879375\beta^{22 }
  -1295009629962512841156893011456/49308808782358125\beta^{23 }+ \ldots ; \]

\normalsize
  for  spin $S=3/2$ we have
\scriptsize

\[ \chi^{sq}_4(\beta;3/2 )=-34/81 -2720/729\beta^{ } -118664/6561\beta^{2 }
 -11887360/177147\beta^{3 } -334443704/1594323\beta^{4 } 
  -8399226944/14348907\beta^{5 }\]\[ -2908139546848/1937102445\beta^{6 } 
  -88177037915392/24407490807\beta^{7 }
 -1296528376807112/156905298045\beta^{8 } \]\[
  -322763959089785408/17793060798303\beta^{9 }
 -22010581967844185824/571919811374025\beta^{10 }\]\[ 
  -1257802911046040816384/15853617171287973\beta^{11 }  
  -1707728267339110983792928/10701191590619381775\beta^{12 } \]\[
  -78686888444506333988053504/250407883220493533535\beta^{13 } 
  -526453195633954664118374336/866796518840169923775\beta^{14 } \]\[
  -2459427224845762914894566483456/2129719046790297502715175\beta^{15 }\]\[ 
  -29627038494967515178515576360392/13691051015080483946026125\beta^{16 }\]\[ 
 -690461009408208832377887854932928/172507242790014097719929175\beta^{17 }\]\[ 
  -8690173400932855609900595158023301408/
1187712366609247062801712369875\beta^{18 } \]\[
  -537565305880315815098619649864360604416/
40619762938036249547818563049725\beta^{19 } \]\[
  -4421841426795056707947352805385022632352/
   186519319613431758127738299718125\beta^{20 } \]\[
  -223736370364790968750978665644014178666624/
   5314939750584589267756875057506325\beta^{21 } \]\[
  -1811369273935165648771349995432520935167516608/
   24429740925008451384296779282180858125\beta^{22 } \]\[
  -48302893144670451094209939317945489591931894272/
   372617837898286800587852980840842772875\beta^{23 }+ \ldots ; \]

\normalsize
 for  spin $S=5/2$ we have
\scriptsize

\[ \chi^{sq}_4(\beta;5/2 )=-518/1875 -58016/28125\beta^{ }
 -29744456/3515625\beta^{2 } -281109248/10546875\beta^{3 } 
  -280605816968/3955078125\beta^{4 }\]\[
 -250525717282624/1483154296875\beta^{5 } 
  -617648336544012704/1668548583984375\beta^{6 } \]\[
  -19071340704710341376/25028228759765625\beta^{7 } 
  -2800959282962631933176/1877117156982421875\beta^{8 } \]\[
  -395102319136173893031616/140783786773681640625\beta^{9 } \]\[
  -9431054390975818445446031456/1847787201404571533203125\beta^{10 } \]\[
  -393025157153444285614939533568/43554984033107757568359375\beta^{11 }\]\[ 
 -50866458118973669195762235471968/3266623802483081817626953125\beta^{12 }\]\[ 
  -83808389264789713946772796081517056/
3184958207421004772186279296875\beta^{13 } \]\[
  -52144037697072081548771211824452837696/
1194359327782876789569854736328125\beta^{14 } \]\[
  -44691861160456557650672561467167427836416/
   627038647086010314524173736572265625\beta^{15 }\]\[ 
  -1510310917425298480866445347247265371647112/
   13167811588806216605007648468017578125\beta^{16 } \]\[
  -10925720515554477678377968064902937181514923328/
   59960570627599736326374113559722900390625\beta^{17 } \]\[
  -2144664077367427274106339436164919219904617356256/
   7495071328449967040796764194965362548828125\beta^{18 } \]\[
  -135691148056059633291036957615452199268350347090176/
   305156475515462943803868256509304046630859375\beta^{19 }\]\[ 
  -1644767015685674369915130854664960381738776543478866336/
   2403107244684270682455462520010769367218017578125\beta^{20 }\]\[ 
  -\frac{263518985808208435106381199066354135829984512558200329856}
{252326260691848421657823564601130783557891845703125}\beta^{21 } \]\[
  -\frac{4430037093106077592970332390270058851101332090472535634447168}
{2802277222106585836680636703022173605859279632568359375}\beta^{22 } \]\[
  -\frac{853008862497126193849388938190698381275778890948135200954115584}
{359091809747086785071790160372984246350824832916259765625}\beta^{23 }+
 \ldots ; \]

\normalsize
 for  spin $S=7/2$ we have
\scriptsize

\[ \chi^{sq}_4(\beta;7/2 )=-78/343 -3744/2401\beta^{ } -4848696/823543\beta^{2 } 
-689772288/40353607\beta^{3 } 
  -580959055848/13841287201\beta^{4 }\]\[
 -44687096772672/484445052035\beta^{5 } 
  -4431819438902048/23737807549715\beta^{6 }\]\[
 -413013288144302336/1163152569936035\beta^{7 } 
  -256385235412851040648/398961331488060005\beta^{8 } \]\[
  -196585430976082399516864/175941947186234462205\beta^{9 } 
  -566933032584956420824929056/301740439424392102681575\beta^{10 } \]\[
  -71372994126137386900518317312/23234013835678191906481275\beta^{11 }\]\[ 
  -16745498964227865232420206946016/
3415400033844694210252747425\beta^{12 } \]\[
  -183229426028987417908427889279488/
23907800236912859471769231975\beta^{13 }\]\[ 
  -1254033470910060182544120679109155648/
106604881256394440384619005376525\beta^{14 } \]\[
  -9745103345620005200203713730723432006144/
   548482114064149395778864782662221125\beta^{15 } \]\[
  -4976893501600111043675824059923262645557176/
   188129365124003242752150620453141845875\beta^{16 }\]\[ 
  -870610032991546823828745567723292825563751744/
   22387394449756385887505923833923879659125\beta^{17 } \]\[
  -557920429282990478167324376812123173422272111328/
   9872840952342566176390112410760430929674125\beta^{18 } \]\[
  -106709066697312807116514337686475848059850446149376/
   1313087846661561301459884950631137313646658625\beta^{19 } \]\[
  -260686398144734955554067339500001400882061599318876256/
   2251945657024577632003702690332400492904019541875\beta^{20 } \]\[
  -\frac{378807977852706601725720695968755758572738179622424259968}
{2317252081078290383331810068352040107198236108589375}\beta^{21 } \]\[
  -\frac{2002324918586151534510861273604520355270051388804870160237632}
{   8742992101908389616310919387892247324458944837707711875}\beta^{22 }\]\[ 
  -\frac{448339653983815338569973257050509186459266858954187477536233984}
{   1407621728407250728226058021450651819237890118870941611875}\beta^{23 }
+ \ldots ; \]

\normalsize
 for  spin $S=1$ we have
\scriptsize

\[ \chi^{sq}_4(\beta;1 )=-2/3 -64/9\beta^{ } -368/9\beta^{2 } 
-4832/27\beta^{3 } -53224/81\beta^{4 } -2608952/1215\beta^{5 }
 -70378412/10935\beta^{6 } 
  -4150020604/229635\beta^{7 }\]\[
 -66380469083/1377810\beta^{8 } -509079717631/4133430\beta^{9 } 
  -37705033932347/124002900\beta^{10 }
 -2983197637134839/4092095700\beta^{11 }\]\[ 
  -83703985380256369/49105148400\beta^{12 }
 -7469497726136509157/1915100787600\beta^{13 } \]\[
  -704373891916979471609/80434233079200\beta^{14 } 
  -23330297521940602506671/1206513496188000\beta^{15 } \]\[
  -2436456639573557209685803/57912647817024000\beta^{16 } 
  -266797165559161332281981479/2953545038668224000\beta^{17 } \]\[
  -87086708500076352713438173/454391544410496000\beta^{18 } 
  -135450847185509549386128405773/336704134408177536000\beta^{19 } \]\[
  -50677978880769439391995533717187/60606744193471956480000\beta^{20 }\]\[ 
  -2192443457123465605985012126743157/
1272741628062911086080000\beta^{21 } \]\[
  -295672144676717551421473703987598409/
84000947452152131681280000\beta^{22 }\]\[ 
  -41374351091687562771726183318618508573/
5796065374198497086008320000\beta^{23 }+ \ldots ; \]

\normalsize
 for  spin $S=2$ we have
\scriptsize

\[ \chi^{sq}_4(\beta;2 )=-13/40 -13/5\beta^{ } 
-4559/400\beta^{2 } -91991/2400\beta^{3 }
 -20891591/192000\beta^{4 } -212034143/768000\beta^{5 }\]\[ 
  -59394696641/92160000\beta^{6 } -729017115311/516096000\beta^{7 } 
  -1215661085848381/412876800000\beta^{8 }\]\[
 -87593156642074631/14863564800000\beta^{9 } 
  -968526844215225451/84934656000000\beta^{10 } \]\[
  -561039895927506171767/26159874048000000\beta^{11 } 
  -35306579709341593798247/896909967360000000\beta^{12 }\]\[ 
  -1776701360631501784007887/25113479086080000000\beta^{13 } 
  -11410627673869075133791388389/91413063873331200000000\beta^{14 }\]\[ 
  -237699003573854970267659063731/1096956766479974400000000\beta^{15 } \]\[
  -18592571699981818139291726036227/50146595039084544000000000\beta^{16 }\]\[ 
  -14948339414669868427619677265762927/
23869779238604242944000000000\beta^{17 } \]\[
  -256695635580835347179708577654920179/
245517729311357927424000000000\beta^{18 } \]\[
  -1128021006332486943632526721512465772507/
   653077159968212086947840000000000\beta^{19 }\]\[ 
  -29532033370218363955331428888636030944661/
   10449234559491393391165440000000000\beta^{20 } \]\[
  -4023731442881361201523624642032255931024463/
   877735702997277044857896960000000000\beta^{21 } \]\[
  -156524116178970984292124102023713790812824743/
   21219984028505598886674432000000000000\beta^{22 } \]\[
  -2093022924688913489017338663508342456586923131431/
   177653706286648873879238344704000000000000\beta^{23 }+ \ldots ; \]

\normalsize
 for  spin $S=3$ we have
\scriptsize

\[ \chi^{sq}_4(\beta;3 )=-20/81 -1280/729\beta^{ } 
-105152/15309\beta^{2 } -25608832/1240029\beta^{3 } 
-1367079136/26040609\beta^{4 } 
  -83956032224/703096443\beta^{5 }\]\[ -1550532656/6200145\beta^{6 }
 -420365977194704/854262178245\beta^{7 } 
  -16527246890223386/17939505743145\beta^{8 }\]\[
 -1444550673339547982/871859979116847\beta^{9 }
  -1319016777450415832531/457726489036344675\beta^{10 }  \]\[
  -662417016420874795312781/135944767243794368475\beta^{11 } 
  -13120462511518678479415897/1631337206925532421700\beta^{12 }  \]\[
  -893317845174858374082305261/68711923155703425602004\beta^{13 }\]\[ 
  -1488758043407224391206507289251/72147519313488596882104200\beta^{14 }\]\[ 
  -941691747456862349205522853460381/
29219745321962881737252201000\beta^{15 } \]\[
  -487136722880575877439636720625200317/
9817834428179528263716739536000\beta^{16 } \]\[
  -9710193824987759497663438098955567153/
128753885786697242087028098486400\beta^{17 }\]\[ 
  -9191425849728984414927794320649105052599/
   81114948045619262514827702046432000\beta^{18 } \]\[
  -21032693916453702522183043024311313955275067/
   124835905042208045010319833449458848000\beta^{19 } \]\[
  -13010250319768441757042921979491059176093614779/
   52431080117727378904334330048772716160000\beta^{20 } \]\[
  -23671436201017036429396963574876296398127764259/
   65337192146706426019247395906932153984000\beta^{ 21} \]\[
  -1029638136916378630435155199834345661471265249931859/
   1962075880165593973357999299085172584139520000\beta^{22 } \]\[
  -\frac{919377218958702576728330343279046975025226274701194669}
{1218449121582833857455317564731892174750641920000}\beta^{23 }+ \ldots ; \]

\normalsize
 for  spin $S=4$ we have
\scriptsize

\[ \chi^{sq}_4(\beta;4 )=-41/192 -205/144\beta^{ } -48191/9216\beta^{2 }
 -3267595/221184\beta^{3 } -1499521679/42467328\beta^{4 } 
  -153998743607/2038431744\beta^{5 }\]\[
 -437016783314621/2935341711360\beta^{6 } 
  -54389323944214871/197254963003392\beta^{7 } \]\[
  -184054448262481836869/378729528966512640\beta^{8 } 
  -2992010459088589045439/3635803478078521344\beta^{9 }\]\[ 
  -11760889747422927650252909/8725928347388451225600\beta^{10 } \]\[
  -1977811179682733568619971967/921458033484220449423360\beta^{11 } \]\[
  -268346490297861896433033135197/80418155649531966495129600\beta^{12 } \]\[
  -112184262899969479136429722953713/
22079608815135496720902782976\beta^{13 } \]\[
  -216763313683496979681246062410630091/
28533648314944334223935904153600\beta^{14 } \]\[
  -39802833470615634685761544893987507323/
3560999309705052911147200838369280\beta^{15 } \]\[
  -1108468746916378778705445757376620304651029/
   68371186746337015894026256096690176000\beta^{16 } \]\[
  -259059836285657924437946725096369817682188999/
   11158177677002200993905084994979836723200\beta^{17 } \]\[
  -58677942505727462297504864574299074492343841823/
   1785308428320352159024813599196773875712000\beta^{18 } \]\[
  -14994356030844133093898410429979751709614941008261/
   325640257325632233806126000493491554929868800\beta^{19 } \]\[
  -299639536653618245508229985944601411638257039773730541/
   4689219705489104166808214407106278390990110720000\beta^{20 } \]\[
  -\frac{27702199593350936292763777936898639672174661534656905189}
{   315115564208867800009512008157541907874535440384000}\beta^{21 } \]\[
  -\frac{1396064114632151524400563852178338195101317887490158353521}
{11635036216942811077274289531970778136905923952640000}\beta^{22 }\]\[ 
  -\frac{4594223393669593726037556997535045906424141416749485870148409}
{28259175963710699544483794415250625938917108096172032000}\beta^{23 }+
 \ldots ; \]

\normalsize
 for  spin $S=5$ we have
\scriptsize

\[ \chi^{sq}_4(\beta;5 )=-122/625 -3904/3125\beta^{ } -1722096/390625\beta^{2 }
 -70139744/5859375\beta^{3 } 
  -6715499144/244140625\beta^{4 }\]\[ -1036239334424/18310546875\beta^{5 } 
  -2700555079983604/25177001953125\beta^{6 } \]\[
 -505146519035330756/2643585205078125\beta^{ 7} 
  -261705504631755035981/807762145996093750\beta^{8 }  \]\[
  -172672772057302989026539/327143669128417968750\beta^{9 } \]\[
  -113375319128888964472675351/136309862136840820312500\beta^{10 } \]\[
  -33842671438716155379460693/26553869247436523437500\beta^{11 } \]\[
  -1949141366875611418580802358453/1022323966026306152343750000\beta^{12 } \]\[
  -111380627324544938006297306000461/
39870634675025939941406250000\beta^{13 }\]\[ 
  -4721358644723268014004063517024919/
1174639910459518432617187500000\beta^{14 } \]\[
  -148911754629669748695991350566744855153/
   26165104005485773086547851562500000\beta^{15 } \]\[
  -1524334746733748359794464761477522669540423/
   191877429373562335968017578125000000000\beta^{16 } \]\[
  -76529758001040785050318247553082766663920573/
   6989820641465485095977783203125000000000\beta^{17 } \]\[
  -6021392479577463001963565928306898547013577788541/
   403662142044631764292716979980468750000000000\beta^{18 } \]\[
  -462772065823731558689353036599647072684333375739253/
   23008742096544010564684867858886718750000000000\beta^{19 } \]\[
  -858454417599183866265072264827944968601089997361011631/
   31956586245200014673173427581787109375000000000000\beta^{20 } \]\[
  -\frac{1074173762253296979432260148599154698418441508236936513977}
{30198974001714013866148889064788818359375000000000000}\beta^{21 } \]\[
  -\frac{117590255912387644245139933212221167595705726419304655875671}
{2516581166809501155512407422065734863281250000000000000}\beta^{22 } \]\[
  -\frac{7557825268987590875168127739127614930351505593747676177122381}
{124031500364182556950254365801811218261718750000000000000}\beta^{23 }
+ \ldots ; \]

\normalsize
 for  spin $S=\infty$ we have
\scriptsize

\[ \chi^{sq}_4(\beta;\infty )=-2/15 -32/45\beta^{ } -9896/4725\beta^{2 }
 -67328/14175\beta^{3 } -13576408/1488375\beta^{4 } 
  -70063552/4465125\beta^{5 }\]\[ -6108870496/245581875\beta^{6 }
 -5462835712/147349125\beta^{7 } 
  -1745672153646728/33186706678125\beta^{8 }\]\[
 -7140636903358016/99560120034375\beta^{9 } 
  -12856806647316286432/135899563846921875\beta^{10 } \]\[
  -49477558217898451456/407698691540765625\beta^{11 } 
  -15802177775783777989024/103963166342895234375\beta^{12 } \]\[
  -58160278464939584874496/311889499028685703125\beta^{13 } 
  -182809108697250013788775744/813671719965990432421875\beta^{14 }\]\[ 
  -8453768486491081135237270528/31733197078673626864453125\beta^{15 } \]\[
  -216871433264966881244349125790472/
696385009891492741540423828125\beta^{16 } \]\[
  -107280274565471715198510534640192/
298450718524925460660181640625\beta^{17 } \]\[
  -92913791572759544016582858438229472/
226524095360418424641077865234375\beta^{18 }\]\[ 
  -11017528365941614628191445823821080576/
   23785030012843934587313175849609375\beta^{19 } \]\[
  -2866331829674808121798702043492636757088/
   5531340868542483899027386339248046875\beta^{20 } \]\[
  -72556458435014167808399763522045153664/
   126243620837402422040354548740234375\beta^{21 } \]\[
  -49585427910097170120263910266781391786720832/
   78406756811589709268713201358841064453125\beta^{22 } \]\[
  -124262416331867054825308406132907754221093376/
   179874324450117568322342050176164794921875\beta^{23}+ \ldots ; \]

\normalsize

\section{The nearest-neighbor correlation 
 function  on the sc lattice}

The HT  expansion coefficients of the nearest-neighbor 
correlation  function  
$G^{sc}(\beta;S) =  \sum_{r=1}^\infty h^{sc}_r(S) \beta^r $ 
 for the spin $S=1/2$ Ising model on the sc lattice are

\scriptsize

\[ G^{sc}(\beta;1/2)=\beta^{ }+ 11/3\beta^{3 }
+ 542/15\beta^{5 }+ 
123547/315\beta^{7 }+ 14473442/2835\beta^{9 }+ 
  11336607022/155925\beta^{11 }+\]\[
 605337636044/552825\beta^{13 }+ 
10976336338579019/638512875\beta^{15 }+ 
   3022947654230404442/10854718875\beta^{17 }+\]\[ 
  8582760723898537620322/1856156927625\beta^{19 }+ 
  15262009695163033631128084/194896477400625\beta^{21 }+\]\[ 
  464794136914310402285931058/344816844631875\beta^{23}+\]\[ 
86983799739588855480442796083084/3698160658676859375\beta^{25}+\ldots ;\]

\normalsize
 for  spin $S=3/2$ we have  
\scriptsize

\[ G^{sc}(\beta;3/2)=25/81\beta^{ }+ 13403/19683\beta^{3 }+
 3701590/1594323\beta^{5 }+ 
133019549371/13559717115\beta^{7 }+ \]\[
  92099486893450/1977006755367\beta^{9 }+ 
10565915091540390286/44037825475799925\beta^{11 }+ \]\[
  7279131165849218160580/5564619627122078523\beta^{13 }+\]\[ 
  676343800187042849627103479
/91013634478217841996375\beta^{15 }+ \]\[
  2838094804561969299462218652218
/65169402831783103583084355\beta^{17 }+ \]\[
  5901615433219500977578268961759337666
/22566534965575694193232535027625\beta^{19 }+ \]\[
  12296360953479982851784833797400112704148/
   7677135195288851164537708416398025\beta^{21 }+ \]\[
  39213291332741115864756042794887004483801977262/
   3933188288926360672871781464431118158125\beta^{23 }+ \]\[
 60124047822007255575582319922300928928742017155916/
  955764754209105643507842895856761712424375\beta^{25}+ \ldots ;\]
\normalsize
 for  spin $S=5/2$ we have  
\scriptsize

\[ G^{sc}(\beta;5/2)=49/225\beta^{ }+ 
3829987/10546875\beta^{3 }+
 32535779594/35595703125\beta^{5 }+ 
  4712279576859799/1668548583984375\beta^{7 }+ \]\[
733481349877404562/75084686279296875\beta^{9 }+ 
   742278877905204315967675642
/20325659215450286865234375\beta^{11 }+ \]\[
  245077783709341332848692491212
/1698644377291202545166015625\beta^{13 }+\]\[ 
  393804276227513961630331175365449481
/663532959879375994205474853515625\beta^{15 }+ \]\[
  8453269034313617257280130165486928812751178/
   3357791955145585234276950359344482421875\beta^{17 }+ \]\[
  74187091275170275367097108008710421337216675782/
   6781255011454732084530405700206756591796875\beta^{19 }+ \]\[
  23299457213950277471359442683272509954318165098663732/
   480621448936854136491092504002153873443603515625\beta^{21 }+ \]\[
 \frac{182910807738952635387374265328069707288922
034340251864947009918}
   {837880889409869165167510374203629908151
924610137939453125}\beta^{23 }+\]\[
 \frac{5637585397498529761047249158508806848714402818449992529301198884}
{5657959187191493462265601266381054301746189594268798828125}\beta^{25}
+\ldots ;\]

\normalsize
 for  spin $S=7/2$ we have  
\scriptsize

\[G^{sc}(\beta;7/2)=9/49\beta^{ }+ 31065/117649\beta^{3 }+ 
39398759034/69206436005\beta^{5 }+ 
  7131241257143/4747561509943\beta^{7 }+ \]\[
2272722581612582402/512950283341791435\beta^{9 }+ 
  9365603740120794949157126/663828966733662625899465\beta^{11 }+ \]\[
  42992356517060935683282349852
/906126539591449484352769725\beta^{13 }+ \]\[
  10623174639539962958450078661386459
/63962928753836664230771403225915\beta^{15 }+ \]\[
  1916973140105892700210890326439387158454938/
   3198199207108055126786560547703411379875\beta^{17 }+ \]\[
  11875982558762168665840488471944334728244838826/
   5359542231271678781468918165841376790394525\beta^{19 }+ \]\[
  2969913101814360014362026154488970922894787042023836/
   355570366898617520842689898473536919932213611875\beta^{21 }+\]\[ 
  \frac{1286708107397776388410377343541078917637090596271637122336806}
  { 402177636687785922350302291843043376
92511146253455474625}\beta^{23 }+\]\[ 
 \frac{ 91790329674857779107668337538165936124577645450696686650049069644}
{739001407413806632318680461261592205099892312407244346234375}
\beta^{25}+\ldots ;\]

\normalsize
 for  spin $S=1$ we have  
\scriptsize

\[ G^{sc}(\beta;1)=4/9\beta^{ }+ 34/27\beta^{3 }
+ 42307/7290\beta^{5 }+ 
10283419/306180\beta^{7 }+ 4840468147/22044960\beta^{9 }+  \]\[
  34153748058211/21824510400\beta^{11 }+ 
120804322009036849/10213870867200\beta^{13 }+  \]\[
  66628192670985178451/714970960704000\beta^{15 }+ 
  35804777970555508386199249/47256720618691584000\beta^{17 }+ \]\[
  222304341666969605402292907/35210889872750592000\beta^{19 }+ \]\[
  364257960338020916013779645552929
/6787955349668859125760000\beta^{21 }+ \]\[
  843006747005511708281177523968372641
/1818373450728940262277120000\beta^{23 }+\]\[
 225920413734594721438126830337886041060843
/55642227592305572025679872000000\beta^{25}+\ldots ;\]
\normalsize
 for  spin $S=2$ we have  
\scriptsize

\[ G^{sc}(\beta;2)=1/4\beta^{ }+ 4489/9600\beta^{3 }+ 
817133/614400\beta^{5 }+ 9616936921/2064384000\beta^{7 }+ 
  43502816085073/2378170368000\beta^{9 }+ \]\[
8133154759676677729/104639496192000000\beta^{11 }+ 
  18242677369609968215429/52236036499046400000\beta^{13 }+ \]\[
  1434217472941210735092162961
/877565413183979520000000\beta^{15 }+ \]\[
  30124855299945216721220354348617
/3819164678176678871040000000\beta^{17 }+ \]\[
  101891712967049369234669271906116739389
/2612308639872848347791360000000000\beta^{19 }+ \]\[
   690559974824775992212799810521503725720401/
   3510942811989108179431587840000000000\beta^{21 }+ \]\[
  37699141897783597047002353484898771845398642747/
   37400780270873447132471230464000000000000\beta^{23 }+\]\[
 1428780602399874320245923502117481159963939624105089/
  272876092856292670278510097465344000000000000\beta^{25}+\ldots ;\]
\normalsize
 for  spin $S=3$ we have  
\scriptsize

\[ G^{sc}(\beta;3)=16/81\beta^{ }+ 5960/19683\beta^{3 }+ 
272584418/390609135\beta^{5 }
+ 37407890291/18983603961\beta^{7 }+\]\[ 
  3445320751714327/553561891502760\beta^{9 }+ 
  1026224764594882577171/48335917242237997680\beta^{11 }+ \]\[
  4768267428714219666406567
/62323739823767279457600\beta^{13 }+ \]\[    
  60853484779090615987912953107
/212025362880456284714755200\beta^{15 }+\]\[ 
  8893766949171382474511585792592625201
/8011352893394495063192859461376000\beta^{17 }+  \]\[
   1115673833729174909171856203481037812395/
   253634854688930631132078391770329088\beta^{19 }+ \]\[
  19168496743516259561226545342479033587873034903/
   1078582219564677508889163361003324446720000\beta^{21 }+  \]\[
\frac{31610613015008271315155320461037317593828439596092883}
{433226354340563149317446245238006106578006016000}\beta^{23 }+\]\[
\frac{ 1880293964737539469605198279907832987379254754575245823111}
  {6192588476750402663772908093696204935203262464000000}\beta^{25}\ldots ;\]

\normalsize
 for  spin $S=4$ we have  
\scriptsize

\[ G^{sc}(\beta;4)=25/144\beta^{ }
+ 52481/221184\beta^{3 }+ 2969269415/6115295232\beta^{5 }+ 
  399438588499387/328758271672320\beta^{7 }+ \]\[
  4117670365952485175/1211934492692840448\beta^{9 }+ 
  2249545200565209916268869/219394769877195345100800\beta^{11 }+  \]\[
  21861476136002172614999987627
/669079055004105961239478272\beta^{13 }+ \]\[
  27485416902303746468750635707494591
/253632429466171859768319148032000\beta^{15 }+  \]\[
  118129239519631442453276613933557099235871/
   318805076485777171254430999856566763520\beta^{17 }+ \]\[
  2898547306586480176745491032872431072066908587/
   2233472272466613400590713309283206823936000\beta^{19 }+  \]\[
  19472053586382642064163307096390269403082582152142901/
   4201540856118237333460160108767225438327139205120\beta^{21 }+ \]\[
  \frac{133729879342475856323332651837556997041686515761222243935573}
   {79517161802649154562
40028731564028943851134312775680000}\beta^{23 }+\]\[
\frac{698066723530411867663671708073956922641179519775484152392156808067}
{11285584512867504970085048137674489974965936289287262699520000}
\beta^{25}\ldots ;\]
\normalsize
 for  spin $S=5$ we have  
\scriptsize

\[ G^{sc}(\beta;5)=4/25\beta^{ }
+ 237842/1171875\beta^{3 }+ 112715689/292968750\beta^{5 }+ 
  171293409213721/192260742187500\beta^{7 }+ \]\[
  19300791971337918941/8374877929687500000\beta^{9 }+ 
  42076741500363453718466531
/6542873382568359375000000\beta^{11 }+ \]\[ 
  43947309495817385783309430793
/2319746017456054687500000000\beta^{13 }+ \]\[
  4862896456088603873588219951058058931
/83728332817554473876953125000000000\beta^{15 }+ \]\[ 
   104507102248323336525915713775619485719603/
   569352663159370422363281250000000000000\beta^{17 }+ \]\[
  218743156027178223182822003014739322457063619919851/
   368139873544704169034957885742187500000000000000\beta^{19 }+ \]\[ 
  \frac{7254180650494697627261195565536847609522345702707327}
{3699013083463534712791442871093750000000000000000}\beta^{21 }+ \]\[
  \frac{1827326131285031480513554800917533290
9444671770786279008439441}
   {277830560815768927568569779396057128
9062500000000000000000}\beta^{23 }+ \]\[
 \frac{54805826409652084317464525469249051894096686704569745945344192257}
{2451446124845019949134439229965209960937500000000000000000000}
\beta^{25}+\ldots ;\]
\normalsize
 for  spin $S=\infty$ we have  
\scriptsize

\[ G^{sc}(\beta;\infty)=1/9\beta^{ }
+ 67/675\beta^{3 }+ 117898/893025\beta^{5 }
+ 2867887/13395375\beta^{7 }+ 
  5652187874/14587563375\beta^{9 }+\]\[
 102724979659728674/135899563846921875\beta^{11 }+ 
  7768216350963076/4992228876009375\beta^{13 }+  \]\[
  17670715939428236221637/5302121483487656953125\beta^{15 }+ \]\[
  55894852614138922509983871
998/7596927380634466271350078125\beta^{17 }+  \]\[
  273071283083797403359592097265118/1641478
9518870900336309990234375\beta^{19 }+ \]\[
  817349455340605120760574237692018204
/21313858063457551773046871865234375\beta^{21 }+  \]\[
   274665071744389366176147944345024748945084466/
   3057863515651998661479814852994801513671875\beta^{23 }+ \]\[
 451402436106050024568984817614708798725624956/
  2116982433912922150255256436688708740234375\beta^{25}+\ldots ;\]

\section{The susceptibility on the sc lattice}

\normalsize

On the sc lattice, the HT expansion coefficients of the susceptibility 
$\chi^{sc}(\beta;S) = \sum_{r=0}^\infty c_r^{sc}(S) \beta^r $ 
  for the spin $S=1/2$ Ising model are

\scriptsize
\[ \chi^{sc}(\beta;1/2)=1  + 6 \beta^{ } + 30 \beta^{ 2} + 148 \beta^{ 3} 
+ 706 \beta^{ 4} + 16804/5 \beta^{ 5} + 47260/3 \beta^{ 6}
+  7744136/105 \beta^{7 } + 35975026/105 \beta^{ 8} + 
  1502899924/945 \beta^{ 9} +\]\[  6942884236/945 \beta^{ 10}  
+ 1763022244376/51975 \beta^{ 11} + 24340522634492/155925 \beta^{ 12}   + 
  1455564288731288/2027025 \beta^{ 13} 
+\]\[ 9352060224330104/2837835 \beta^{ 14} + 
  3217856261632544032/212837625 \beta^{15 } 
+   14736837331613648866/212837625 \beta^{ 16} +\]\[ 
  1147018945297343112964/3618239625 \beta^{ 17} 
+   9439914937963264249708/6512831325 \beta^{ 18} + \]\[
  4099376147364845437656056/618718975875 \beta^{ 19} +   
  93592219478518291774477772/3093594879375 \beta^{ 20} +\]\[ 
  8972803527064109944099241768/64965492466875 \beta^{ 21} +   
  5296430224856866468505272024/8407299025125 \beta^{22 } + \]\[
  47221618622049399307213422740992/16436269594119375\beta^{23} + \]\[
  37975296352037116774213386661036/2900518163668125\beta^{24}+ \]\[
73538934029908819825899053186296808/1232720219558953125\beta^{25}+
\ldots ;\]

\normalsize

 for $S=3/2$ we have:

\scriptsize
\[ \chi^{sc}(\beta;3/2)=5/9  + 50/27 \beta^{ } + 1330/243 \beta^{ 2} 
+   105556/6561 \beta^{ 3} + 2711290/59049 \beta^{ 4} 
+ 69531380/531441 \beta^{ 5} + 
  5278498156/14348907 \beta^{6 } + \]\[ 4671279598472/4519905705 \beta^{7 } 
+   3349792395278/1162261467 \beta^{8 } + 
  5293926492810020/659002251789 \beta^{ 9} 
+\]\[   132188730779640812/5931020266101 \beta^{ 10} + 
  907346485616242567832/14679275158599975 \beta^{ 11} + \]\[
  13545817801712579975068/79268085856439865 \beta^{ 12} +   
  79640437505437552258888/168624837185517531 \beta^{ 13} +\]\[ 
  108706432080095643203144744/83469294406831177845 \beta^{ 14} +\]\[
  1416010847006865786556690756384/394392416072277315317625 \beta^{ 15} + \]\[
  7014559632454359217771387527122/709906348930099167571725 \beta^{ 16} +\]\[   
  590602695850026035381830146909316
/21723134277261034527694785 \beta^{ 17} + \]\[ 
   657262450937939225540161232862286588
/8797869382290718983716387925 \beta^{ 18} + \]\[  
  1543902692575978655680834099461710066552
/7522178321858564731077511675875 \beta^{ 19} + \]\[  
  57343728048721143970509128811331678028
/101803917137935462525861060275 \beta^{ 20} + \]\[  
  12170230017577253032427294508499850567816
/7873984815680872989269444529639 \beta^{ 21} + \]\[  
  5367657952423726157541896730707167970989877624/
   1266727307222660442148721888705674125 \beta^{ 22} +\]\[   
  1384446864685107522702787844738225614759476329984/
   119187523906859414329447923164579338125\beta^{23} +\]\[
    225260489727098034111887814958749117313788724672396/
   7079738920067449211169206635976012684625\beta^{24}+ \]\[
 27763426872993735153388476341822564567848361319504872/
   318588251403035214502614298618920570808125\beta^{25}+ 
  \ldots ;\]

\normalsize
 for $S=5/2$ we have:
\scriptsize
\[ \chi^{sc}(\beta;5/2)=7/15  + 98/75 \beta^{ } + 30674/9375 \beta^{ 2} 
+ 28668724/3515625 \beta^{ 3} + 1042154302/52734375 \beta^{ 4} +
  189148520836/3955078125 \beta^{ 5} +\]\[
 56487949053172/494384765625 \beta^{ 6} + 
  1873016820604328/6866455078125 \beta^{ 7} 
+\]\[   5394142636212804278/8342742919921875 \beta^{ 8} + 
  191691733078119852676/125141143798828125 \beta^{ 9} + \]\[
  18837453687910795200196/5214214324951171875 \beta^{ 10} +\]\[ 
  19235641376235265409855777368/2258406579494476318359375 \beta^{ 11} + \]\[
  96878980663984811198646470708/4839442670345306396484375 \beta^{ 12} + \]\[  
  133165279256820200816712298336744
/2831073962152004241943359375 \beta^{ 13} + \]\[
  39024031357078803744682426325858552/
353884245269000530242919921875 \beta^{ 14} + \]\[  
  15590925900759498027802248840633324896
/60321178170852363109588623046875 \beta^{ 15} + \]\[  
  1685412314812381666488574393172609955058/
   2786838431493379175662994384765625 \beta^{ 16} + \]\[  
  58650693253165501125802687597025788760349548/
   41454221668464015237987041473388671875 \beta^{ 17} + \]\[  
  1694007677886501189727566665037360132863502604/
   512483509637604583986103534698486328125 \beta^{ 18} + \]\[  
  52360967705038452246156824563273621516126340653112/
   6781255011454732084530405700206756591796875 \beta^{ 19} + \]\[  
  27490767680237224355989926029693414012156395285507804/
   1525782377577314719019341282546520233154296875 \beta^{ 20} + \]\[  
  1346927125506726909676123683947176975846741585251408968/
   32041429929123609099406166933476924896240234375 \beta^{ 21} + \]\[  
 \frac{  50367389785177197740627539099383786380230579161318750874744}
   {513997938446357895969640594557859003543853759765625} \beta^{ 22} + \]
\[\frac{ 21263415029634715912911002539387850474338242967132363182961664512}
   {93097876601096573907501152689292212016880512237548828125}\beta^{23} + \]\[
\frac{ 15160525284980695591202138266166263789549761913406189662302334964}
   {28499349979927522624745250823252717964351177215576171875}\beta^{24}+\]\[
\frac{328263988539263809337847401509057234627656638362395118673996811256}
   {264973295267921181428416588232721551321446895599365234375}\beta^{25}+
\ldots ;\]
\normalsize
 for $S=7/2$ we have:
\scriptsize

\[ \chi^{sc}(\beta;7/2)=3/7  + 54/49 \beta^{ } + 6102/2401 \beta^{ 2} 
+ 686484/117649 \beta^{ 3} + 525949398/40353607 \beta^{ 4} + 
  2011996304964/69206436005 \beta^{ 5} 
+\]\[  31021513597932/484445052035 \beta^{ 6} + 
  3345656592405336/23737807549715 \beta^{ 7} 
+ 358279789805571918/1163152569936035 \beta^{ 8} +\]\[ 
  115046933858124853204/170983427780597145 \beta^{ 9} + 
  429074326243464390508724/293236578643724103675 \beta^{ 10} + \]\[
  3519311452670951549243616728/1106381611222771043165775 \beta^{ 11} + \]\[  
  160170895492158676690362527924/23234013835678191906481275 \beta^{ 12} + \]\[ 
  644487350840609247271710915992/43148882837688070683465225 \beta^{ 13} + \]\[ 
  3345211530169280495341594303328392
/103600467693289057711000005225 \beta^{ 14} + \]\[
  37198753442696820491691023610785733664
/533024406281972201923095026882625 \beta^{ 15} + \]\[ 
  192517808523848768891532740474257928474
/1278513086396618638179171987557625 \beta^{ 16} + \]\[ 
  31481927942865547977152073394800997270389836/
   96915127488122882629895774172830647875 \beta^{ 17} + \]\[ 
  3615588451686609834123079813684219555907704868/
   5166321796097627512501367038597818382875 \beta^{ 18} + \]\[ 
  13465865748299235980392477170790465026500762439992/
   8932570385452797969114863609735627983990875 \beta^{ 19} + \]\[ 
  7098999538663046347665193153197221390373628209946172/
   2188479744435935502433141584385228856077764375 \beta^{ 20} + \]\[ 
 \frac{15716353545637701914128109513899433042447016946486269608}
 {2251945657024577632003702690332400492904019541875} \beta^{ 21} + \]\[ 
 \frac{127466829205636020747510827707483063339271874832078744063528}
   {8496590963953731405549970250624147059726865731494375} \beta^{ 22} + \]\[ 
 \frac{ 2161364795747642776200557693208480476998729321955337108182442496}
   {67029606114630987058383715307173896154185243755759124375}\beta^{23} +\]\[
\frac{97485098493672257823056603774678550971241534858954449708840035844}
 {1407621728407250728226058021450651819237890118870941611875}\beta^{24}+\]\[ 
\frac{36637329684854174534294306849535555939294016747165917237712490642408}
{246333802471268877439560153753864068366630770802414782078125}\beta^{25}+
\ldots ;\]
\normalsize
 for $S=1$ we have:
\scriptsize

\[ \chi^{sc}(\beta;1)=2/3  + 8/3 \beta^{ } + 28/3 \beta^{ 2} 
+ 292/9 \beta^{ 3} 
+ 2951/27 \beta^{ 4} + 148849/405 \beta^{ 5} + 987151/810 \beta^{ 6} 
+  2543651/630 \beta^{ 7} + \]\[
  2322545251/174960 \beta^{ 8} + 160264448467/3674160 \beta^{ 9} 
+ 1747122407693/12247200 \beta^{ 10} + 
  565427906556473/1212472800 \beta^{ 11} 
+\]\[  22110005692656437/14549673600 \beta^{ 12} + 
  8426712545527651649/1702311811200 \beta^{ 13} 
+ 54773315847684951739/3404623622400 \beta^{ 14} + \]\[ 
  18686236921729743979777/357485480352000 \beta^{ 15} + 
  20343875021824485883513/119995126272000 \beta^{ 16} + \]\[
  14579302882745988616902697/26518922906112000 \beta^{ 17} + \]\[
  9348066688118373225121476439/5250746735410176000 \beta^{ 18} + \]\[
  1725189093312846058617382641673/299292563918380032000 \beta^{ 19} + \]\[ 
  1576652687149998226160608896049/84572467041300480000 \beta^{ 20} + \]\[
  68201424753418481045118928725493649
/1131325891611476520960000 \beta^{ 21} + \]\[ 
  124297426165308377769874467941986187
/638183836293653422080000 \beta^{ 22} + \]\[ 
  1080509130967664124865236696363688300531
/1717352703466221358817280000\beta^{23} +\]\[
  83708348865349977176067428458367045555399
/41216464883189312611614720000\beta^{24}+\]\[
  60789701511213968879520941087443488811248203/
   9273704598717595337613312000000\beta^{25}+
\ldots ;\]
\normalsize
 for $S=2$ we have:
\scriptsize

\[ \chi^{sc}(\beta;2)=1/2  + 3/2 \beta^{ } + 321/80 \beta^{ 2} 
+ 17089/1600 \beta^{ 3} + 707317/25600 \beta^{ 4} 
+ 292325/4096 \beta^{ 5} + 
  2235984829/12288000 \beta^{ 6} +\]\[ 797526050117/1720320000 \beta^{ 7} 
+ 12909939253837/11010048000 \beta^{ 8} + 
  5874566480368757/1981808640000 \beta^{ 9} \]\[
+ 591342196242924017/79272345600000 \beta^{ 10}  + 
  65453335356157737509/3487983206400000 \beta^{ 11} + \]\[   
  39394233999882369383269/837115969536000000 \beta^{ 12}  + 
  5135686107070361004225961/43530030415872000000 \beta^{ 13} + \]\[   
  3596988000168580765751186537/12188408516444160000000 \beta^{ 14} + \]\[ 
  49065668403883700664236766103/66482228271513600000000 \beta^{ 15} + \]\[   
  6633529964407547354986934776061
/3600268361780428800000000 \beta^{ 16} + \]\[ 
  14636945936219058038227655763477293
/3182637231813899059200000000 \beta^{ 17} + \]\[ 
  1010373074444910394406358044292134113
/88134569496384897024000000000\beta^{ 18} + \]\[ 
  12439712599363673590484486746705780413461/
   435384773312141391298560000000000 \beta^{19 } + \]\[ 
  495524134743535225645283946705647277518749/
   6966156372994262260776960000000000 \beta^{ 20} + \]\[ 
  103614032364300142790663330999721523508452417/
   585157135331518029905264640000000000 \beta^{ 21} + \]\[ 
  113386771977238589303561447991666381936148664321/
   257469139545867933158316441600000000000 \beta^{ 22} + \]\[ 
  5188234765259308554230152403062712986959679329841/
   4737432167643969970113022525440000000000\beta^{23} + \]\[
\frac{2380163846439981475292980090941013357753180082897737}
 {874602861718886763713173389312000000000000}\beta^{24}+\]\[
\frac{1215525069659546421230133670623750609094278898334581}
{179760271973842338786897297408000000000000}\beta^{25}+
\ldots ;\]
\normalsize
 for $S=3$ we have:
\scriptsize

\[ \chi^{sc}(\beta;3)=4/9  + 32/27 \beta^{ } + 688/243 \beta^{ 2} 
+ 44176/6561 \beta^{ 3} + 6437644/413343 \beta^{ 4} 
+ 4684124356/130203045 \beta^{ 5} + \]\[ 
  288448191434/3515482215 \beta^{ 6} + 5916528613694/31639339935 \beta^{ 7} + 
  481985643744571/1139016237660 \beta^{ 8}  +\]\[ 
88302340232526487/92260315250460 \beta^{ 9} + 
  125261247274529390677/58123998607789800 \beta^{ 10} + \]\[
  195387802071297208477183/40279931035198331400 \beta^{ 11} + \]\[
  47351416276247226554394619/4350232551801419791200 \beta^{ 12} + \]\[
  161405228226285178488632339/6610093617672287215200 \beta^{ 13} + \]\[ 
  500725183630226977287626472037/9161589754093790080267200 \beta^{ 14} 
+  \]\[  
  151270170985614286761762481678009/1236814616802661660836072000 \beta^{ 15} + 
\]\[ 
  340641340671904583620507283562374059
/1246709133737082954122760576000 \beta^{ 16} + \]\[ 
  814871640535718936854774147913411588401/
1335225482232415843865476576896000 \beta^{ 17} + \]\[ 
  3235846307307931153100443723695579579583/
   2376994814523641392375903356672000 \beta^{ 18} + \]\[ 
  112300676437766466657843666668253260132199001/
   36988416308802383706761432133172992000 \beta^{ 19} + \]\[ 
  45034730936338914492979176624009418921726948093/
   6657914935584429067217057783971138560000 \beta^{ 20} + \]\[ 
  2708607224449504320444504924244042584938099706583/
   179763703260779584814860560167220741120000 \beta^{ 21} + \]\[ 
  8355312743594652919908853714829310781489437213371893/
   249152492719440504553396736391767947192320000 \beta^{ 22} + \]\[ 
  2449298974327285033782330052503164425680524286447016157/
   32820178359133571917988351911970159589242880000\beta^{23} +\]\[ 
\frac{12941157750116557751929822978008762693955508098885299923611}
  {77980743781301366877140324142841099184041082880000}\beta^{24} +\]\[ 
\frac{340755803681450985674150304723702935576284190418986062024333}
   {923456176357516186702977522744170911389960192000000}\beta^{25}+
 \ldots ;\]
\normalsize
 for $S=4$ we have:
\scriptsize

\[ \chi^{sc}(\beta;4)=5/12  + 25/24 \beta^{ } + 1795/768 \beta^{ 2} 
+ 192481/36864 \beta^{ 3} + 80332535/7077888 \beta^{ 4}
 + 8370294605/339738624 \beta^{ 5} + \]\[ 
  574170547931/10871635968 \beta^{ 6} 
+ 229585691283841/2029372047360 \beta^{ 7} + 
  1012604068654494737/4208105877405696 \beta^{ 8} + \]\[
  103333875650399513015/201989082115473408 \beta^{ 9} + 
  2332900263447364811981/2154550209231716352 \beta^{ 10} +\]\[ 
  195461901256071907671681281/85320188285575967539200 \beta^{ 11} + \]\[
  3166395185656481692639101485/655259046033223430701056 \beta^{ 12} + \]\[
  12499493079144197211519097088153
/1226644934174194262272376832 \beta^{ 13} + \]\[ 
  29455413202931535900450406057002977
/1373842326275097573745062051840 \beta^{ 14} + \]\[
  74353954887426127153070199781761574969
/1648610791530117088494074462208000 \beta^{ 15} + \]\[ 
  23983863169261854841259330096064490543531/
   253226617579025984792689837395148800 \beta^{ 16} +\]\[ 
  2739441018816629071941624910695000630381851/
   13775527996299013572722327154296094720 \beta^{ 17} + \]\[ 
  8271664378599655675655738428038736292575323347/
   19836760314670579544720151102186376396800 \beta^{ 18} + \]\[ 
  79078625688753384321771470869218859170816184645681/
   90455627034897842723923889025969876369408000 \beta^{ 19} + \]\[ 
  8672160629872986710839928915053386075789438730143019/
   4736585561100105218998196370814422617161728000 \beta^{ 20} + \]\[ 
  1032588145638103894658735541046945448296034317819081901/
   269329542058861367529497442869693938354303795200 \beta^{ 21} + \]\[ 
\frac{16476419470025685003673974660303040564365880313220048307867}
 {2054086640768916029691633830952865769848823611392000} \beta^{ 22} + \]\[ 
\frac{570787870641742705511077174252686184409471984651064693943173307}
{34015674771133249451693456240579457148696519004651520000}\beta^{23} +\]\[
\frac{91636766215752367920118789526228052047192899914564256740983718523}
{2612403822423033557890057439276502309019892659557236736000}\beta^{24}+\]\[
\frac{137909107451703040523523627490875598211619923822097673406089570644227}
{1880930752144584161680841356279081662494322714881210449920000}\beta^{25}+
 \ldots ;\]
\normalsize
 for $S=5$ we have:
\scriptsize

\[ \chi^{sc}(\beta;5)=2/5  + 24/25 \beta^{ } + 6468/3125 \beta^{ 2} 
+ 1735684/390625 \beta^{ 3} + 3626269/390625 \beta^{ 4} 
+ 945741529/48828125 \beta^{ 5} + \]\[ 
  1461530112857/36621093750 \beta^{ 6} 
+ 2633175463970881/32043457031250 \beta^{ 7} + 
  338410617772913029/2014160156250000 \beta^{ 8} + \]\[ 
  2396351149332443767729/6979064941406250000 \beta^{ 9} + 
  1218885547718940026600989/1744766235351562500000 \beta^{ 10} + \]\[ 
  1549396936739197450147067051/1090478897094726562500000 \beta^{ 11} +\]\[ 
  188498652280425311835844467331/65428733825683593750000000 \beta^{ 12} + \]\[ 
  24836851641149342219357934126227/4252867698669433593750000000 \beta^{ 13} + 
\]\[ 
  12558764916105293289648394225171189
/1063216924667358398437500000000 \beta^{ 14} + \]\[ 
  333319249327682863851094935805398979211/
   13954722136259078979492187500000000 \beta^{ 15} + \]\[ 
  53830958131542181318301148040847542353379/
   1116377770900726318359375000000000000 \beta^{ 16} + \]\[ 
  9235351696951963921518260196861028010096243/
   94892110526561737060546875000000000000 \beta^{ 17} + \]\[ 
  460740695972857341388361311848681029286992973521/
   2348579735532402992248535156250000000000000 \beta^{ 18} + \]\[ 
  24259073265477950253517129387883674269566117441260691/
   61356645590784028172492980957031250000000000000 \beta^{ 19} + \]\[ 
  697710470196309108073005252862857042502459834672999741/
   876523508439771831035614013671875000000000000000 \beta^{ 20} + \]\[ 
\frac{1032297498986558065955299862999529946253266753798215145787}
   {644244778703232295811176300048828125000000000000000} \beta^{ 21} + \]\[ 
\frac{519047880698181714782360810398731715649854305312457549796849}
   {161061194675808073952794075012207031250000000000000000} \beta^{ 22} + \]\[ 
 \frac{3000950111375958778335159867047099355795511694502472512280893241}
{463050934692948212614282965660095214843750000000000000000}\beta^{23} + \]\[
 \frac{13157502568683448679598139611779526765816600873147251913403408007}
{1010292948420977918431162834167480468750000000000000000000}\beta^{24} +\]\[
 \frac{181760490403185105841390278618466635980262526044099355369120164272689}
{6945764020394223189214244484901428222656250000000000000000000}\beta^{25}+
\ldots ;\]
\normalsize
 for $S=\infty$ we have:
\scriptsize

\[ \chi^{sc}(\beta;\infty)=1/3  + 2/3 \beta^{ } + 6/5 \beta^{ 2}
 + 484/225 \beta^{ 3} + 2534/675 \beta^{ 4} + 129844/19845 \beta^{ 5} 
+ 16764212/1488375 \beta^{ 6} + \]\[ 
  144194632/7441875 \beta^{ 7} + 4877332994/147349125 \beta^{ 8}
 + 196306025276/3473229375 \beta^{ 9} + 
  50614099608844/526773121875 \beta^{ 10} 
+\]\[  493118157570944296/3019990307709375 \beta^{ 11} + 
  12532040821579469324/45299854615640625 \beta^{ 12} + \]\[
  27291561298536715784/58242670220109375 \beta^{ 13} + 
  27444006557193974207032/34654388780965078125 \beta^{ 14} + \]\[
  1313870752936334263132096/981874348794010546875 \beta^{ 15} + \]\[
  7958773125136339617868227346/3525910786519291873828125 \beta^{ 16} + \]\[ 
  5355733833932968587631869467452
/1406838403821197457657421875 \beta^{ 17} + \]\[ 
  637901662382461481341800483669916
/99483572841641820220060546875 \beta^{ 18} + \]\[ 
  137865369452122460437523968096056856
/12767058514677366928241103515625 \beta^{ 19} + \]\[ 
  144028358015402280728889758747927000836
/7928343337614644862437725283203125 \beta^{ 20} + \]\[ 
  75639750685886657866662449206819857272
/2475365114458871020399108798828125 \beta^{ 21} + \]\[ 
  105641606804713559197921012577980309104632/
   2057110240367039466580432936083984375 \beta^{ 22} + \]\[ 
  308636780166090205629928590295416776829902368/
   3576448556318127089450075851455908203125\beta^{23} +  \]\[ 
147704683376025485951434745749854246026829179596/
   1019287838550666220493271617664933837890625\beta^{24} +\]\[ 
 2232010725208829820398769811211763679699326982312/
   9173590546955995984439444558984404541015625\beta^{25} +
\ldots ;\]

\normalsize

\section{ The second correlation moment on the sc lattice}

The HT  expansion coefficients of the second moment of the 
correlation function
$ \mu_{2}^{sc}(S;\beta)=  \sum_{r=1}^\infty b^{sc}_r(S) \beta^r $
 for the spin $S=1/2$ Ising model on the sc lattice are

\scriptsize
\[\mu^{sc}_2(\beta;1/2 )=  6\beta^{ }+  72\beta^{2 }+ 
 580\beta^{3 }+  3984\beta^{4 }+  124804/5\beta^{5 }+
  738256/5\beta^{6 }+  17603848/21\beta^{7 }+
  161679008/35\beta^{8 }+ \]\[ 
  23502612244/945\beta^{9 }+  207084686896/1575\beta^{10 }+ 
 7118350956184/10395\beta^{11 }+  
  183105409684576/51975\beta^{12 }+\]\[
  36357148765588568/2027025\beta^{13 }+  
  428105990895623072/4729725\beta^{14 }+
  3858980041412441056/8513505\beta^{15 }+\]\[  
  159972662342929823168/70945875\beta^{16 }+
  40346953458333412225924/3618239625\beta^{17 }+ \]\[ 
  595480220035499088584176/10854718875\beta^{18 }+  
  33241579832993565724372216/123743795175\beta^{19 }+\]\[  
  1350819222602054183075867296/1031198293125\beta^{20 }+\]\[  
  413420004249138569994431044328/64965492466875\beta^{21 }+ \]\[ 
  7338889399155531947061431193632/238206805711875\beta^{22 }+  \]\[
  37598813360082690796022724976448/252865686063375\beta^{23 }+ \]\[
 11761287986182312586072257247131456/16436269594119375\beta^{24}+ \]\[
  4233897510636895869927021649138716008/1232720219558953125\beta^{25 }+ 
\ldots ; \]

\normalsize 
 for $S=3/2 $ we have 
\scriptsize

\[\mu^{sc}_2(\beta;3/2 )=  50/27\beta^{ }+  
1000/81\beta^{2 }+  375556/6561\beta^{3 }+  
4528240/19683\beta^{4 }+  451613780/531441\beta^{5 }+  
  14243337040/4782969\beta^{6 }+   \]\[
45408151777832/4519905705\beta^{7 }+  
89414169586592/2711943423\beta^{8 }+  
  69776448057180260/659002251789\beta^{9 }+   \]\[
660920350555374320/1977006755367\beta^{10 }+
  15277888715620145292152/14679275158599975\beta^{11}+ \]\[  
  84634563642513813788512/26422695285479955\beta^{12 }+  
  18105994764322415546655704/1854873209040692841\beta^{13 }+ \]\[  
  1149101790508953636180379168/38952337389854549661\beta^{14 }+\]\[   
  34906646119238111683402872082144
/394392416072277315317625\beta^{15 }+  \]\[
  12488287530408065393737831392448
/47327089928673277838115\beta^{16 }+  \]\[
  16993954269998506619808788348389636
/21723134277261034527694785\beta^{17 }+  \]\[
  270719536632016688382382435950932464
/117304925097209586449551839\beta^{18 }+  \]\[
  4635036642399425615883048539184325220872
/683834392896233157370682879625\beta^{19 }+\]\[  
  447443318454989062992890540007979879696288/
   22566534965575694193232535027625\beta^{20 }+\]\[  
  147894041126453257810747624719054407262142568/
   2559045065096283721512569472132675\beta^{21 }+  \]\[
  14178938412891452201123257278578865597158593696/
   84448487148177362809914792580378275\beta^{22 }+  \]\[
  637604863364869213033249376368691918172955989615424/
   1311062762975453557623927154810372719375\beta^{23 }+\]\[
 3315217387442266958331431109429245203762971217827136/
   2359912973355816403723068878658670894875\beta^{24}+\]\[
  1289543300432109170161989065081515350286734609755682152/
   318588251403035214502614298618920570808125\beta^{25}+
\ldots ; \] 

\normalsize 
 for $S=5/2 $ we have 
\scriptsize

\[\mu^{sc}_2(\beta;5/2 )=  98/75\beta^{ }+  
2744/375\beta^{2 }+  100698724/3515625\beta^{3 }+  
1712052944/17578125\beta^{4 }+  
  1205314338916/3955078125\beta^{5 }+ \]\[  
9946748307664/10986328125\beta^{6 }+  
  160109483824949512/61798095703125\beta^{7 }+  
743176365375509408/102996826171875\beta^{8 }+  \]\[ 
  2461512686383827849668/125141143798828125\beta^{9 }+  
  164970177315896499343888/3128528594970703125\beta^{10 }+\]\[  
  314855453558774832727473557048
/2258406579494476318359375\beta^{11 }+  \]\[ 
  1763643297959881528573426011488
/4839442670345306396484375\beta^{12 }+  \]\[
  2670870314135399721983576302613032
/2831073962152004241943359375\beta^{13 }+ \]\[  
  171437201293602650306807300531983328
/70776849053800106048583984375\beta^{14 }+  \]\[
  1365662176020944739680437197947521276832
/221177653293125331401824951171875\beta^{15 }+ \]\[  
  31136264401554455812285980847287658747328/
   1990598879638127982616424560546875\beta^{16 }+\]\[  
  1633524276236739195588220584606747501771697388/
   41454221668464015237987041473388671875\beta^{17 }+ \]\[  
  131632073125141696266050173161714049666065670544/
   1332457125057771918363869190216064453125\beta^{18 }+\]\[  
  1672087312241595464083443264786487933011861645372952/
   6781255011454732084530405700206756591796875\beta^{19 }+ \]\[ 
  311782615567834264335441885635950905135271623860215328/
   508594125859104906339780427515506744384765625\beta^{20 }+  \]\[
   \frac{243298473213276718174613983085971366941493187114764361512}
{160207149645618045497030834667384624481201171875}\beta^{21 }+  \]\[
 \frac{56483006343771153450257925343867271826746931690109297888}
 {15062210650442722226216574199497699737548828125}\beta^{22 }+  \]\[
  \frac{859509864124157148499210797240388
617182524390381488676931897893312}
   {930978766010965739075011526892922
12016880512237548828125}\beta^{23 }+ \]\[
 \frac{347861611113785351918851074397845001500203406359662158704638111936}
{15345803835345589105632058135597617365419864654541015625}\beta^{24}+ \]\[
\frac{1780012692685487171739030815734683277515258704344031241467884149753016}
   {32061768727418462952838407176159307709895074367523193359375}\beta^{25}+
\ldots ; \]

\normalsize 
 for $S=7/2 $ we have 
\scriptsize

\[\mu^{sc}_2(\beta;7/2 )=  54/49\beta^{ }+  
1944/343\beta^{2 }+  2401092/117649\beta^{3 }+  
52611120/823543\beta^{4 }+  
  12739972785444/69206436005\beta^{5 }+ \]\[  
244062083037168/484445052035\beta^{6 }+  
  31531460371912248/23737807549715\beta^{7 }+ 
113290997586012000/33232930569601\beta^{8 }+   \]\[
  1464396134058428539924/170983427780597145\beta^{9 }+  
  42213373803442962108048/1994806657440300025\beta^{10 }+\]\[  
  57056920238747799061305429944
/1106381611222771043165775\beta^{11 }+  \]\[ 
  64158770806068507593879276768
/516311418570626486810695\beta^{12 }+  \]\[
  89567842680144504078228520224104
/302042179863816494784256575\beta^{13 }+  \]\[ 
  8079541997544199137850406162352032
/11511163077032117523444445025\beta^{14 }+ \]\[ 
  879199266102049307825658233367745549792
/533024406281972201923095026882625\beta^{15 }+ \]\[  
  84657089600367191362333864640383424
/21975856783424951635666667775\beta^{16 }+  \]\[
  867141308136271431771322733789881799863577036/
   96915127488122882629895774172830647875\beta^{17 }+ \]\[  
  154329168213712455968699214158752811180678700784/
   7462464816585461962501974611307959886375\beta^{18 }+  \]\[
  425114625293869549879921805664907721816469396384536/
   8932570385452797969114863609735627983990875\beta^{19 }+  \]\[
  5305444390691073148026083478411594870861914805516512/
   48632883209687455609625368541893974579505875\beta^{20 }+  \]\[
 \frac{561129491831070233223181006961945231389211803201636944488}
   {2251945657024577632003702690332400492904019541875}\beta^{21 }+ \]\[ 
 \frac{  10930120065320893377975909724580893336036606532846462592032}
{19266646176765830851587234128399426439289944969375}\beta^{22 }+  \]\[
 \frac{ 863205761708654907370114442656390
10091759978924478879787614282048}
 {670296061146309870583837153071738
96154185243755759124375}\beta^{23 }+ \]\[
\frac{91196502091357509792625252470446299964224019216662501007729863232}
 {31280482853494460627245733810014484871953113752687591375}\beta^{24}+ \]\[ 
\frac{1621820285462163874754453426780912371135146620661125311148034432939368}
{246333802471268877439560153753864068366630770802414782078125}\beta^{25}+
\ldots ; \] 

\normalsize 
 for $S=1 $ we have 
\scriptsize

\[\mu^{sc}_2(\beta;1 )=  8/3\beta^{ }+  64/3\beta^{2 }+  
1060/9\beta^{3 }+  5056/9\beta^{4 }+  995569/405\beta^{5 }+
  458656/45\beta^{6 }+   46098943/1134\beta^{7 }+  
  148965776/945\beta^{8 }+ \]\[  2196751115987/3674160\beta^{9 }+  
2559023824766/1148175\beta^{10 }+  
  1986093263628389/242494560\beta^{11 }+   \]\[
3381221952418327/113669325\beta^{12 }+  
  182077068489220585409/1702311811200\beta^{13 }+\]\[  
142005677634975023711/372380708700\beta^{14 }+  
  32164388280039908038831/23832365356800\beta^{15 }+ \]\[ 
  14459362809367839706901/3046751253000\beta^{16 }+  
  440048543032072993941341737/26518922906112000\beta^{17 }+ \]\[  
  4736286184376297655753969643/82042917740784000\beta^{18 }+  
  478712340650944684826881624381
/2394340511347040256\beta^{19 }+   \]\[
  96753053465026900131372413762131
/140293389336740640000\beta^{20 }+  \]\[
  2681397057918937589388436874283800849
/1131325891611476520960000\beta^{21 }+  \]\[
  350806050315113818974245636302787659
/43210363915716117120000\beta^{22 }+  \]\[
  9522577753051135592808663370131739683671
/343470540693244271763456000\beta^{23}+ \]\[
 15201188211243902404003996311272643952597
/161001815949958252389120000\beta^{24}+  \]\[
  2974200332500711101744264985456112085493799243/
   9273704598717595337613312000000\beta^{25}+
\ldots ; \]

\normalsize 
 for $S=2 $ we have 
\scriptsize

\[\mu^{sc}_2(\beta;2 )=  3/2\beta^{ }+  9\beta^{2}+  
60289/1600\beta^{3}+  54777/400\beta^{4}+  
46875869/102400\beta^{5}+  92960863/64000\beta^{6}+  \]\[ 
  7635641717381/1720320000\beta^{7}+  
189203628553/14336000\beta^{8}+  
  76118339567511413/1981808640000\beta^{9} + \]\[  
9076031646565957/82575360000\beta^{10} + 
  5409527297580243704953/17439916032000000\beta^{11}+ \]\[  
  1886587812730952255999/2179989504000000\beta^{12 }+ 
  104083010757681844969183273/43530030415872000000\beta^{13 }+ \]\[  
  415924338916828608109269011/63481294356480000000\beta^{14 }+\]\[  
  13034873324901918665844049675709/731304510986649600000000\beta^{15 }+\]\[   
  1468304441651190798116789126209/30471021291110400000000\beta^{16 }+\]\[  
  412306443769501533394976371824941357
/3182637231813899059200000000\beta^{17 }+ \]\[  
  2068206463847388914567462166030241129/
5967444809651060736000000000\beta^{18 }+ \]\[  
  401913267170119808819756511921466327957973/
   435384773312141391298560000000000\beta^{19 }+  \]\[
  44428240541843317500367886785633003792049/
18141032221339224637440000000000\beta^{20 }+\]\[  
  3788322496744379738436373427759773999474577857/
   585157135331518029905264640000000000\beta^{21 }+  \]\[
  1203957988378130390421872856288572026746948293/
   70578163252705025536819200000000000\beta^{22 }+  \]\[
  279346088074102514358500907939486609402474580305907/
   6233463378478907855411871744000000000000\beta^{23 }+ \]\[
 5381400066315570749095817495572281722602563984749/
   45834289547639028348616704000000000000\beta^{24}+ \]\[ 
  2790935071426336534134238626070765321199242347750660077/
   9095869761876422342617003248844800000000000\beta^{25}+
\ldots ; \] 

\normalsize 
 for $S=3 $ we have 
\scriptsize

\[\mu^{sc}_2(\beta;3 )=  32/27\beta^{ }+  
512/81\beta^{2 }+  154768/6561\beta^{3 }+  
1505792/19683\beta^{4 }+  29733488836/130203045\beta^{5 }+  \]\[ 
  758569398016/1171827405\beta^{6 }+ 
55930281404222/31639339935\beta^{7 }+  
  445971449623424/94918019805\beta^{8 }+  \]\[ 
1127841966748294807/92260315250460\beta^{9 }+   
  75753838592962827856/2421833275324575\beta^{10 }+  \]\[ 
  3179612472771635417296639/40279931035198331400\beta^{11 }+   
  8925427370318340470950792/45314922414598122825\beta^{12 }+  \]\[
  247730819552859545299126685623/508977208560766115570400\beta^{13 }+ \]\[  
  56922168941393453404420037342/47716613302571823334725\beta^{14 }+\]\[  
  3590300504234464425425723981902777
/1236814616802661660836072000\beta^{15 }+\]\[  
  2848946657479635150403018911101891
/405829796138373357461836125\beta^{16 }+  \]\[
  22542842681815973655691034559408343279921/
   1335225482232415843865476576896000\beta^{17 }+  \]\[
  22760887815562610707152493322234847087049/
   563298250316800434130747930878000\beta^{18 }+  \]\[
  209487324417602647393969720230171393816148169/
   2175789194635434335691848949010176000\beta^{19 }+  \]\[
  116534543227065784496199379708432843606532209/
   509950592492679922427777097424260000\beta^{20 }+  \]\[
  97154545841875007760872827896741442704760279405783/
   179763703260779584814860560167220741120000\beta^{21 }+\]\[  
  63589340196765532818924025697977834662636895696063/
   49910355112067408764702871873350950960000\beta^{22 }+  \]\[
 \frac{ 1081118101264635857780839432289127041238668850456033706111}
   {361021961950469291097871871031671755481671680000}\beta^{23 }+\]\[
 \frac{ 356375621322438882451092769861339076326503647411315674153}
   {50768713399284744060638231863828840614610080000}\beta^{24}+\]\[
 \frac{16940275614897581006934617209480542541467898467761322446946311}
{1032098079458400443962151348949367489200543744000000}\beta^{25}+
\ldots ; \] 

\normalsize 
 for $S=4 $ we have 
\scriptsize

\[\mu^{sc}_2(\beta;4 )=  25/24\beta^{ }+  
125/24\beta^{2 }+  672481/36864\beta^{3 }+  1023905/18432\beta^{4 }+  
52911568205/339738624\beta^{5 }+  
  3913828525/9437184\beta^{6 }+ \]\[  
6477872812330883/6088116142080\beta^{7 }+  
  808881739511803/304405807104\beta^{8 }+   \]\[
1312229284857372282295/201989082115473408\beta^{9 }+ 
  788817409783096081715/50497270528868352\beta^{10 }+  \]\[ 
  3160853893631712353097707521/85320188285575967539200\beta^{11 }+ \]\[  
  1111816447643401403601363871/12798028242836395130880\beta^{12 }+  \]\[  
  247486135154355113495077631754905
/1226644934174194262272376832\beta^{13 }+ \]\[  
  997650087836403949664696038170499
/2146628634804839958976659456\beta^{14 }+ \]\[ 
  584118470663472307100357975615726090003
/549536930510039029498024820736000\beta^{15 }+\]\[  
  119490692365267473746026588902024104147
/49458323745903512654822233866240\beta^{16 }+  \]\[
  75231516928503202526968115197795794283230491/
   13775527996299013572722327154296094720\beta^{17 }+  \]\[
  380774901867875373055241939962215212798269391/
   30994937991672780538625236097166213120\beta^{18 }+  \]\[
  2488844034442803467342809373436528820987506306771761/
   90455627034897842723923889025969876369408000\beta^{19 }+\]\[  
  4164315947826337334759742811812526633925480445655387/
   67841720276173382042942916769477407277056000\beta^{20 }+  \]\[
\frac{  477762250350299448644651284870992097520157125075079561929}
{   3501284046765197777883466757306021198605949337600}\beta^{21 }+ \]\[ 
\frac{  323475485199189675060464172067459007298321694722493809311}
{   1069836792067143765464392620287950921796262297600}\beta^{22 }+  \]\[
  \frac{22722642803893232625438051419884395680648330384903392842545560507}
 {34015674771133249451693456240579457148696519004651520000}\beta^{23 }+ \]\[
 \frac{395259066522446341063795241951655099367981774139430908846294287}
   {268544800824736179881790444004574661700235676352512000}\beta^{24}+ \]\[
\frac{6084728860422066811333915921599524536570223529745573662895730522918147}
{1880930752144584161680841356279081662494322714881210449920000}\beta^{25}+
\ldots ; \] 

\normalsize 
 for $S=5 $ we have 
\scriptsize

\[\mu^{sc}_2(\beta;5 )= 24/25\beta^{ }+  
576/125\beta^{2 }+  6055684/390625\beta^{3 }+ 
 88583232/1953125\beta^{4 }+  5965944409/48828125\beta^{5 }+\]\[   
  381664545376/1220703125\beta^{6 }+   
24702345110160001/32043457031250\beta^{7 }+  
  49408024338787216/26702880859375\beta^{8 }+\]\[   
  30343101939335494566577/6979064941406250000\beta^{9 }+   
  7304688833140062896962/726985931396484375\beta^{10 }+\]\[ 
  24976583739183936582070618091/1090478897094726562500000\beta^{11 }+ \]\[  
  8796191915026154058040506541/170387327671051025390625\beta^{12 }+\]\[  
  8911059985666272515756028738821/77324867248535156250000000\beta^{13 }+\]\[  
  1978174836712939909453389680953153
/7752623409032821655273437500\beta^{14 }+\]\[  
  7827824088050927263381616299523403063371/
   13954722136259078979492187500000000\beta^{15 }+  \]\[
  178149953358339962348811011702738350763
/145361688919365406036376953125000\beta^{16 }+\]\[  
  252691976966731504434728564798576211177607283/
   94892110526561737060546875000000000000\beta^{17 }+  \]\[
  14086827680132981392155593552763142868513162237/
   2446437224512919783592224121093750000000\beta^{18 }+  \]\[
  760605956581255747499404677330273218538012759813429971/
   61356645590784028172492980957031250000000000000\beta^{19 }+\]\[  
  424766294604243617876871747409397674470116931765777859/
   15978293122600007336586713790893554687500000000\beta^{20 }+  \]\[
 \frac{  36597138298248146037023604780746892627367629352652423313851}
   {644244778703232295811176300048828125000000000000000}\beta^{21 }+  \]\[
\frac  {20299999966890101083247086095539288379392846407921044319247}
   {167772077787300077034160494804382324218750000000000}\beta^{22 }+  \]\[
 \frac{ 118988080243362194099399355687537253797147924730985196695439678841}
{463050934692948212614282965660095214843750000000000000000}\beta^{23 }+\]\[
\frac{39377920649210934440137802729203719357377831744561104304645072839}
{72351708545773158220981713384389877319335937500000000000}\beta^{24}+\]\[
\frac{7986797002170728087084830071643809156476089180682220421406874219943729}
{6945764020394223189214244484901428222656250000000000000000000}\beta^{25}+
\ldots ; \] 

\normalsize 
 for $S=\infty $ we have 
\scriptsize

\[\mu^{sc}_2(\beta;\infty )= 2/3\beta^{ }+  
8/3\beta^{2 }+  1684/225\beta^{3 }+  4112/225\beta^{4 }+
  4078436/99225\beta^{5 }+  14524432/165375\beta^{6 }+  
  1345827496/7441875\beta^{7 }+  \]\[ 1618729312/4465125\beta^{8 }+  
17298972613348/24312605625\beta^{9 }+  
  77922055600432/56729413125\beta^{10 }+ \]\[  
39494620955213052712/15099951538546875\beta^{11 }+  
  223093911068868088096/45299854615640625\beta^{12 }+ \]\[  
  340400824294138981864/37063517412796875\beta^{13 }+ 
  34631469408991021610336/2038493457703828125\beta^{14 }+ \]\[  
  214243030952262108916736/6866254187370703125\beta^{15 }+  
  1509913688905145083698261056/26510607417438284765625\beta^{16 }+ \]\[  
  5387350957208867401226859506164/52105126067451757691015625\beta^{17 }+\]\[  
  150032723154909173772497570866832
/803212237475770174787109375\beta^{18 }+ \]\[ 
  12868716540039518248876100881725639464
/38301175544032100784723310546875\beta^{19 }+\]\[  
  69172167192347970975285665601805186976
/114903526632096302354169931640625\beta^{20 }+  \]\[
  587981535915526074005698709538086238132952/
   547055690295410495508203044541015625\beta^{21 }+  \]\[
  14427886247033138339243473184986233280530016/
   7542737548012478044128254098974609375\beta^{22 }+  \]\[
  50140033308638033160521231008630786609014351328/
   14772287515227046673815530690796142578125\beta^{23 }+\]\[
  6122530233223105667416176223956400321808439321536/
   1019287838550666220493271617664933837890625\beta^{24}+\]\[ 
  19454620833972223897493892950557959025743730206728/
1834718109391199196887888911796880908203125\beta^{25}+
\ldots ; \]

\normalsize
\section{The second field derivative of the susceptibility on the sc lattice}

The HT  expansion coefficients of the second field derivative of the 
susceptibility 
$\chi_4^{sc}(\beta;S) =    \sum_{r=0}^\infty e^{sc}_r(S) \beta^r $ 
 for the spin $S=1/2$ Ising model on the sc lattice are

\scriptsize

\[ \chi^{sc}_4(\beta;1/2 )=- 2 - 48\beta^{ } - 
636\beta^{ 2} - 6464\beta^{ 3} - 55892\beta^{ 4} 
 - 2174432/5\beta^{ 5} - 47009464/15\beta^{ 6} - 
2239468288/105\beta^{ 7}   -\]\[  14570710772/105\beta^{ 8}  - 
823130010272/945\beta^{ 9} - 25080975789304/4725\beta^{ 10} - 
1640401398782848/51975\beta^{ 11} -\]\[ 
 28654566671774104/155925\beta^{ 12}    - 
2130434175575247424/2027025\beta^{ 13} - 
83969257269976828688/14189175\beta^{ 14}   -  \]\[ 
6995762565293277161216/212837625\beta^{ 15} - 
38389375874347206695732/212837625\beta^{ 16}   -  \]\[ 
272537955948789968719904/278326125\beta^{ 17}    - 
171450770247965944104542584/32564156625\beta^{ 18}  - \]\[ 
17364135415904212387813053568/618718975875\beta^{ 19}  - 
459165603261532162714762705624/3093594879375\beta^{20 }   - \]\[ 
50632615742924634090269478798784/64965492466875\beta^{21 }- 
2905534181681774015819695033964048/714620417135625\beta^{22 }- \]\[ 
26656952206944051765111802118842112/1264328430316875\beta^{23} + \ldots ;\]

\normalsize 
 for $S=3/2 $ we have 
\scriptsize

\[ \chi^{sc}_4(\beta;3/2   )=- 34/81 - 1360/243\beta^{ } -
 93332/2187\beta^{ 2} -
 14964800/59049\beta^{ 3} -
 686460020/531441\beta^{ 4} 
  - 28475741152/4782969\beta^{ 5}\]\[  - 
 16466501209144/645700815\beta^{ 6} 
  - 841246690421504/8135830269\beta^{ 7}  -
 29398318450235332/73222472421\beta^{ 8}\]\[    -  
8931263353244728480/5931020266101\beta^{ 9} -
 7324380876091964274424/1334479559872725\beta^{ 10}\]\[    - 
 516105814602829923152768/26422695285479955\beta^{ 11} -
 1388354744967134063721256/20383222077370251\beta^{ 12}\]\[   - 
 19482083790341476018437586496/83469294406831177845\beta^{ 13} \]\[   - 
 20711880288958153025618003387408/26292827738151821021175\beta^{ 14}\]\[   - 
 1862378117325556144984434269357824/709906348930099167571725\beta^{15} \]\[ - 
 55165347490194223586385335564809348
/6389157140370892508145525\beta^{ 16}\]\[    -
 1616965985438372498998669979001649568
/57502414263338032573309725\beta^{ 17} \]\[   - 
 732999948971904423452768940631161229816
/8079675963328211311576274625\beta^{ 18} \]\[   - 
 3929438071221989991526510624427800744055936/
   13539920979345416515939521016575\beta^{ 19} \]\[  - 
 43177317734309278235169045677054162343662744/
   46868957236195672555175265057375\beta^{ 20} \]\[   - 
 334403074421217846422253081480694505576587663808/
   115157027929332767468065626245970375\beta^{ 21} \]\[ - 
518438247051003931686289185513315884873853315680528/
      57002728825019719896692484991755335625\beta^{ 22}\]\[  - 
13365689587949165281017549839168414068849068375508736.0q0
/471982594671163280744613775731734178975\beta^{ 23} +\ldots ;\]

\normalsize 
 for $S=5/2 $ we have 
\scriptsize

\[ \chi^{sc}_4(\beta;5/2 )=- 518/1875 -
 29008/9375\beta^{ } - 69998684/3515625\beta^{ 2} -
 5272471232/52734375\beta^{ 3} 
  - 2844414630364/6591796875\beta^{ 4}\]\[  -
 833326652939872/494384765625\beta^{ 5} 
  - 378394915909716328/61798095703125\beta^{ 6}\]\[  - 
 58578552151074882304/2780914306640625\beta^{ 7} 
  -  72395571975800617857164/1042842864990234375\beta^{ 8} 
\]\[   -  2074743403700493228686432/9385585784912109375\beta^{ 9} 
  -  46823729015078926554382853992/68436563014984130859375\beta^{ 10} 
\]\[   -  400323116623696911029093612416/193577706813812255859375\beta^{ 11} 
\]\[   -  11116926395144690608476154517119544
/1814791001379489898681640625\beta^{ 12} 
 \]\[  -  6310855570247946288284041456287174848
/353884245269000530242919921875\beta^{ 13} 
\]\[   -  6786154374042242820656528025503036845616/
   132706591975875198841094970703125\beta^{ 14} 
 \]\[  -  1120190603336157033995947846635927899377408/
   7741217865259386599063873291015625\beta^{ 15} 
\]\[   -  74014613749903645129496649485963153762034847828/
   182886272066753008402884006500244140625\beta^{ 16} 
\]\[   -  191342478421101937387616373722580820134697643104/
   170827836545868194662034511566162109375\beta^{ 17} 
 \]\[  -  7672730502501429100182383421597889902154977881425928/
   2498357109483322346932254731655120849609375\beta^{ 18} 
\]\[   - \frac{ 283104980072078766849510370119039113064055926373107328}
   {33906275057273660422652028501033782958984375}\beta^{ 19} 
\]\[   -\frac{  30075272411361118946434461870227831961876710887255503717352}
   {1335059580380150379141923622228205204010009765625}\beta^{ 20} 
\]\[   -\frac{ 497691882640699590455209554389369868921427443544885316644672}
   {8245956231759752341758940019644796848297119140625}\beta^{ 21}\]\[  -
\frac{ 24088203725627269038762691093177859165434444552997006068058380432}
  {149916065380187719657811840079375542700290679931640625}\beta^{ 22}\]\[  -
\frac{3142078978083460712632106641114905081247261326222588639282709590784}
{7388720365166394754563583546769223175942897796630859375}\beta^{ 23}
  +\ldots ;\]
\normalsize 
 for $S=7/2 $ we have 
\scriptsize

\[ \chi^{sc}_4(\beta;7/2 )=- 78/343 -
 5616/2401\beta^{ } - 11400804/823543\beta^{ 2} -
 2581797312/40353607\beta^{ 3} 
  - 3519239144436/13841287201\beta^{ 4}\]\[  -
 443170531912608/484445052035\beta^{ 5} 
  - 72673924888074072/23737807549715\beta^{ 6}\]\[  - 
 11288026400816504064/1163152569936035\beta^{ 7} 
  - 11759049722495518107396/398961331488060005\beta^{ 8} 
 \]\[  -  5072943196157009398046432/58647315728744820735\beta^{ 9} 
  - 24819539572507636762254033208/100580146474797367560525\beta^{ 10} 
 \]\[  -  5324565839162940631591070893696
/7744671278559397302160425\beta^{ 11} 
 \]\[  -  2137219819143075408548160596079784
/1138466677948231403417582475\beta^{ 12} 
 \]\[  -  40149461844237709041872875837831744
/7969266745637619823923077325\beta^{13 } 
\]\[   -  473261103532483552996490192025092034704
/35534960418798146794873001792175\beta^{ 14} 
\]\[   -  6352290455096091688274123754314184933354752/
   182827371354716465259621594220740375\beta^{ 15} 
 \]\[  - 5618009167218016013327255626188062961343496564/
   62709788374667747584050206817713948625\beta^{ 16} 
 \]\[  -  22154870311834904888176642586051217005527325536/
   96915127488122882629895774172830647875\beta^{ 17} 
 \]\[  -  1901796212064607254613013754768784769455820558754184/
   3290946984114188725463370803586810309891375\beta^{ 18} 
 \]\[  -  634041889213121098792819026927419875060165564988449408/
   437695948887187100486628316877045771215552875\beta^{ 19} 
 \]\[  - \frac{ 2704958237057956159645421041805759360045727641468691422824}
  { 750648552341525877334567563444133497634673180625}\beta^{ 20} 
 \]\[  - \frac{528915257972566299669138770259567203716043997911337714069568}
   {59416720027648471367482309444924105312775284835625}\beta^{ 21}\]\[  -
\frac{4898439705317883749255529027247456117941283298534801926092066512}
  {224179284664317682469510753535698649345101149684813125}\beta^{ 22}\]\[  - 
\frac{25019088729086793512641417450671862464426876185331371583472846422272}
{469207242802416909408686007150217273079296706290313870625}\beta^{ 23}
+\ldots ;\]

\normalsize 
 for $S=1 $ we have 
\scriptsize

\[ \chi^{sc}_4(\beta;1 )=- 2/3 -
 32/3\beta^{ } - 872/9\beta^{2 } -
 18448/27\beta^{ 3} - 111424/27\beta^{ 4} -
 9117356/405\beta^{ 5} - 46179004/405\beta^{ 6} 
  - 13939286114/25515\beta^{ 7}\]\[  -
 383567190061/153090\beta^{ 8} -
 30574095401411/2755620\beta^{ 9} 
  - 219223908317179/4592700\beta^{ 10} -
 182291985575347033/909354600\beta^{ 11} 
 \]\[  -  4500048414442433423/5456127600\beta^{ 12} -
 1418927265266308826939/425577952800\beta^{ 13} 
 \]\[  - 118622262250443454071619/8937137008800\beta^{ 14} 
  -   1553092173114849639288953/29790456696000\beta^{ 15} 
 \]\[  -  1302075236595453962967411101/6434738646336000\beta^{ 16} 
  - 170021167715692957237957386611/218781113975424000\beta^{ 17} 
  \]\[ -  529194602562505786119523326547/179002729616256000\beta^{ 18} 
 \]\[  -  64179321630139176457147251907567/5755626229199616000\beta^{ 19} 
 \]\[  - 281033242505941699193085862470138829/6734082688163550720000\beta^{20} 
\]\[   -  43864969814703433492340864003743736939
/282831472902869130240000\beta^{21}\]\[  -
11644612005725977172750113663870487881/20334288901513466880000\beta^{22}\]\[ -
100273501512367064172773159728409309239433/47704241762950593300480000
\beta^{23} +\ldots ;\]
\normalsize 
 for $S=2 $ we have 
\scriptsize

\[ \chi^{sc}_4(\beta;2 )=- 13/40 - 39/10\beta^{ } -
 21477/800\beta^{2 } - 230579/1600\beta^{ 3} - 42532427/64000\beta^{ 4} 
  - 1419749291/512000\beta^{ 5} \]\[ -
 82615915409/7680000\beta^{ 6} - 67977029300407/1720320000\beta^{ 7} 
  - 4783529929225183/34406400000\beta^{ 8}\]\[  -
 4682984548608524819/9909043200000\beta^{ 9} 
  - 17267421029437387/11059200000\beta^{ 10}\]\[  -
 87848945215859563479059/17439916032000000\beta^{ 11} 
\]\[   -  1666399743942784751609809/104639496192000000\beta^{ 12} 
\]\[   -  828396566937595719558618139/16742319390720000000\beta^{ 13} 
\]\[   -  576738143333743247072069745107/3808877661388800000000\beta^{ 14} 
\]\[  -  334405167359231981319240028114783/731304510986649600000000\beta^{ 15} 
\]\[ -  1140641423822750220197630425402637/835776583984742400000000\beta^{ 16} 
\]\[   -  64143594253808431001893608829875376091
/15913186159069495296000000000\beta^{ 17} 
\]\[   -  2111353332495710849305926755501089171823
/179023344289531822080000000000\beta^{ 18} 
\]\[   -  2979418906150370209954631332141631918662619/
   87076954662428278259712000000000\beta^{ 19} 
\]\[   -  428885819891780515463677322708246468041306697/
   4353847733121413912985600000000000\beta^{ 20} 
\]\[   -  823975815817245901255497238099461346159023776839/
   2925785676657590149526323200000000000\beta^{ 21}\]\[  -
9195463554810292281801691207025791575582325306079/
11494158015440532730281984000000000000\beta^{22}\]\[  -
\frac{267580430135326655242364954499946616716917346286306147}
{118435804191099249252825563136000000000000}\beta^{23}
+\ldots ;\]
\normalsize 
 for $S=3 $ we have 
\scriptsize

\[ \chi^{sc}_4(\beta;3 )=- 20/81 -
 640/243\beta^{ } - 247328/15309\beta^{ 2} -
 31978304/413343\beta^{ 3} - 8294219312/26040609\beta^{ 4} \]\[
  - 278207263984/234365481\beta^{ 5} - 
 130188156863272/31639339935\beta^{ 6} 
  - 3846688382034088/284754059415\beta^{ 7} \]\[- 
 762246528787926677/17939505743145\beta^{ 8} 
  - 37529424937696612639/290619993038949\beta^{ 9} 
  \]\[-  349244458584999871190603/915452978072689350\beta^{10} 
  -  99754174970944850522437589/90629844829196245650\beta^{11} 
  \]\[ - 30462120537058915585927342979/9788023241553194530200\beta^{12} 
  \]\[ - 1980882441801111839651481763753/229039743852344752006680\beta^{ 13} 
  \]\[
 - 3416070435951176156484691321980443/144295038626977193764208400\beta^{ 14} 
  \]\[-  177969404461604209023950514141987827
/2782832887805988736881162000\beta^{ 15} 
  \]\[-  3352709929726227069971432070176196668933
/19635668856359056527433479072000\beta^{ 16} 
  \]\[-  2978768119531043384378227221364967818141
/6602763373676781645488620435200\beta^{ 17} 
  \]\[-  52247148682750608595111596705984898517146471/
   44244517115792325008087837479872000\beta^{ 18} 
  \]\[-  36440086386306899792766173955130384805746374389/
   11889133813543623334316174614234176000\beta^{ 19} 
  \]\[-  118263164885165746432905343009318732601250915111413/
   14980308605064965401238380013935061760000\beta^{ 20} 
  \]\[-  1633486247106443891632018585384344493653425715872713/
   80893666467350813166687252075249333504000\beta^{ 21}\]\[ 
-201394796873926905141402120920780016165537409831722033227/
     3924151760331187946715998598170345168279040000\beta^{22}\]\[-
\frac{105336245070893769981170884208359872012583122139466881339061}
{812299414388555904970211709821261449833761280000}\beta^{23}
+\ldots ;\]
\normalsize 
 for $S=4 $ we have 
\scriptsize

\[ \chi^{sc}_4(\beta;4 )=- 41/192 -
 205/96\beta^{ } - 226573/18432\beta^{ 2} -
 24446905/442368\beta^{ 3} -
 6049302757/28311552\beta^{ 4} 
   \]\[- 1016492393407/1358954496\beta^{ 5} - 
 176540607539593/72477573120\beta^{ 6} 
  - 329343378547431893/43834436222976\beta^{ 7} 
    \]\[- 1868850280479231542729/84162117548113920\beta^{ 8} 
 - 21959134152044708405185/346266997912240128\beta^{ 9} 
    \]\[-    113798549110115996809925531/646365062769514905600\beta^{ 10} 
\]\[- 97749441328565056960615208333/204768451885382322094080\beta^{ 11} 
 \]\[-249362989111033971789818242916107/196577713809967029210316800\beta^{ 12} 
-\]\[81279874617041213809806704733943663
/24532898683483885245447536640\beta^{13 } 
 \]\[-702609784938753363201046974134895893531
/82430539576505854424703723110400\beta^{ 14} 
  \]\[-  1359161270482536594655831798210237563707
/62804220629718746228345693798400\beta^{ 15} 
  \]\[-  75010690674803149194457051363209980151368939/
   1381236095885596280687399113064448000\beta^{ 16} 
  \]\[-  111471313009065587082868341767281238644333263319/
   826531679777940814363339629257765683200\beta^{ 17} 
  \]\[-  394921949354307587219788663525303335185032580468353/
   1190205618880234772683209066131182583808000\beta^{ 18} 
  \]\[-  11715733226398719272120352048117306052590740400952819/
   14472900325583654835827822244155180219105280\beta^{ 19} 
 \]\[- \frac{2042229717548707363389655189079742907112350105940250266321}
   {1042048823442023148179603201579172975775580160000}\beta^{ 20} 
  \]\[- \frac{47136257290173792745465620310586508288511838640010398785737}
   {10003668705043422222524190735160060567445569536000}\beta^{ 21} \]\[-
\frac{154147105819335515194960398749307387642331443403201481033451313}
  {13693910938459440197944225539685771798992157409280000}\beta^{22} \]\[-
\frac{4254552151267762591160574306225874314486527370326768755423587359}
{159137659747991810300320263113822021748287808208896000}\beta^{23} 
+\ldots ;\]
\normalsize 
 for $S=5 $ we have 
\scriptsize

\[ \chi^{sc}_4(\beta;5 )=- 122/625 - 5856/3125\beta^{ } -
 4047144/390625\beta^{ 2} - 87397744/1953125\beta^{ 3} 
  - 40584191432/244140625\beta^{ 4} \]\[ -
 3413049286972/6103515625\beta^{ 5} 
  - 44058466165396136/25177001953125\beta^{ 6} \]\[- 
 4571248791029577802/881195068359375\beta^{ 7} 
  -  17853589974526075203947/1211643218994140625\beta^{ 8} 
  \]\[-   252031228974639763438877/6231307983398437500\beta^{ 9} 
  - 668689069694544043183176007/6195902824401855468750\beta^{ 10} 
  \]\[-  383387899946671699082377218533/1363098621368408203125000\beta^{ 11} 
 \]\[- 367217198173699190695886498820157/511161983013153076171875000\beta^{12} 
  \]\[-  239687891427420811767428875020945797
/132902115583419799804687500000\beta^{ 13} 
  \]\[-  259319255048541927767469193901287957147
/58144675567746162414550781250000\beta^{ 14} 
  \]\[-  5424596188423034454801577680733462874647/
   498382933437824249267578125000000\beta^{ 15} 
  \]\[-  2518828080881691284486074999199611686164652107/
   95938714686781167984008789062500000000\beta^{ 16} 
  \]\[-  157258154774830832101306528135422257588073231763/
   2509166384115815162658691406250000000000\beta^{ 17} 
  \]\[-  29915144637921215468736780313259726491741571788515127/
   201831071022315882146358489990234375000000000\beta^{ 18} 
  \]\[-  26658496527852247125461323703953765358353974830413062077/
   76695806988480035215616226196289062500000000000\beta^{ 19} 
  \]\[- \frac{  38775456956865097363208131318415322617883651170393491792957}
   {47934879367800022009760141372680664062500000000000}\beta^{ 20} 
 \]\[- \frac{28753222365151170088763353539204996897737509003470890878529}
  {15380175198224605992436408996582031250000000000000}\beta^{21} \]\[-
\frac{5402077114786914370993088397829005478307160351921083839023464897}
  {1258290583404750577756203711032867431640625000000000000}\beta^{22} \]\[-
\frac{436407711706972712141306636704571306723558709975508196038542335923}
{44524128335860405059065669775009155273437500000000000000}\beta^{23}
+\ldots ;\]
\normalsize 
 for $S=\infty $ we have 
\scriptsize

\[ \chi^{sc}_4(\beta;\infty )=- 2/15 - 16/15\beta^{ } -
 7748/1575\beta^{ 2} - 11968/675\beta^{ 3} -
 9092812/165375\beta^{ 4} - 76615264/496125\beta^{ 5} 
  \]\[- 99106074968/245581875\beta^{ 6} -
 81786489088/81860625\beta^{ 7} 
  - 672319943649364/283647065625\beta^{ 8} \]\[-
  25764440028690656/4740958096875\beta^{ 9} 
  -  12193817882683009016/1006663435903125\beta^{ 10} 
   \]\[- 1194590824410496479872/45299854615640625\beta^{ 11} 
  -  129997205856238775124712/2310292585397671875\beta^{ 12} 
   \]\[- 50488733360875712401472/427831960258828125\beta^{ 13} 
  \]\[-  286548113139162254251640555536/1175303595506430624609375\beta^{ 14} 
  \]\[-  1752763532689588607191898646784/3525910786519291873828125\beta^{ 15} 
\]\[-15503241575922027536554940207047988/
 15475222442033172034231640625\beta^{16} 
  \]\[-  22085835316358113859252394975695072
/11053730315737980024451171875\beta^{ 17} 
  \]\[-  10433546123086912576147612692622341422792/
   2642781112538214954145908427734375\beta^{ 18} 
  \]\[-  314507739514931860192037192099709746816
/40658170962126383909937052734375\beta^{ 19} 
  \]\[-  1523768885597614160339166407392658110425688/
   101306609313964906575593156396484375\beta^{ 20} 
  \]\[-  2409815264657411364526851725244390060023467328/
   82970113028137258485410795088720703125\beta^{ 21} \]\[- 
6311543202855300900050155432370857084191792968336/
      113254204283407357832585735296103759765625\beta^{ 22} \]\[-
 36119515406225638631265501401866401556394139821824/
339762612850222073497757205888311279296875\beta^{ 23} 
+\ldots ;\]

\normalsize
\section{The nearest-neighbor correlation 
 function  on the bcc lattice}

The HT  expansion coefficients of the nearest-neighbor 
correlation  function  
$G^{bcc}(\beta;S) =  \sum_{r=1}^\infty h^{bcc}_r(S) \beta^r $ 
 for the spin $S=1/2$ Ising model on the bcc lattice are

\scriptsize


\[ G^{bcc}(\beta;1/2 )=\beta^{ } +  35/3\beta^{3 } +  2972/15\beta^{5 } +
  279011/63\beta^{7 } +   46439636/405\beta^{9 } +  
  100877055128/31185\beta^{11 } +\]\[
  587703506650264/6081075\beta^{13 } +  
  10981652882712713/3648645\beta^{15 } + 
 1049923978894758374012/10854718875\beta^{17 } +  \]\[
  1182698210781462071363672/371231385525\beta^{19 } +  
  2980059927747623321534851312/27842353914375\beta^{21 } + \]\[ 
  36044188353849211843538057776648/9861761756471625\beta^{23 }+ \]\[ 
 467893739905922667033437721565324744/3698160658676859375\beta^{25 }+
\ldots ;\]

\normalsize
 for $S=3/2 $ we have
\scriptsize
 
\[ G^{bcc}(\beta;3/2 )=25/81\beta^{ } +  30803/19683\beta^{3 } +
  17273740/1594323\beta^{5 } +  1218510986191/13559717115\beta^{7 } + \]\[ 
  1651602952704700/1977006755367\beta^{9 } + 
 368904784004965493176/44037825475799925\beta^{11 } + \]\[ 
  492889791165597167305240/5564619627122078523\beta^{13 } + \]\[ 
  104752019905133255226297955477/107561568019711995086625\beta^{15 } +  \]\[
719014231933437715774242521760188/65169402831783103583084355\beta^{17 }
 +\]\[  
  2888120726856975537171983027607683599096
/22566534965575694193232535027625\beta^{19 } +  \]\[
  11617675870256847978556677962637188857759888/
   7677135195288851164537708416398025\beta^{21 } + \]\[ 
  71498997365439540870374840531122363238535911174312/
   3933188288926360672871781464431118158125\beta^{23 }+\]\[ 
19227095690267333783342031962840464588953920714653016/
  86887704928100513046167535986978337493125\beta^{25}+
\ldots ;\]

\normalsize
 for $S=5/2 $ we have
\scriptsize
 
\[ G^{bcc}(\beta;5/2 )=49/225\beta^{ } +  8563387/10546875\beta^{3 } + 
 146907158324/35595703125\beta^{5 } +  
  124910824116740977/5005645751953125\beta^{7 } +  \]\[ 
  12633396780233047532/75084686279296875\beta^{9 } +  
  918574580902148479587206776/752802193164825439453125\beta^{11 } + \]\[  
  5275113437837300626182091150232/566214792430400848388671875\beta^{13 }
 +\]\[  
  147100017398418477862743995770700779703
/1990598879638127982616424560546875\beta^{15 } + \]\[ 
  2026761533354127394524744740327247095199773468/
   3357791955145585234276950359344482421875\beta^{17 } + \]\[ 
  923930682673189622203654344267935410123007926667704/
   183093885309277766282320953905582427978515625\beta^{19 } +  \]\[
  20662278930711838725176710189703801891756120628157824272/
   480621448936854136491092504002153873443603515625\beta^{21 } +  \]\[
\frac{55006945972957908791809250974348782458333571208842207966898086712}
 {147861333425271029147207713094758219085633754730224609375}\beta^{23 }+\]\[
\frac{ 941196584075234797031460043737709817862544435614126927707209027677064}
  {288555918546766166575545664585433769389055669307708740234375}\beta^{25 }+
\ldots ;\]

\normalsize
 for $S=7/2 $ we have
\scriptsize
 
\[ G^{bcc}(\beta;7/2 )=9/49\beta^{ } +  68865/117649\beta^{3 } +  
175962774804/69206436005\beta^{5 } +  
  62241299130587/4747561509943\beta^{7 } +\]\[
  38606448815320511372/512950283341791435\beta^{9 } +   
308204981225047172923470488/
663828966733662625899465\beta^{11 } +  \]\[
2730746463220656739313579150632/906126539591449484352769725\beta^{13 }
 + \]\[ 
  6496994179077337023343099283347158739
/319814643769183321153857016129575\beta^{15 } + \]\[ 
  450905998614188301340432725756645978807946748/
   3198199207108055126786560547703411379875\beta^{17 } +  \]\[
  5366676396560031762281412001852813674512791888472/
   5359542231271678781468918165841376790394525\beta^{19 } + \]\[ 
  3765796384261059100878995038308177013507318917254069968/
   519679767005671761231623697769015498362466048125\beta^{21 } +  \]\[
\frac{2142033783842827619596099807233542244889637000649050914927896136}
{40217763668778592235030229184304337692511146253455474625}\beta^{23 }+ \]\[
\frac{ 293099299372112401566171075016839192895870885684480329564114554853064}
  {739001407413806632318680461261592205099892312407244346234375}\beta^{25}+
\ldots ;\]
\normalsize
 for $S=1 $ we have
\scriptsize
 
\[ G^{bcc}(\beta;1 )=4/9\beta^{ } +  82/27\beta^{3 } +  207547/7290\beta^{5 } +
  297746761/918540\beta^{7 } +  91974431627/22044960\beta^{9 } + \]\[ 
  47009808483889/808315200\beta^{11 } + 
 2916275405988118963/3404623622400\beta^{13 } + \]\[ 
  28168557867000207712369/2144912882112000\beta^{15 } + 
  9810790857180605510951828569/47256720618691584000\beta^{17 } +  \]\[
  4179569959982894370718261061173/1243215265507117056000\beta^{19 } +\]\[  
  28984786765873359044135303189199533/522150411512989163520000\beta^{21 } 
+ \]\[ 
  86372796723290657962386506799241569262843
/92737045987175953376133120000\beta^{23 }+\]\[ 
 2643833135030372720467318889023139446639475209/
  166926682776916716077039616000000\beta^{25}+
\ldots ;\]

\normalsize
 for $S=2 $ we have
\scriptsize
 
\[ G^{bcc}(\beta;2 )=1/4\beta^{ } +  10129/9600\beta^{3 } + 
 149261/24576\beta^{5 } +  430230920069/10321920000\beta^{7 } +  
  108558817003663/339738624000\beta^{9 } + \]\[
 39423939340616154319/14948499456000000\beta^{11 } + 
  5988802675563907448672593/261180182495232000000\beta^{13 } +  \]\[
  129866236304405726743544159291/626832437988556800000000\beta^{15 } + \]\[ 
  5259755282482369240925068910783027
/2727974770126199193600000000\beta^{17 } + \]\[
  47983312063727669635816212091131676582949/
   2612308639872848347791360000000000\beta^{19 } +  \]\[
  1376119314267494850255491078495999134558127/
   7716357828547490504245248000000000\beta^{21 } + \]\[ 
  25507668903508543610665395577689698404232588748313/
   14502343370338683581978640384000000000000\beta^{23}+ \]\[ 
 191854056242268773025225551422414559088071122898016769/
  10915043714251706811140403898613760000000000\beta^{25}+
\ldots ;\]
\normalsize
 for $S=3 $ we have
\scriptsize
 
\[ G^{bcc}(\beta;3 )=16/81\beta^{ } +  13256/19683\beta^{3 } + 
 1222614338/390609135\beta^{5 } +  
  1640348357359/94918019805\beta^{7 } +\]\[
  58844218109237287/553561891502760\beta^{9 } +  
  169867965161102491738831/241679586211189988400\beta^{11 } +  \]\[
  2133842737061995320806250049/436266178766370956203200\beta^{13 } + \]\[ 
  23849804093356086063236045304923/674626154619633633183312000\beta^{15 } 
+\]\[  
  2107761689901071265184828127128355696161/
   8011352893394495063192859461376000\beta^{17 } +\]\[  
  444717028609291924024149398236783506463015177/
   221930497852814302240568592799037952000\beta^{19 } + \]\[ 
  16773972664614757224085175117024144401973299248103/
   1078582219564677508889163361003324446720000\beta^{21 } + \]\[ 
  \frac{265546887558934784450078237181469760513035848741779329999}
   {2166131771702815746587231226190030532890030080000}\beta^{23 }+ \]\[ 
 \frac{ 9370064222720584576907550089006633357481691400238464854004597}
 {9570364009523349571285403417530498536223223808000000}\beta^{25}+
\ldots ;\]

\normalsize
 for $S=4$ we have
\scriptsize
 
\[ G^{bcc}(\beta;4 )=25/144\beta^{ } +  116081/221184\beta^{3 } +
  13223059415/6115295232\beta^{5 } +  
  10424618130411361/986274815016960\beta^{7 } +  \]\[
  9955302687794205025/173133498956120064\beta^{9 } + 
  19114277844080002854176329/56880125523717311692800\beta^{11 } +\]\[  
  5068917011938583328405199127779/2453289868348388524544753664\beta^{13 }
 +\]\[  
  18641401431966060853031287921429026607
/1413094964168671790137778110464000\beta^{15 } + \]\[ 
  193503805687923259688151928337023449711663337/
   2231635535400440198781016998995967344640\beta^{17 } + \]\[ 
  2848834394460535871581778940070193965717386722656969/
   4884603859884483507091890007402373323948032000\beta^{19 } +\]\[  
  479825273481996516714712320858777414070162275981757943/
   120044024460521066670290288821920726809346834432\beta^{21 } + \]\[ 
\frac{511039696724592472116256952078407325418
93738695201059609016868163}
{1836846437641195470391446636991290686029612026251182080000}\beta^{23 }+ \]\[ 
\frac{ 288805185168435131621544594246122739474411592317309014280299512388687}
{1472032762547935430880658452740150866299904733385295134720000}\beta^{25}+
\ldots ;\]
\normalsize
 for $S=5 $ we have
\scriptsize
 
\[ G^{bcc}(\beta;5 )=4/25\beta^{ } +  524642/1171875\beta^{3 } +
  500170369/292968750\beta^{5 } +  
  1484108245918561/192260742187500\beta^{7 } + \]\[ 
  325144369669136707061/8374877929687500000\beta^{9 } +  
  1372040503341682786740867611/6542873382568359375000000\beta^{11 } +  
\]\[ 30401579771402271011896174072043/25517206192016601562500000000\beta^{13 }
 +\]\[  
  588465852321007941758629542606857571851/
   83728332817554473876953125000000000\beta^{15 } + \]\[ 
  24299687649360225254081036686448116936239323/
   569352663159370422363281250000000000000\beta^{17 } +  \]\[
  97632025778098774927156025495568077779536857640091811/
   368139873544704169034957885742187500000000000000\beta^{19 } + \]\[ 
\frac{  118003725622221044713115666605161721665477406410634692389}
   {70281248585807159543037414550781250000000000000000}\beta^{21 } + \]\[ 
\frac{1764357411506629334309654188029032976732413888232089919925422473}
   {163429741656334663275629281997680664062500000000000000000}\beta^{23 }+\]\[ 
\frac{2930780718549491261010275964790486329673412530593353390686668295287049}
{41674584122365339135285466909408569335937500000000000000000000}\beta^{25}+
\ldots ;\]
\normalsize
 for $S=\infty $ we have
\scriptsize
 
\[ G^{bcc}(\beta;\infty )=1/9\beta^{ } +  49/225\beta^{3 } + 
 103892/178605\beta^{5 } +  369689933/200930625\beta^{7 } +  
  31446417476/4862521125\beta^{9 } + \]\[ 
 40932678003454168/1677772393171875\beta^{11 } +  
  5636458876589702552/58242670220109375\beta^{13 } +  \]\[
  10564051571495116864184173/26510607417438284765625\beta^{15 } + \]\[ 
  64147649756693472660139600804132/37984636903172331356750390625\beta^{17 }
 +\]\[  
  582518103635561888035633411254389288
/79548595360682055475963798828125\beta^{19 } + \]\[ 
  107253483031834250506702779872074490416
/3315489032093396942473957845703125\beta^{21 } +\]\[  
  1331335124354772430341534709286267330742274099112/
   9173590546955995984439444558984404541015625\beta^{23 }+\]\[ 
\frac{ 837797563114290463414089968588071049963673640856}
 {1270189460347753290153153862013225244140625}\beta^{25}+
\ldots ;\]

\normalsize
\section{The susceptibility on the bcc lattice}

The HT  expansion coefficients of the susceptibility 
$\chi^{bcc}(S;\beta) =   \sum_{r=0}^\infty a^{bcc}_r(S) \beta^r $ 
 for the spin $S=1/2$ Ising model on the bcc lattice are

\scriptsize
\[ \chi^{bcc}(\beta;1/2 )=1 + 8\beta^{ } + 56\beta^{ 2} + 1168/3\beta^{ 3} 
+ 7832/3\beta^{ 4} +
 262096/15\beta^{ 5} +
 5186272/45\beta^{ 6} +
 239256896/315\beta^{ 7} + 
  223501256/45\beta^{ 8} +\]\[ 
 92020145776/2835\beta^{ 9} +
 2992356296032/14175\beta^{ 10} + 
  213982290222656/155925\beta^{ 11} +
 594224833766272/66825\beta^{ 12} +\]\[ 
 350264122225879712/6081075\beta^{ 13} + 
  205770248144874592/552825\beta^{ 14} +
 1535420477610758667232/638512875\beta^{ 15} + \]\[ 
  9901906655322559820312/638512875\beta^{ 16} +
 1085334591165630782296336/10854718875\beta^{ 17} + \]\[ 
  62829959108514755867552/97594875\beta^{ 18} + 
  7692533427793876337979291776/1856156927625\beta^{ 19} + \]\[ 
  35332545614795796034757603392/1325826376875\beta^{ 20} + 
  2568708718024303514802517007264/14992036723125\beta^{ 21} + \]\[ 
  2359516416887631689794514525990624/2143861251406875\beta^{ 22} + \]\[
  348549803690378595986956983640658272/49308808782358125\beta^{ 23} +\]\[
 87152802098662246655746762503038432/1921122420091875\beta^{ 24}+ \]\[
1076583145712120200365204858124445508992/3698160658676859375\beta^{25}+
 \ldots ;\]

\normalsize
 for $S=3/2 $ we have
\scriptsize

\[ \chi^{bcc}(\beta;3/2 )=5/9 + 200/81\beta^{ } +
 2440/243\beta^{ 2} + 801424/19683\beta^{ 3} + 
3163000/19683\beta^{ 4} + 
1010101520/1594323\beta^{ 5} + \]\[  
  35400746528/14348907\beta^{ 6} +  
130204164968768/13559717115\beta^{ 7} + 
  301948781090024/8135830269\beta^{ 8} + \]\[  
283495979372602160/1977006755367\beta^{ 9} +  
  40389460803306016/73222472421\beta^{ 10} +  \]\[ 
93415848776861307917888/44037825475799925\beta^{ 11} +  
  92096152524709358158208/11324012265205695\beta^{ 12} +  \]\[  
  173473051288457117219498144/5564619627122078523\beta^{ 13} + \]\[ 
  69664410672221682779351217824/584285060847818244915\beta^{ 14} +  \]\[
  41496110554011756003070418696032/91013634478217841996375\beta^{ 15} + \]\[ 
411871084151951614364830730865304/236635449643366389190575\beta^{ 16} +\]\[   
  39359563977305553345098085639228464
       /5924491166525736689371305\beta^{ 17} + \]\[ 
  31828782656286425620428732552152668832
       /1256838483184388426245198275\beta^{ 18} + \]\[ 
  2178177256627788620012581292039952279321728/
   22566534965575694193232535027625\beta^{ 19} + \]\[ 
  24879296500926926422959861637620672437849152/
   67699604896727082579697605082875\beta^{ 20} + \]\[ 
  10740599724478721796654153328817461487222678816/
   7677135195288851164537708416398025\beta^{ 21} + \]\[ 
  2246917041666430505758828570520774276537818706144/
   422242435740886814049573962901891375\beta^{ 22} + \]\[ 
  79602251118823849570544123390839294442512041505388896/
   3933188288926360672871781464431118158125\beta^{ 23} +\]\[ 
 15558445562856960287199566619565043453628559166207584/
   202278254859069977461977332456457505275\beta^{ 24}+\]\[ 
279362973202137530013748502902169433533435378858176569728/
955764754209105643507842895856761712424375\beta^{ 25}
\ldots ;\]

\normalsize
 for $S=5/2 $ we have
\scriptsize

\[ \chi^{bcc}(\beta;5/2 )=7/15 + 392/225\beta^{ } +
 505288/84375\beta^{2 } +
 649706288/31640625\beta^{ 3} + 
32577735736/474609375\beta^{ 4} +  \]\[
  8160011552752/35595703125\beta^{ 5} + 
10099670388779552/13348388671875\beta^{ 6} + \]\[
  12494523230238728896/5005645751953125\beta^{ 7} +
  614159080049110361528/75084686279296875\beta^{ 8} + \]\[
  30179019899482853442064/1126270294189453125\beta^{ 9} + \]\[ 
  36915221676049728781254368/422351360321044921875\beta^{ 10} + \]\[ 
17380505171728061296686387684928/60976977646350860595703125\beta^{ 11} + \]\[ 
24219210719679938252145722148736/26132990419864654541015625\beta^{ 12} + \]\[ 
  6978512313860176517864558250560096
        /2316333241760730743408203125\beta^{ 13} + \]\[ 
  93375602261136669238165758371525282912
/9554874622263014316558837890625\beta^{ 14} + \]\[ 
  567822392635496562210430517115390596447456/
   17915389916743151843547821044921875\beta^{ 15} + \]\[ 
  14849655866975127724753953310212102448372024/
   144701226250617764890193939208984375\beta^{ 16} + \]\[ 
  1115573091651392808684015389247489223490978875024/
   3357791955145585234276950359344482421875\beta^{ 17} + \]\[ 
  193225710769057315886763570821534876546559569781152/
   179881711882799208979122340679168701171875\beta^{ 18} + \]\[ 
  635822597235056234493116860806163011779199781348972672/
   183093885309277766282320953905582427978515625\beta^{ 19} + \]\[ 
  154001339904638270067556763919367493364044437223490511936/
   13732041398195832471174071542918682098388671875\beta^{ 20} + \]\[ 
\frac{  52215360999264913223602876893614085697216443429068250008672}
{   1441864346810562409473277512006461620330810546875}\beta^{ 21} + \]\[ 
 \frac{ 374184010155588930870812023448639519937707060664611332836678944}
  { 3202602539550383813349299089168198406696319580078125}\beta^{ 22} + \]\[ 
\frac{86131874101125888946014602302961851386192201109343502108757848284512}
  {228512969839055226863866465691899065859615802764892578125}\beta^{23} + \]\[
\frac{6544940552103506666937148197112556563902575888935654758410232301332832}
 {5386377146206301776076852405594763695262372493743896484375}\beta^{24}+ \]\[ 
\frac{161459550794658695445567359635238165307463513468831674902019534293334656}
{41222274078109452367935094940776252769865095615386962890625}\beta^{25}
\ldots ;\]

\normalsize
 for $S=7/2 $ we have
\scriptsize

\[ \chi^{bcc}(\beta;7/2 )=3/7 + 72/49\beta^{ } + 
11160/2401\beta^{2 } + 1725744/117649\beta^{ 3} + 
1822209240/40353607\beta^{ 4} + 
  9611766494832/69206436005\beta^{ 5} + \]\[ 
40910930327968/96889010407\beta^{ 6} + 
  30459052389751616/23737807549715\beta^{ 7} + 
901122336202962680/232630513987207\beta^{ 8} + \]\[ 
  5996542593162563484016/512950283341791435\beta^{ 9} + 
  6181044640299261251574304/175941947186234462205\beta^{ 10} + \]\[ 
  350335801693087392158111564864/3319144833668313129497325\beta^{ 11} + \]\[ 
  4407803183274341269705755561088/13940408301406915143888765\beta^{ 12} + \]\[ 
  858110034307517221737387299709536
       /906126539591449484352769725\beta^{ 13} + \]\[ 
  35185889422806784045986611605376416
       /12432056123194686925320000627\beta^{ 14} + \]\[ 
  13523795172591494765233059750248099122912/
   1599073218845916605769285080647875\beta^{ 15} + \]\[ 
  2767662145666708400555539887505625949833768/
   109696422812829879155772956532444225\beta^{ 16} + \]\[ 
  240691238453763993101681230137884581151770606864/
   3198199207108055126786560547703411379875\beta^{ 17} + \]\[ 
  9034466759489565587993498234804649133189672359136/
   40297310009561494597510662901062983386425\beta^{ 18} + \]\[ 
  17895663275426806388957441770425447524364467726082176/
   26797711156358393907344590829206883951972625\beta^{ 19} + \]\[ 
  2609236512686093763843848697309519715935274775674524352/
   1313087846661561301459884950631137313646658625\beta^{ 20} + \]\[ 
 \frac{  3072439973764562246761973532170973927293741160281298721184}
{   519679767005671761231623697769015498362466048125}\beta^{ 21} + \]\[ 
 \frac{ 89597824035898117188357344936884761198795678948273719263084064}
  { 5097954578372238843329982150374488235836119438896625}\beta^{ 22} + \]\[ 
 \frac{10505261625813496811684507762698361329814776334121557662276939597408}
{201088818343892961175151145921521688462555731267277373125}\beta^{ 23} +  \]\[ 
\frac{18722432417055350869665075266751570901505460494470447343165147814496}
{120653291006335776705090687552913013077533438760366423875}\beta^{ 24} + \]\[ 
\frac{48656656548896432011649694618405535739010322859110992235826169613951616}
{105571629630543804616954351608798886442841758915320620890625}\beta^{ 25} +
 \ldots ;\]
\normalsize
 for $S=1 $ we have
\scriptsize
 
\[ \chi^{bcc}(\beta;1 )=2/3\beta^{ } + 32/9\beta^{ } + 
464/27\beta^{2 } + 6704/81\beta^{ 3} + 
94052/243\beta^{ 4} + 6589228/3645\beta^{ 5} + 
91122502/10935\beta^{ 6} + \]\[
  8815758946/229635\beta^{ 7} + 
483832129771/2755620\beta^{ 8} +  
6635894630401/8266860\beta^{ 9} + \]\[
  129423968404717/35429400\beta^{ 10} + 
136015769938958297/8184191400\beta^{ 11} + 
  7402346489685050273/98210296800\beta^{ 12} + \]\[
1308970217819257777787/3830201575200\beta^{ 13} + 
  248705271355907245516633/160868466158400\beta^{ 14} + \]\[
  16873212392338754585674433/2413026992376000\beta^{ 15} + 
  522411230694300253583828813/16546470804864000\beta^{ 16} + \]\[
  841941252787133972984605734169/5907090077336448000\beta^{ 17} + \]\[ 
  22774394808866366311312767606059/35442540464018688000\beta^{ 18} + \]\[ 
  5851614730985264166632373177487337/2020224806449065216000\beta^{ 19} + \]\[ 
  1580917022736817140044087163547783699/121213488386943912960000\beta^{ 20} + 
\]\[   149471732527607885838924219561534052267
       /2545483256125822172160000\beta^{ 21} + \]\[ 
  6339403392731735441879765201819841545359
       /24000270700614894766080000\beta^{ 22} + \]\[ 
  13771912004892938532233358601488508268424179
/11592130748396994172016640000\beta^{ 23} + \]\[
	1485528000774137331413153549266755015208472011/
278211137961527860128399360000\beta^{ 24}+\]\[
	500700711408879795838781278653136636958446930729/
20865835347114589509629952000000\beta^{ 25} + \ldots ;\]

 \normalsize
 for $S=2 $ we have
\scriptsize
 
\[ \chi^{bcc}(\beta;2 )=1/2 + 2\beta^{ } + 
147/20\beta^{ 2} + 32329/1200\beta^{ 3} + 615787/6400\beta^{ 4} + 
26365109/76800\beta^{ 5} + 1239288849/1024000\beta^{ 6} + \]\[ 
  5502200300813/1290240000\beta^{ 7} + 
29343263492329/1966080000\beta^{ 8} + 
  77612240856842861/1486356480000\beta^{ 9} + \]\[ 
3604820351921046047/19818086400000\beta^{ 10} + 
  8285806075209581133697/13079937024000000\beta^{ 11} + \]\[ 
  153439650446591514621641/69759664128000000\beta^{ 12} +  
  22662553119616798929896387/2967956619264000000\beta^{ 13} + \]\[ 
  11513465523796555577015411713/435300304158720000000\beta^{ 14} + \]\[ 
  50242534931710882764880197341381/548478383239987200000000\beta^{ 15} + \]\[ 
  3705953570334177123940072296342983
        /11700872175786393600000000\beta^{ 16} + \]\[ 
  2613607017302541692720960666752989989
/2386977923860424294400000000\beta^{ 17} + \]\[ 
  360944378749992022967523656567921794657
/95479116954416971776000000000\beta^{ 18} + \]\[ 
  4261431754404240912450087700937225846036093/
   326538579984106043473920000000000\beta^{ 19} + \]\[ 
  2239317753380036506403617253449242530645569/
   49758259807101873291264000000000\beta^{ 20} + \]\[ 
  5238735562646801603162926994137911479344202653/
   33759065499895270956072960000000000\beta^{ 21} + \]\[ 
  34411892855018109702323404036200931551870383067311/
   64367284886466983289579110400000000000\beta^{ 22} + \]\[ 
  14871809843018572845211912123453777210765334349154739/
   8075168467574948812692652032000000000000\beta^{ 23}  + \]\[ 
2002240928788513679827675275042696810973759752927467411/
315828811176264664674201501696000000000000\beta^{ 24}  +\]\[ 
744317994741895592012780004297934249341361653039046500409/
34109511607036583784813762183168000000000000\beta^{ 25} + \ldots ;\]
\normalsize
 for $S=3 $ we have
\scriptsize

\[ \chi^{bcc}(\beta;3 )=4/9 + 
128/81\beta^{ } + 
3776/729\beta^{ 2} + 
333376/19683\beta^{ 3} + 
9568208/177147\beta^{ 4} + 
67220141584/390609135\beta^{ 5} + \]\[ 
  5715508118248/10546446645\beta^{ 6} + 
161915791123064/94918019805\beta^{ 7} + 
  4556563080673571/854262178245\beta^{ 8} + \]\[ 
1153702190558804647/69195236437845\beta^{ 9} + 
  323189298083171443403/6227571279406050\beta^{ 10} + \]\[ 
  4878729685652556882670447/30209948276398748550\beta^{ 11} + 
  1634806467913538111304526499/3262674413851064843400\beta^{ 12} + \]\[ 
  593343409331869814387982953863/381732906420574586677800\beta^{ 13} + \]\[ 
  4722752458277405438269463148923/981598902224334651457200\beta^{ 14} + \]\[ 
  13812555159452970087683367290205961/927610962601996245627054000\beta^{ 15} + 
\]\[ 
  3309189999629106495293071383607954463
/71925526946370170430159264000\beta^{ 16} + \]\[ 
  1848576574350671390526080576463387475493
/13005443008757297180507888736000\beta^{ 17} + \]\[ 
  923978906111238215533527700788697214635537/
   2106881767418682143242277975232000\beta^{ 18} + \]\[ 
  37532706783772993121805965447447371685559958249/
   27741312231601787780071074099879744000\beta^{ 19} + \]\[ 
  20820356817785093863998436668839675420038570026773/
   4993436201688321800412793337978353920000\beta^{ 20} + \]\[ 
  1732270153545812627990779646555002318195108021817383/
   134822777445584688611145420125415555840000\beta^{ 21} + \]\[ 
  7392096825048586474155341597145321312941459712805206509/
   186864369539580378415047552293825960394240000\beta^{ 22} + \]\[ 
\frac{  4710748818025156061123183393022391458898740412057177171609}
  { 38680924494693138331914843324821973801607680000}\beta^{ 23}+\]\[ 
\frac{1288934376524071598990004582735957897817706772171065320029523}
{3440326931528001479873837829831224964001812480000}\beta^{ 24}+\]\[ 
\frac{2166554672715550652991773534643901939971022006367674892176963201}
{1879892930442086522931061385586347926758133248000000}\beta^{25}+
 + \ldots ;\]

\normalsize
 for $S=4 $ we have
\scriptsize
 
\[ \chi^{bcc}(\beta;4 )=5/12 + 25/18\beta^{ } + 7385/1728\beta^{ 2} + 
1088243/82944\beta^{ 3} + 
625804685/15925248\beta^{ 4} + 
  89889576215/764411904\beta^{ 5} +  \]\[ 
25527116804779/73383542784\beta^{ 6} + 
  126805148497492481/123284351877120\beta^{ 7} + \]\[
4086652186627741469/1352605460594688\beta^{ 8} +   
  4032208318564807605365/454475434759815168\beta^{ 9} + \]\[
  1131917027324055917899447/43629641736942256128\beta^{ 10} + \]\[ 
  43680615504096107519305439609/575911270927637780889600\beta^{ 11} +  \]\[ 
  4889296626942198834126626327281/22114992803621290786160640\beta^{ 12} + \]\[ 
  1778308429675262169560212244815547/2759951101891937090112847872\beta^{ 13} + 
\]\[ 
  225667017874605721819781853394188949
/120434229900739073023106088960\beta^{ 14} + \]\[ 
  4663832140290914810894054911998982017077
/856009449448330026718077124608000\beta^{ 15} + \]\[ 
  27034040535463153750833196837500712453197211/
   1709279668658425397350656402417254400\beta^{ 16} + \]\[ 
  12805862108068838902569265411169300376260674537/
   278954441925055024847627124874495918080\beta^{ 17} + \]\[ 
  17818305758510008504812908702485523000512292124643/
   133898132124026411926861019939758040678400\beta^{ 18} + \]\[ 
  235504871022832837079895782379876873282556475609643689/
   610575482485560438386486250925296665493504000\beta^{ 19} + \]\[ 
  3740662012739628014309204777333699909481332499555292033/
   3349442646777931547720153147933055993564364800\beta^{ 20} + \]\[ 
\frac{  25471812533287169484752551517677332431928085394998426043083}
  { 7877889105221695000237800203938547696863386009600}\beta^{ 21} + \]\[ 
 \frac{ 389040139471180505541006635308632692527813671407924248240506689}
  { 41595254475570549601255585076795531839438678130688000}\beta^{ 22} + \]\[ 
 \frac{6211520737954471941481241438754077144787289516210155079259513276963}
{229605804705149433798930829623911335753701503281397760000}\beta^{ 23} + \]\[ 
\frac{125348196418625156057916587295604555779959427093384657567165243876769}
{1603065981941406955977989792283308235080388677455577088000}\beta^{ 24} 
+ \]\[  
\frac{956404580968344494410646350181557198768701802066390656968648043170007961}
{4232094192325314363781893051627933740612226108482723512320000}\beta^{ 25}+
 \ldots ;\]

\normalsize
 for $S=5 $ we have
\scriptsize

\[ \chi^{bcc}(\beta;5 )=2/5 + 32/25\beta^{ } + 
11824/3125\beta^{ 2} + 13077136/1171875\beta^{ 3} + 
188173684/5859375\beta^{ 4} + 
  13526899396/146484375\beta^{ 5} + \]\[  
14419545545834/54931640625\beta^{ 6} + 
  35849948440838162/48065185546875\beta^{ 7} + 
2023967531150679349/961303710937500\beta^{ 8} + \]\[
  31099700478224528402473/5234298706054687500\beta^{ 9} + 
  21848574095101059112755209/1308574676513671875000\beta^{ 10} + \]\[
  38364615546894072568261137251/817859172821044921875000\beta^{ 11} + \]\[ 
6448231317426608780911750931207/49071550369262695312500000\beta^{ 12} + \]\[  
106718644121359439885564732934961/289968252182006835937500000\beta^{ 13} 
+\]\[  
  5741420042662110477307412103076581863
/5581888854503631591796875000000\beta^{ 14} + \]\[ 
  30081763075825461239986893284676806784611/
   10466041602194309234619140625000000\beta^{ 15} + \]\[ 
  1342750204671933454477789358306175112872667/
   167456665635108947753906250000000000\beta^{ 16} + \]\[ 
  1591845365507280894249992321263500367221076763/
   71169082894921302795410156250000000000\beta^{ 17} + \]\[ 
  767532560443105358296394835174869405153553787907/
   12317725885659456253051757812500000000000\beta^{ 18} + \]\[ 
  7987275318930075762055868344470533163132698578420140491/
   46017484193088021129369735717773437500000000000\beta^{ 19} + \]\[ 
 \frac{  2222572497937148317438628484221769035596851597187635109583}
 {  4601748419308802112936973571777343750000000000000}\beta^{ 20} + \]\[ 
 \frac{ 49948012995617460773193297915966844969501652454256317586703}
  { 37167968002109555527567863464355468750000000000000}\beta^{ 21} + \]\[ 
\frac{  41025376255356987936696536525801182682756832754382602786610279}
  { 10981445091532368678599596023559570312500000000000000}\beta^{ 22} + \]\[ 
\frac{3606541399815139961660065856084585228625345924124341940594684760641}
{347288201019711159460712224245071411132812500000000000000}\beta^{ 23} +\]\[ 
\frac{1202175135959558914045523893624257432237946963651511075815861872355981} 
{41674584122365339135285466909408569335937500000000000000000}\beta^{ 24} +\]\[ 
\frac{417389600493233221539249837694526848207390067230216532700872347045096489}
 {5209323015295667391910683363676071166992187500000000000000000}\beta^{ 25} +
 \ldots ;\]

\normalsize
 for $S=\infty $ we have
\scriptsize

\[ \chi^{bcc}(\beta;\infty)=1/3 + 8/9\beta^{ } + 
296/135\beta^{ 2} + 10928/2025\beta^{ 3} + 
551368/42525\beta^{ 4} + 27795632/893025\beta^{ 5} + 
  989620736/13395375\beta^{ 6} + \]\[ 
35218619648/200930625\beta^{ 7} + 
2739580321064/6630710625\beta^{ 8} + 
  213043482548432/218813450625\beta^{ 9} + \]\[
 682431678417159872/298680360103125\beta^{ 10} + 
  2185510965979723549568/407698691540765625\beta^{ 11} + \]\[
  1394385908519458231424/111190552238390625\beta^{ 12} +  
  107626809815644272096992/3669288223866890625\beta^{ 13} + \]\[
  64055961266381160958056928/935668497086057109375\beta^{ 14} + \]\[ 
  38118371922464507007900011168/238595466756944562890625\beta^{ 15} + \]\[ 
  3221384468534024615436806439848/8654508294183716417578125\beta^{ 16} + \]\[ 
  361943339485349237796955232692496
/417413592342553091832421875\beta^{ 17} + \]\[ 
  1806278429374132965947828374768937344
/895352155574776381980544921875\beta^{ 18} + \]\[ 
  4853299923370537386980355544223380307264/
   1034131739688866721187529384765625\beta^{ 19} + \]\[ 
  155645504470355669048254630091827537177792/
   14271018007706360752387905509765625\beta^{ 20} + \]\[ 
  124777376875164180242010595944553082668578848/
   4923501212658694459573827400869140625\beta^{ 21} + \]\[ 
  395437305772922404637310628959035520961123917536/
   6720579155279117937318274402186376953125\beta^{ 22} + \]\[ 
  179013658775060249036669346814891247279401725345056/
   1310512935279427997777063508426343505859375\beta^{ 23} +\]\[ 
379175744572610259570928155210224120442515024068064/
1196555288733390780579057985954487548828125\beta^{ 24} +\]\[ 
8670085435876162491988057673185203657287330133364736/
11794616417514851979993571575837091552734375\beta^{ 25}+
 \ldots ;\]

\normalsize

\section{ The second correlation moment on the bcc lattice}

The HT  expansion coefficients of the second correlation moment
$ \mu_{2}^{bcc}(S;\beta)=  \sum_{r=1}^\infty c^{bcc}_r(S) \beta^r $
for the spin $S=1/2$ Ising model on the bcc lattice are:

\scriptsize

\[\mu^{bcc}_2(\beta;1/2  )=  8\beta^{1} +  128\beta^{2} +  
4240/3\beta^{3} +  40448/3\beta^{4} +  1767376/15\beta^{5} +
  43862656/45\beta^{6} +  488322112/63\beta^{7}  +  \]\[
  18867219968/315\beta^{8} +  1282950553456/2835\beta^{9} +
  47619744077056/14175\beta^{10} +  
  766490179148608/31185\beta^{11} +  \]\[83121960882597376/467775\beta^{12} +  
  7733317943660749472/6081075\beta^{13} +
  384090193673880221312/42567525\beta^{14} +  \]\[
  8114589875096001385568/127702575\beta^{15} +
  283873458550660118340608/638512875\beta^{16} +  \]\[
  2582349138528384648119632/834978375\beta^{17} +  
  2091109673217277557956275456/97692469875\beta^{18} +  \]\[
  995381021756635176086923648/6749661555\beta^{19} +  
  9390942009659704730085453139456/9280784638125\beta^{20} +  \]\[
  103702876796861733353936997485984/14992036723125\beta^{21} +   \]\[
  101035886864436355362008305187606912/2143861251406875\beta^{22} +  \]\[
  3156805250924212020953392925068240352/9861761756471625\beta^{23}+ \]\[
 320727270586132663582409995829854660096/147926426347074375\beta^{24}+ \]\[
54168825753838137402817958723519884679552/3698160658676859375\beta^{25}+
\ldots ;\]   
\normalsize 
 for $S=3/2 $ we have 
\scriptsize
\[\mu^{bcc}_2(\beta;3/2  )=  200/81\beta^{} +  16000/729\beta^{2} +
  2721424/19683\beta^{3} +  134755840/177147\beta^{4} + 
 6154011920/1594323\beta^{5} +  \]\[
  801722089600/43046721\beta^{6} +
  1174773128491328/13559717115\beta^{7} +  
  9579774490138112/24407490807\beta^{8} +\]\[
  3442568724281034800/1977006755367\beta^{9} +  
  135227751939098397440/17793060798303\beta^{10} +  \]\[
  205868311610908678737344/6291117925114275\beta^{11} +  
  33135825967715686933468672/237804257569319595\beta^{12} +  \]\[
  3270343097850137001370624160/5564619627122078523\beta^{13} +  \]\[
  862011460830182040665928114304/350571036508690946949\beta^{14} +  \]\[
12086504494080672591124031353502176/1183177248216831945952875\beta^{15} +  \]\[
  17966460146543498115584746908095488/425943809358059500543035\beta^{16} +
  \]\[
  1612694029436910204216007723774521712
/9309914690254729083297765\beta^{17} +  \]\[
  3737216681035259388640991219954952870656
/5278721629374431390229832755\beta^{18} +  \]\[
  65015896214897016070122944833127664003594368/
   22566534965575694193232535027625\beta^{19} +  \]\[
  182463388568006454344368138735361198186077696/
   15622985745398557518391755019125\beta^{20} +  \]\[
  362192291613534236767772263627636610158900891424/
   7677135195288851164537708416398025\beta^{21} +  \]\[
  144384839150089152257673364235640252832996572075904/
   760036384333596265289233133223404475\beta^{22} +  \]\[
  428560759073526184824375077595857649494948170031232928/
   561884041275194381838825923490159736875\beta^{23}+\]\[
 \frac{64867332126865549618677928089688847271789189429133989376}
{  21239216760202347633507619907928038053875}\beta^{24}+\]\[
\frac{ 11659782488710451609895385323507391669036798169540945614208}
{  955764754209105643507842895856761712424375}\beta^{25}+ \ldots ;\]   
\normalsize 
 for $S=5/2 $ we have 
\scriptsize

\[\mu^{bcc}_2(\beta;5/2  )=  392/225\beta^{} +
  43904/3375\beta^{2} +  728782096/10546875\beta^{3} +
  152356890112/474609375\beta^{4} +  \]\[
  49010565623792/35595703125\beta^{5} + 
  5000633194083968/889892578125\beta^{6} +  
  110733100808086004416/5005645751953125\beta^{7} + \]\[ 
  6370442147252574780928/75084686279296875\beta^{8} +   
  23933568725292400739824/75084686279296875\beta^{9} +\]\[  
  99539404546807568828482816/84470272064208984375\beta^{10} +  \]\[ 
262105467997845155997484363485248/60976977646350860595703125\beta^{11} +  \]\[
675464298437519845609125400575488/43554984033107757568359375\beta^{12} +  \]\[ 
  282461301514497845320820252691061664
/5095933131873607635498046875\beta^{13} +  \]\[
  375583250506932117863530120935976121984
/1910974924452602863311767578125\beta^{14} +  \]\[
  4132806535775700716029447588854094367623072/
   5971796638914383947849273681640625\beta^{15} + \]\[ 
  130183322600728077721069577166493571476794368/
   53746169750229455530643463134765625\beta^{16} +  \]\[
  28313566551841857025065917254271389260687075269264/
   3357791955145585234276950359344482421875\beta^{17} + \]\[ 
  10616910503565739063500331978310606482616093944576/
   363397397743028705008327960968017578125\beta^{18} +  \]\[
  18454807409148017168418258633393289199780328860754965632/
   183093885309277766282320953905582427978515625\beta^{19} +\]\[  
\frac{  4756985010870413135466418128711568978899144450148099203584}
{   13732041398195832471174071542918682098388671875}\beta^{20} +  \]\[
\frac{  21119596139411632125511001860716557844151155289204306230368}
  { 17800794405068671721892314963042736053466796875}\beta^{21} +  \]\[
\frac{370651215979942000672613315182721356472388070339352541734080384}
{91502929701439537524265688261948525905609130859375}\beta^{22} +  \]\[
\frac{34662778360468610900614883674971899174533608084510396446207381685036832}
 {2513642668229607495502531122610889724455773830413818359375}\beta^{23}+ \]\[
\frac{ 84066097593880996132659410873879008982978449759001348507592242764517888}
  {1795459048735433925358950801864921231754124164581298828125}\beta^{24}+  \]\[
\frac{4576172371531058925083215774348009798541107911165867405683183
8073783856512}
{288555918546766166575545664585433769389055669307708740234375}
\beta^{25}+\ldots ;\]   
\normalsize 
 for $S=7/2 $ we have 
\scriptsize

\[\mu^{bcc}_2(\beta;7/2  )=  72/49\beta^{} +
  3456/343\beta^{2} +  5790000/117649\beta^{3} +  173219328/823543\beta^{4} +  
  57457204190832/69206436005\beta^{5} + \]\[ 
  1510978979356032/484445052035\beta^{6} +  
  53667241536926272/4747561509943\beta^{7} + \]\[ 
  6633277870477963776/166164652848005\beta^{8} +  
  70842759404383222700656/512950283341791435\beta^{9} +  \]\[
  19697097655650733374529792/41890939806246300525\beta^{10} +  \]\[
  1048604095303392390900156471616/663828966733662625899465\beta^{11} +  \]\[
  121958167478198649661264771810816
/23234013835678191906481275\beta^{12} +  \]\[ 
  15658199447511227845534685944670816
/906126539591449484352769725\beta^{13} +  \]\[
  5847124947823268441784677948808758912
/103600467693289057711000005225\beta^{14} +  \]\[
  58545430160519729773818747857323404843872/
   319814643769183321153857016129575\beta^{15} +  \]\[
  15414966599595769156657187681150339347487744/
   26118195907816637894231656317248625\beta^{16} +  \]\[
  6053199140424121723511968616383775834082753155344/
   3198199207108055126786560547703411379875\beta^{17} +  \]\[
  405728846720983382270881860385060705642925629183744/
   67162183349269157662517771501771638977375\beta^{18} +  \]\[
  14700223544731525481531460072643081904025479230904704/
   765648890181668397352702595120196684342075\beta^{19} +  \]\[
  12094169239888196808562691720289861999099722840253828608/
   198952704039630500221194689489566259643433125\beta^{20} +  \]\[
\frac{  1295782097205221523861387944583214957619651778341093729498912}
  { 6755836971073732896011108070997201478712058625625}\beta^{21} +  \]\[
\frac{732261827140175599003213272941786384786037209796893177359366784}
  { 1213798709136247343649995750089163865675266533070625}\beta^{22} +  \]\[
\frac{15218950957027337148452755358743841877641243057833000176179778883936}
{8043552733755718447006045836860867538502229250691094925}\beta^{23}+ \]\[
\frac{ 8331066185083183687329415665609534025038247701242814971451912591894016}
{1407621728407250728226058021450651819237890118870941611875}\beta^{24}+ \]\[
\frac{ 136484472377454714637430295529179994049829130
75421898705646976276313187712}
{739001407413806632318680461261592205099892312407244346234375}\beta^{25}+
\ldots ;\]   

\normalsize 
 for $S=1 $ we have 
\scriptsize

\[\mu^{bcc}_2(\beta;1  )=  32/9\beta^{} +  1024/27\beta^{2} +
  7696/27\beta^{3} +  453632/243\beta^{4} +  41036588/3645\beta^{5} +
  235090304/3645\beta^{6} +  \]\[
  81741284962/229635\beta^{7} +  1317351871424/688905\beta^{8} +
  27710133756587/2755620\beta^{9} +  
  1612223805027128/31000725\beta^{10} +\]\[
  2170999288133506009/8184191400\beta^{11} +  
  1368605669268695668/1023023925\beta^{12} +\]\[
  25592476184504211275387/3830201575200\beta^{13} +  
  83198544446064314230967/2513569783725\beta^{14} +  \]\[
  10072536940221588986131159/61872486984000\beta^{15} +  \]\[
  180007489565953181359748503/226221280535250\beta^{16} +  \]\[
  22847810086448612685982551733529/5907090077336448000\beta^{17} +  \]\[
  1151250220070296170223053377659/61532188305588000\beta^{18} +  \]\[
  182035358432682251612963887125670889/2020224806449065216000\beta^{19} +  \]\[
  409336772772878527900425756955252723/946980378022999320000\beta^{20} +  \]\[
  21643901001542514660920444250280438889
/10475239737143301120000\beta^{21} +  \]\[
  25842538827908898853101666876076718871547
/2625029607879754115040000\beta^{22} +  \]\[
  542133598549200607440741338426363948434866163
/11592130748396994172016640000\beta^{23}+ \]\[
80264698614082395089928457150727458225378007/
362254085887406067875520000\beta^{24}+ \]\[
 21848929536143669104474159748357398927699421236009/
  20865835347114589509629952000000\beta^{25}+\ldots ;\]   
\normalsize 
 for $S=2 $ we have 
\scriptsize

\[\mu^{bcc}_2(\beta;2  )=  2\beta^{} +  16\beta^{2} +
  109129/1200\beta^{3} +  33889/75\beta^{4} +
  31826161/15360\beta^{5} +  1301601859/144000\beta^{6} +\]\[  
  49064563507469/1290240000\beta^{7} +
  25134739981769/161280000\beta^{8} +  
  929837568644858477/1486356480000\beta^{9} +\]\[
  2295302569773862409/928972800000\beta^{10} +  
  5037463613838811141033/523197480960000\beta^{11} + \]\[ 
  182014040360322673576391/4904976384000000\beta^{12} +  
  4625274127509466950444652897/32647522811904000000\beta^{13} +\]\[  
  76641494593141781309989367507/142832912302080000000\beta^{14} +\]\[  
  1106886233630435373362619668696261/548478383239987200000000\beta^{15} +  \]\[
  517234962093115418227963055453641/68559797904998400000000\beta^{16} +  \]\[
  66957116650566442558653401615101475621
/2386977923860424294400000000\beta^{17} +  \]\[
  126703953457274539130301778570069979147
/1220613711064989696000000000\beta^{18} +  \]\[
  124895279533955807554801457435603719054422653/
   326538579984106043473920000000000\beta^{19} +\]\[  
  57304705268347743317483087465719800092989577/
   40817322498013255434240000000000\beta^{20} +  \]\[
  2253700468192132480306411514153692375712152805113/
   438867851498638522428948480000000000\beta^{21} +  \]\[
  56495597374025760901251464592595346562659470521023/
   3017216479053139841699020800000000000\beta^{22} +  \]\[
  355704571388834055337249908994883603139734650218532001/
   5225109008430849231742304256000000000000\beta^{23}+ \]\[
 432766449965591365899684422307172549118349438327808453/
  1753161575197192834334588928000000000000\beta^{24}+ \]\[
\frac{ 30456646773493468661511818951598779672718002428361285656633}
{  34109511607036583784813762183168000000000000}\beta^{25}+
\ldots ;\]   
\normalsize 
 for $S=3 $ we have 
\scriptsize

\[\mu^{bcc}_2(\beta;3 )=  128/81\beta^{} +  8192/729\beta^{2} +
  1119808/19683\beta^{3} +  44638208/177147\beta^{4} +
  402572441104/390609135\beta^{5} +  \]\[
  42308316276736/10546446645\beta^{6} +
  204252754066376/13559717115\beta^{7} +  
  47077202752848896/854262178245\beta^{8} +\]\[
  1952372990346922081/9885033776835\beta^{9} +  
  15182889718356698755456/21796499477921175\beta^{10} +  \]\[
  73232061822853309717922671/30209948276398748550\beta^{11} +  
  3403045460026530465243790528/407834301731383105425\beta^{12} +  \]\[
  1551693528484006417360628016769/54533272345796369525400\beta^{13} + \]\[  
  41340139729467825073204741371152/429449519723146410012525\beta^{14} +  \]\[
  300000599802059974522798410027468553
/927610962601996245627054000\beta^{15} +  \]\[
  3944802088269032077349289642746860424
/3652468165245360217156525125\beta^{16} +  \]\[
  3592596882572065406946388895289250421256481/
   1001419111674311882899107432672000\beta^{17} +  \]\[
  1073710924945861688279339969963407197542969/
   90530075943771498342441631748250\beta^{18} +  \]\[
  1083108242017244157909560617528566019567878040297/
   27741312231601787780071074099879744000\beta^{19} +  \]\[
  1248736226891764438331143830399803134416548151379/
   9752805081422503516431236988238972500\beta^{20} +  \]\[
  56414795147963025827857167083658138788745260233682983/
   134822777445584688611145420125415555840000\beta^{21} +  \]\[
  10934565603589614340243445733753347041547038418026343/
   8021307071582262122898675836788545690000\beta^{22} +  \]\[
\frac{63106272618266036825024593769047847492513100694745410282741}
   {14250866919097472017021258067039674558487040000}\beta^{23}+ \]\[
\frac{819433480458963940159663878385306302913947961336267333799459}
{  57114802574195337068218010846807445691436340000}\beta^{24}+ \]\[ 
\frac{ 87171765652161901535171047603945942527443803439188878391252874881}
{  1879892930442086522931061385586347926758133248000000}\beta^{25}+
\ldots ;\]   
\normalsize 
 for $S=4 $ we have 
\scriptsize

\[\mu^{bcc}_2(\beta;4  )=  25/18\beta^{} +  250/27\beta^{2} +
  1216081/27648\beta^{3} +  5687215/31104\beta^{4} +
  536666715415/764411904\beta^{5} + \]\[ 
  245167441205/95551488\beta^{6} +
  1115347452769077761/123284351877120\beta^{7} +  
  287401602203829997/9246326390784\beta^{8} + \]\[ 
  15849593657830393025575/151491811586605056\beta^{9} +  
  59063627629407820739725/170428288034930688\beta^{10} + \]\[ 
  652382208059850685361212393849/575911270927637780889600\beta^{11} +  \]\[
  52732022030115081268893140327/14397781773190944522240\beta^{12} +  \]\[
  32378193059642164291925217551302331
/2759951101891937090112847872\beta^{13} +  \]\[
  270097149738688312677879677850660389
/7244871642466334861546225664\beta^{14} +  \]\[
  436433419882569190816670064086199368857147/
   3709374280942763449111667539968000\beta^{15} +  \]\[
  61610184550833135511727259738522133476313/
   166921842642424355210025039298560\beta^{16} +  \]\[
  321271817516109017699493383001923755922244305897/
   278954441925055024847627124874495918080\beta^{17} +  \]\[
  41573389111215193580783924559875421797347800697/
   11623101746877292701984463536437329920\beta^{18} +  \]\[
  6753679965733657038248347723991761674052583674878323369/
   610575482485560438386486250925296665493504000\beta^{19} +  \]\[
  600187215451870866096988593666553116540901964420094269/
   17612754302468089568840949545922019196928000\beta^{20} +  \]\[
\frac{  6105064893922221283372606430615609442759035671482775941277}
  { 58354734112753296298057779288433686643432488960}\beta^{21} +  \]\[
\frac{10405059100265244516426632598927808370752943080436409877009921}
   {32496292559039491875980925841246509249561467289600}\beta^{22} +  \]\[
\frac{224358055736534058640840825524795444118316555410338842484351527664163}
{ 229605804705149433798930829623911335753701503281397760000}\beta^{23}+ \]\[
\frac{ 17072288046388622667968515123161346868207376858047387177692567152781}
{  5740145117628735844973270740597783393842537582034944000}\beta^{24}+ \]\[
\frac{ 382191378723522522986938985966049707875576833
01676421652021903361325845401}
  {4232094192325314363781893051627933740612226108482723512320000}\beta^{25}+
\ldots ;\]   
\normalsize 
 for $S=5 $ we have 
\scriptsize

\[\mu^{bcc}_2(\beta;5  )=  32/25\beta^{} +  1024/125\beta^{2} +
  43797136/1171875\beta^{3} +  874415104/5859375\beta^{4} +  
  80634247876/146484375\beta^{5} + \]\[
 21233889802624/10986328125\beta^{6} +  
  314709454315771922/48065185546875\beta^{7} +
  5194468528550902336/240325927734375\beta^{8} +  \]\[
  73180581269425236006101/1046859741210937500\beta^{9} +  
  7278880232341022250734792/32714366912841796875\beta^{10} +  \]\[
  81649404181963743798402571973/116837024688720703125000\beta^{11} +  \]\[ 
  3329435830643625397425077741476/1533485949039459228515625\beta^{12} +  \]\[
  21315653883384003336997369187945243
/3189650774002075195312500000\beta^{13} +  \]\[
  355977466729485931192772962489949749
/17443402670323848724365234375\beta^{14} +  \]\[
  647744938153396254752219496441894698097251/
   10466041602194309234619140625000000\beta^{15} +  \]\[
  61021413712966122880202860685148333293183/
   327063800068572163581848144531250\beta^{16} +  \]\[
  437531573996766002992742943255170299352552513/
   782077834010124206542968750000000000\beta^{17} +  \]\[
  1838115176634725231410255388156655971169192975037/
   1100896751030813902616500854492187500000\beta^{18} +  \]\[
  228334760120713556181292924208557948992988259695721188811/
   46017484193088021129369735717773437500000000000\beta^{19} +  \]\[
  528126663789565107660814478748990607406098476919836408871/
   35951159525850016507320106029510498046875000000\beta^{20} +  \]\[
 \frac{ 322179948243143285938514937244966791025295924934690022410499}
   {7433593600421911105513572692871093750000000000000}\beta^{21} +  \]\[
\frac{  48119390985062775111582906327315981497307552266902563014227643}
  { 377487175021425173326861113309860229492187500000000}\beta^{22} +  \]\[
\frac{18547327895721062593669938323906646282105991708290241934082607295863}
{ 49612600145673022780101746320724487304687500000000000000}\beta^{23}+ \]\[
\frac{178014611570954451469979486648556288444705238716931112186332433621131}
{  162791344227989605997208855114877223968505859375000000000}\beta^{24}+ \]\[
\frac{ 166222055659610923100019391422809545454875
06831031649361996598800899550249}
 { 5209323015295667391910683363676071166992187500000000000000000}\beta^{25}+
\ldots ;\]

\normalsize 
 for $S=\infty $ we have 
\scriptsize

\[\mu^{bcc}_2(\beta;\infty )=  8/9\beta^{} +  128/27\beta^{2} + 
 12176/675\beta^{3} +  365056/6075\beta^{4} +
  33035504/178605\beta^{5} +  2420177152/4465125\beta^{6} +  \]\[
  43993740032/28704375\beta^{7} +  2545864991744/602791875\beta^{8} +
  277241999503696/24312605625\beta^{9} +   \]\[
  696185787722570624/22975412315625\beta^{10} + 
  6479056413973256514944/81539738308153125\beta^{11} + \]\[  
  84032954257205333103616/407698691540765625\beta^{12} +  
  25242268054018494768736/47653093816453125\beta^{13} +   \]\[
  74339452044961117329053824/55039323358003359375\beta^{14} +  \]\[
  2989224336573820097165677216/873976068706756640625\beta^{15} +  \]\[ 
  6162559741005217097339984339968/715786400270833688671875\beta^{16} +  \]\[
  818828124763084548270884490696240304
/37984636903172331356750390625\beta^{17} +   \]\[
  5344492461714294991014906011034989696
/99483572841641820220060546875\beta^{18} +  \]\[
  8109872778642117605403611851883478175552
/60831278805227454187501728515625\beta^{19} +  \]\[
  1022525312346582803395089505579732087033856/
   3102395219066600163562588154296875\beta^{20} +  \]\[
  1332669037063140248817997367871506998133578592/
   1641167070886231486524609133623046875\beta^{21} +  \]\[
  13403443577222566364510704380489177808912033003904/
   6720579155279117937318274402186376953125\beta^{22} +  \]\[
  2358263177389607900735562196124579055893598302613664/
   482820555102947157075760239946547607421875\beta^{23}+ \]\[
443110921565163117476351485511360564766820670868992/
  37140042700226704390443095380503662109375\beta^{24}+  \]\[
\frac{ 342900630293561763358238734113909407846134587836183552}
{11794616417514851979993571575837091552734375}\beta^{25}+
\ldots ;\]

\normalsize
\section{The second field derivative of the susceptibility on the bcc lattice}

The HT  expansion coefficients of the second field derivative of the 
susceptibility 
$\chi_4^{bcc}(\beta;S) =  \sum_{r=0}^\infty e^{bcc}_r(S) \beta^r $ 
 for the spin $S=1/2$ Ising model on the bcc lattice are

\scriptsize

\[ \chi^{bcc}_4(\beta; 1/2)=-2 -64\beta^{ } -1168\beta^{ 2} -
49664/3\beta^{ 3} -601360/3\beta^{ 4} -32820608/15\beta^{ 5} -
 996463616/45\beta^{ 6} 
  -\]\[66712488448/315\beta^{ 7} -122056132496/63\beta^{ 8} -
48489867797888/2835\beta^{ 9} 
  - 2078558044733696/14175\beta^{ 10} -\]\[
191285725186144768/155925\beta^{ 11} 
  -188087379936809600/18711\beta^{ 12} - 
492034524872707515136/6081075\beta^{ 13} 
  -\]\[27296494302637993572352/42567525\beta^{ 14} -
32011976778677393162
48576/638512875\beta^{ 15} 
  -\]\[ 4945781553886665074906384/127702575\beta^{ 16} 
  -3212941768987291424807915648/10854718875\beta^{ 17} 
  -\]\[ 218891517638961233925349581056/97692469875\beta^{ 18} 
  - 2837416820552973655593858057728/168741538875\beta^{ 19} 
  -\]\[ 232406819903350165473891311994496/1856156927625\beta^{ 20} 
  - 10612802176781701102740105748890368/11464498670625\beta^{ 21}\]\[
-14577540459141181848411708987543436672/2143861251406875\beta^{ 22}
 -\]\[2448111915520417182899850659794357369856/49308808782358125\beta^{ 23}
            +\ldots ;\]

\normalsize 
 for $S=3/2 $ we have 
\scriptsize

\[ \chi^{bcc}_4(\beta; 3/2)=-34/81 -5440/729\beta^{ } -2096/27\beta^{ 2} -
112263680/177147\beta^{ 3} -2367195184/531441\beta^{ 4} 
  -406894972288/14348907\beta^{ 5} -\]\[ 108439778662912/645700815\beta^{ 6} 
  -22994744550875648/24407490807\beta^{ 7} -
205988526310282736/40679151345\beta^{ 8} 
  -\]\[ 467949177919990427008/17793060798303\beta^{ 9} 
  -177186643515203487702272/1334479559872725\beta^{ 10} 
  -\]\[ 7413406834005674899860992/11324012265205695\beta^{ 11} 
  -11284182242274665794011296128/3567063863539793925\beta^{ 12} 
  -\]\[ 3762105689711203193350913590016/250407883220493533535\beta^{ 13} 
  -\]\[ 615985582245155802851832336421888/8764275912717273673725\beta^{ 14} 
-\]\[138209158365245742381264318269825024/425943809358059500543035\beta^{ 15} 
-\]\[47297659110203208067663673873146795504/
31945785701854462540727625\beta^{ 16} 
  -\]\[ 164758616451387199357945605521382599552
/24643891827144871102847025\beta^{ 17} 
  -\]\[ 11840638826406378708579965604712114803342592/
   395904122203082354267237456625\beta^{ 18} 
  -\]\[ 5388958543121650848350155168279438678121232896/
   40619762938036249547818563049725\beta^{ 19} 
  -\]\[ 593065315757651183900760239145795898265994311296/
   1015494073450906238695464076243125\beta^{ 20} 
  -\]\[ 882001091074380883583374848636712368666489711792896/
   345471083787998302404196878737911125\beta^{ 21}
-\]\[ 632144388079683765006468878756566209099655302479999104/
      57002728825019719896692484991755335625\beta^{ 22}
-\]\[ 48435362540657725070442092280582201611133027851066524672/
1011391274295349887309886662282287526375\beta^{ 23}
+\ldots ;\]
\normalsize 
 for $S=5/2 $ we have 
\scriptsize

\[ \chi^{bcc}_4(\beta; 5/2 )=-518/1875 -116032/28125\beta^{ } -
381522736/10546875\beta^{ 2} -39432060416/158203125\beta^{ 3} 
  -\]\[87825896148496/59326171875\beta^{ 4} -
35461804001486464/4449462890625\beta^{ 5} 
  -7404362241428483072/185394287109375\beta^{ 6} 
  -\]\[1582241886367201765888/8342742919921875\beta^{ 7} 
  - 8101360721003943387287696/9385585784912109375\beta^{ 8} 
  -\]\[534567007251031288785164416/140783786773681640625\beta^{ 9} 
  -\]\[ 30008001880338988332119400332032/1847787201404571533203125\beta^{ 10} 
-\]\[2954997454722876203868154634059264/43554984033107757568359375\beta^{ 11} 
  -\]\[ 4537585807261536531981150589034642816
/16333119012415409088134765625\beta^{12 } 
  -\]\[ 3561451392843109883310213082902640132864
/3184958207421004772186279296875\beta^{ 13} 
  -\]\[ 481418138826181138705072868651949718423552/
   108578120707534253597259521484375\beta^{ 14} 
  -\]\[ 10879904520636677050692917629137009179270782976/
   627038647086010314524173736572265625\beta^{ 15} 
  -\]\[ 110470684921277861469909988558511790576920016334192/
   1645976448600777075625956058502197265625\beta^{ 16} 
  -\]\[ 2200803966609780093812152438371989164705537771964032/
   8565795803942819475196301937103271484375\beta^{ 17} 
  -\]\[ 7303586413264891297456026602010195642597101243856477952/
   7495071328449967040796764194965362548828125\beta^{ 18} 
  -\]\[ 5735238148513648024928251605575765991650777246372678144/
   1564905002643399711814709007740020751953125\beta^{ 19} 
  -\]\[ \frac{ 665435539426183010334563484763138199672544516678193333565824}
   {48645895641381997620555921457707881927490234375}\beta^{ 20} 
-\]\[ \frac{63970402136572149703888914194997383842262094927063148772855300864}
   {1261631303459242108289117823005653917789459228515625}\beta^{ 21}
-\]\[
\frac{400206505850393265010194964053595893215827810402991883932551424235904}
     {2142917875728565639814604537605191580951213836669921875}\beta^{ 22}
-\]\[
\frac{12278869072753281485289663600644863580773426
04694519705442975070914337792}
{1795459048735433925358950801864921231754124164581298828125}\beta^{ 23}
+\ldots ;\]
\normalsize 
 for $S=7/2 $ we have 
\scriptsize

\[ \chi^{bcc}_4(\beta; 7/2)=-78/343 -7488/2401\beta^{ } -
20704752/823543\beta^{ 2} -6429866496/40353607\beta^{ 3} 
  -12053319624240/13841287201\beta^{ 4} -\]\[
298629038032512/69206436005\beta^{ 5} 
  - 67504227810231808/3391115364245\beta^{ 6} -
101233647203766057472/1163152569936035\beta^{ 7} 
  -\]\[29104382142758857172464/79792266297612001\beta^{ 8} 
  - 259977082853647172976338048/175941947186234462205\beta^{ 9} 
  -\]\[1756229744821565984931955201792/301740439424392102681575\beta^{ 10} 
-\]\[520328944074524106664992032863744/23234013835678191906481275\beta^{ 11} 
-\]\[11539439713645359617056365235199872/
136616001353787768410109897\beta^{ 12} 
  -\]\[ 7486879678501459983682038149392765696
/23907800236912859471769231975\beta^{ 13} 
  -\]\[ 121932836983139459949364620484742803260416/
   106604881256394440384619005376525\beta^{ 14} 
  -\]\[ 205590140577344763663200952378552099140313088/
   49862010369468126888987707514747375\beta^{ 15} 
  -\]\[ 552791109622852016756251970826697870452534548752/
   37625873024800648550430124090628369175\beta^{ 16} 
  -\]\[ 1159917536997641249039795491826202223546258241813888/
   22387394449756385887505923833923879659125\beta^{ 17} 
  -\]\[ 1787227135923512994470780166935075071647227671435776256/
   9872840952342566176390112410760430929674125\beta^{ 18} 
  -\]\[ 823586351285535214311723875459012661668879436059113042432/
   1313087846661561301459884950631137313646658625\beta^{ 19} 
  -\]\[ \frac{971361307665304167558705213449051075445126848626909842337664}
   {450389131404915526400740538066480098580803908375}\beta^{ 20} 
-\]\[ \frac{17066327522846647546682948684827192127344704686901235547252060416}
   {2317252081078290383331810068352040107198236108589375}\beta^{ 21}
-\]\[\frac{168076812691714558392307349470918212829397
08869661141466134903991168}
    {672537853992953047408532260607095948035303449054439375}\beta^{ 22}
-\]\[
\frac{118680196081319878161950521315989749547878658295116876974003289517257728}
{1407621728407250728226058021450651819237890118870941611875}\beta^{ 23}
+\ldots ;\]
\normalsize 
 for $S=1 $ we have 
\scriptsize

\[ \chi^{bcc}_4(\beta; 1 )=-2/3 -128/9\beta^{ } -4768/27\beta^{ 2} -
139072/81\beta^{ 3} -3486352/243\beta^{ 4} -395367632/3645\beta^{ 5} 
  - 926111816/1215\beta^{ 6} -\]\[388136700776/76545\beta^{ 7} -
22254946400369/688905\beta^{ 8} 
  - 410872707191111/2066715\beta^{ 9} -\]\[
73715954683908977/62001450\beta^{ 10} 
  -14204945654889919999/2046047850\beta^{ 11} -
139332068337333964501/3507510600\beta^{ 12} 
  -\]\[213874428552292960351357/957550393800\beta^{ 13} 
  - 49744620166439077102213259/40217116539600\beta^{ 14} 
-\]\[ 370676214357209362962828101/54841522554000\beta^{ 15} 
- 1056940240155242437010987457889/28956323908512000\beta^{ 16} 
-\]\[ 26187051617585248456343358898669/134252047212192000\beta^{ 17} 
-\]\[ 435685960462954298917571358956477/421935005524032000\beta^{ 18} 
-\]\[ 27650613531569984190291818677713221/5101577794063296000\beta^{ 19} 
-\]\[ 77771964801716097281521149765759362491/2754852008794179840000\beta^{ 20} 
-\]\[92920978332904980874078982501756045924237
/636370814031455543040000\beta^{ 21}
-\]\[31517312296166193707116275173508766312798139/
 42000473726076065840640000\beta^{ 22}
-\]\[11110482715669139318287458527738488522440031093/
2898032687099248543004160000\beta^{ 23}
+\ldots ;\]
\normalsize 
 for $S=2 $ we have 
\scriptsize

\[ \chi^{bcc}_4(\beta; 2)=-13/40 -26/5\beta^{ } -9759/200\beta^{ 2} -
431567/1200\beta^{ 3} -73086519/32000\beta^{ 4} -1009466059/76800\beta^{ 5} 
  -\]\[ 43251204653/614400\beta^{ 6} -92179755748103/258048000\beta^{ 7} 
  -119536938403982081/68812800000\beta^{ 8} -\]\[ 
60674411369998755551/7431782400000\beta^{ 9} 
  -410707681080770020169/11010048000000\beta^{ 10} 
  -\]\[ 311068145925992700998201/1868562432000000\beta^{ 11} 
  -761954784655042970941842821/1046394961920000000\beta^{ 12} 
  -\]\[ 7862120581920073606384238459/2511347908608000000\beta^{ 13} 
  -\]\[ 18364150712094338544946316345851/1385046422323200000000\beta^{ 14} 
-\]\[30396313223519635006488704691613099/548478383239987200000000\beta^{ 15} 
-\]\[4464926948877511071877021412891005459/
19501453626310656000000000\beta^{ 16} 
  -\]\[ 1596032863049308216630774975780144763777
/1704984231328874496000000000\beta^{ 17} 
  -\]\[ 5430545744252985247590021257875643127123341/
   1432186754316254576640000000000\beta^{ 18} 
  -\]\[ 4973358088183865804021334274367088529231632019/
   326538579984106043473920000000000\beta^{ 19} 
  -\]\[ 6219532434928249630347866873033554484103138121/
   102443476073445033246720000000000\beta^{ 20} 
  -\]\[ 105475161188748598372035390732972002029069275661799/
   438867851498638522428948480000000000\beta^{ 21}
-\]\[1193148427671850329152678759800850211146619459718943/
1262103625224842809599590400000000000\beta^{ 22}
-\]\[46913342889018988713048777112456410826908579755539777289/
12689550449046348134231310336000000000000\beta^{ 23}
+\ldots ;\]
\normalsize 
 for $S=3 $ we have 
\scriptsize

\[ \chi^{bcc}_4(\beta; 3)=-20/81 -2560/729\beta^{ } -
1347712/45927\beta^{ 2} -34145024/177147\beta^{ 3} -
85278317632/78121827\beta^{ 4} 
  -\]\[3940538573248/703096443\beta^{ 5} -
2542596266458208/94918019805\beta^{ 6} 
  -103652654777684896/854262178245\beta^{ 7} -\]\[ 
28351205160853208332/53818517229435\beta^{ 8} 
  -275348457113981762500/124551425588121\beta^{ 9} 
  -\]\[ 12386741688573656788726586/1373179467109034025\beta^{ 10} 
  -4887670628769413464133127994/135944767243794368475\beta^{ 11} 
  -\]\[ 1031146761930807496181918333221/7341017431164895897650\beta^{ 12} 
  -\]\[92662252830323239220285787411121/171779807889258564005010\beta^{ 13} 
-\]\[220855523647152628791314050202466253/
108221278970232895323156300\beta^{ 14} 
  -\]\[111329025657015731621409025951297501997
/14609872660981440868626100500\beta^{ 15} 
  -\]\[ 414175111501039319125610657018874096916627/
   14726751642269292395575109304000\beta^{ 16} 
  -\]\[ 46293905968756409476099332182324154734559559/
   450638600253440347304598344702400\beta^{ 17} 
  -\]\[ 135702817437367457884432242050265545683700461291/
   365017266205286681316724659208944000\beta^{ 18} 
  -\]\[ 582357583967087663960271420965572287784058178853/
   436489178469258898637481935137968000\beta^{ 19} 
  -\]\[ 53387092370318339267918835814876829397859065384966307/
   11235231453798724050928785010451296320000\beta^{ 20} 
-\]\[ \frac{7137729813047719430195055370798429888808265508601811511}
   {424691748953591769125108073395059000896000}\beta^{ 21}
-\]\[ \frac{173849594669720112426832095386716293723115436821358332629277}
   {2943113820248390960036998948627758876209280000}\beta^{ 22}
 -\]\[\frac{125747528682555337132810306583389711602271107159604365681793853}
 {609224560791416928727658782365946087375320960000}\beta^{ 23}
+\ldots ;\]
\normalsize 
 for $S=4 $ we have 
\scriptsize

\[ \chi^{bcc}_4(\beta; 4)=-41/192 -205/72\beta^{ } -308573/13824\beta^{ 2} -
45651025/331776\beta^{ 3} -46596502705/63700992\beta^{ 4} 
  -\]\[10781275334101/3057647616\beta^{ 5} -
860188925808523/54358179840\beta^{ 6} 
  -2212919091271591949/32875827167232\beta^{ 7} 
  -\]\[ 10394435808118530818749/37872952896651264\beta^{ 8} 
  -280928685545919178742521/259700248434180096\beta^{ 9} 
  -\]\[ 18087555481424523236724316409/4362964173694225612800\beta^{ 10} 
  -\]\[7150706443424405665569320803151/460729016742110224711680\beta^{ 11} 
  -\]\[ 5038306169727104530832567190012263
/88459971214485163144642560\beta^{ 12} 
  -\]\[11341146864273550393097034599637649637
/55199022037838741802256957440\beta^{ 13} 
  -\]\[ 135423000581086116987935947335050992252883/
   185468714047138172455583376998400\beta^{ 14} 
  -\]\[ 465311521273371874533338024132163825705959/
   181683638250257801589142899916800\beta^{ 15} 
  -\]\[ 60710441493959226178452506060971351184716489613/
   6837118674633701589402625609669017600\beta^{ 16} 
  -\]\[ 33998172981623304579405425471710362217472161031339/
   1115817767700220099390508499497983672320\beta^{ 17} 
  -\]\[ 8406053654278638244763024271075907965771833502697993/
   81150383105470552682946072690762448896000\beta^{ 18} 
  -\]\[ 2105963860139940313404740559990049580859339026703297967/
   6030375135659856181594925935064658424627200\beta^{ 19} 
  -\]\[\frac{ 109583141515551509892888943406869866965205164370458713177889}
 {93784394109782083336164288142125567819802214400}\beta^{ 20} 
-\]\[ \frac{87389554916977286359312128686637361037463008005801821757240579}
   {22508254586347700000679429154110136276752531456000}\beta^{ 21}
 -\]\[ 
\frac{10664680242880663078769323667232598327845527466586913195852375535203} 
{831905089511410992025111701535910636788773562613760000}\beta^{ 22}
-\]\[
\frac{1546007108736521306527424029968358325171267959234348902945639560552977}
{36736928752823909407828932739825813720592240525023641600}\beta^{ 23}
+\ldots ;\]
\normalsize 
 for $S=6 $ we have 
\scriptsize

\[ \chi^{bcc}_4(\beta; 5)=-122/625 -7808/3125\beta^{ } -
7348192/390625\beta^{ 2} -652598464/5859375\beta^{ 3} 
  -416587973392/732421875\beta^{ 4} -\]\[48228931649968/18310546875\beta^{ 5} 
  - 857811544536427144/75531005859375\beta^{ 6} -
11154805291126644824/240325927734375\beta^{ 7} 
  -\]\[660985301540660686165747/3634929656982421875\beta^{ 8} 
  - 16092468082064147071339621/23367404937744140625\beta^{ 9} 
  -\]\[518537549897695794744501522883/204464793205261230468750\beta^{ 10} 
  -\]\[ 1865386843058990642741926927429/204464793205261230468750\beta^{ 11} 
  -\]\[49335014608305603013214421880912697
/1533485949039459228515625000\beta^{ 12} 
  -\]\[1010556140743210699004671099657862971
/9061507880687713623046875000\beta^{ 13} 
  -\]\[ 66433807516510500783383494272451100707913/
   174434026703238487243652343750000\beta^{ 14} 
  -\]\[2399190405152415813481780778336456976209471/
   1868936000391840934753417968750000\beta^{ 15} 
  -\]\[ 1230998788970039711864491954003912262237192136787/
   287816144060343503952026367187500000000\beta^{ 16} 
  -\]\[ 570303959555693293638397024823021452709719851743/
   40436978917568922042846679687500000000\beta^{ 17} 
  -\]\[ 2536294080215135488161499788971111918913675787697228803/
   55044837551540695130825042724609375000000000\beta^{ 18} 
  -\]\[\frac{ 8586960840345784817770878703170590285943799895215862396369}
   {57521855241360026411712169647216796875000000000}\beta^{ 19} 
  -\]\[\frac{6275007093423782656294702414522114795838707528703288083277107}
   {13073148918490915093570947647094726562500000000000}\beta^{ 20} 
-\]\[ \frac{3306172705272744231234815397636627871823958710556073371517733687}
   {2157069571551000990439206361770629882812500000000000}\beta^{ 21}
-\]\[
\frac{1079930080230315689113046290222630116682630124281281523236430959539}
 {222051279424367749015800654888153076171875000000000000}\beta^{ 22}
-\]\[
\frac{512341894449508693974929505534215273293804420157266211388345872833087}
 {33393096251895303794299252331256866455078125000000000000}\beta^{ 23}
+\ldots ;\]
\normalsize 
 for $S=\infty $ we have 
\scriptsize

\[ \chi^{bcc}_4(\beta; \infty)=-2/15 -64/45\beta^{ } -4688/525\beta^{ 2} 
 -208384/4725\beta^{ 3} -93234544/496125\beta^{ 4} -72063104/99225\beta^{ 5} 
  -\]\[ 385125951104/147349125\beta^{ 6} -788304121856/88409475\beta^{ 7} 
  -322590120993239696/11062235559375\beta^{ 8} -\]\[
3059803752362387584/33186706678125\beta^{ 9} 
  - 1426788408907331963008/5033317179515625\beta^{ 10} 
  -\]\[115705968774278431115264/135899563846921875\beta^{ 11} 
  -\]\[ 86871905911543433866631552/34654388780965078125\beta^{ 12} 
  -\]\[653090876086687664244992/90011399431078125\beta^{ 13} 
  -\]\[ 72950164115598663978808849966336/3525910786519291873828125\beta^{ 14} 
  -\]\[ 205318641813685892470016592582656/3525910786519291873828125\beta^{ 15} 
  -\]\[ 37605803886862158082991177881030653712
/232128336630497580513474609375\beta^{ 16} 
  -\]\[ 28240527919566981550929946841675838592
/63307728171953885594583984375\beta^{ 17} 
  -\]\[ 125306376261322605586441062720306276655232/
   102965497891099283927762666015625\beta^{ 18} 
  -\]\[ 78311212815370153410855777412263395234192384/
   23785030012843934587313175849609375\beta^{ 19} 
  -\]\[ 62869842501611694232140969054726115978141016192/
   7111723973840336441606639579033203125\beta^{ 20} 
  -\]\[ 3520569781618809808202917814301470125124794969856/
   149346203450647065273739431159697265625\beta^{ 21}
-\]\[4244562426404164523425537802418003647451513227629184/
67952522570044414699551441177662255859375\beta^{ 22}
-\]\[503195983389079326625596350787555429455021228650089472/
3057863515651998661479814852994801513671875\beta^{ 23}
+\ldots \]

\newpage

\begin{table}
\squeezetable
\caption{ The most extensive HT expansion data, 
published (or obtainable from data in the literature)
 before our  expansions, for the nearest-neighbor correlation $G$, 
for the susceptibility $\chi$, for the second moment of the 
correlation function $\mu_2$  and for 
the second field-derivative of the susceptiblity $\chi_4$
  in the case of  the  Ising   models with general spin.}
\label{tabella1}
\begin{tabular}{lccc}
  Observable &  Lattice  & Order &  Ref. \\
\tableline

$G$ &fcc &14& \cite{mckest} \\

$\chi$ &sq &21& \cite{nr90}  \\

$\chi$ &sc &15& \cite{sack}  \\

$\chi$ &bcc&  21& \cite{nr90}\\

$\chi$ &fcc& 14& \cite{mckest} \\

$\mu_2$ &sq &21& \cite{nr90} \\

$\mu_2$ &sc& 15& \cite{sack} \\

$\mu_2$ &bcc&21& \cite{nr90} \\

$\mu_2$ &fcc& 14& \cite{mckest} \\

$\chi_4$ &sq & 10& \cite{bakin} \\

$\chi_4$ &sc& 14& \cite{lw} \\

$\chi_4$ &bcc& 10& \cite{bakin} \\

$\chi_4$ &fcc &13& \cite{mck} \\
\end{tabular}
\end{table}
 
\acknowledgments 
This work has been partially supported by the Ministry of Education, 
 University and Research.


\normalsize


\begin{references}
\bibitem[*]{pb}Electronic address: butera@mib.infn.it

\bibitem[**]{mc}Electronic address: comi@mib.infn.it

\bibitem{fisherrmp} M. E. Fisher, Rev. Mod. Phys.  {\bf 70}, 653 (1998).

\bibitem{dosy} C. Domb and M. F. Sykes, Phys. Rev.  {\bf 128}, 168 (1962).

\bibitem{univ} M. E. Fisher,  Phys. Rev. Lett.  {\bf 16}, 11 (1966); 
L. P. Kadanoff, Physics {\bf 2}, 263 (1966);
 D. Jasnow and M. Wortis, Phys. Rev.  {\bf 176}, 739 (1968); 
 P. G. Watson, J. Phys. C {\bf 2}, 1883 (1969) and 
{\it ibid.}. C {\bf 2}, 2158 (1969)  ; R. B. Griffiths, 
Phys. Rev. Lett.  {\bf 24}, 1479 (1970); 
 R. B. Griffiths and J. C. Wheeler, Phys. Rev. A
 {\bf 2}, 1047 (1970);
 L. P. Kadanoff, Report at the Newport 
Beach Conference 1970 (unpublished) and  in {\it Proc. E. Fermi 1970 School on critical
 phenomena } edited by M.S. Green  (Academic , London 1971);
 C. Domb, in {\it Statistical Mechanics at the turn of the decade}
 edited by E. G. D. Cohen  (Dekker, New York, 1971).

\bibitem{dg} C. Domb, in {\it Phase Transitions and critical Phenomena}, 
edited by C. Domb and M. S. Green, (Academic, London, 1974),Vol. 3.

\bibitem{dc} C. Domb, {\it The critical point},
(Taylor and Francis, London 1996).

\bibitem{fisherising} M. E. Fisher, Physica A {\bf 106}, 28 (1981).

\bibitem{camp}D. M. Saul, M. Wortis and D. Jasnow, 
Phys. Rev. B {\bf 11}, 2571 (1975);  J.P. Van Dyke and W.J.Camp, {\it ibid}
B {\bf 9}, 3121 (1974).
\bibitem{camp2} W. J. Camp and J. P. Van Dyke, Phys. Rev. B
{\bf 11}, 2579 (1975);  W. J. Camp, D. M. Saul,  J. P. Van Dyke and M. Wortis, 
{\it ibid.} {\bf 14}, 3990 (1976).


\bibitem{saul} D. M. Saul, M. Wortis and D. Jasnow,  Phys. Rev. B {\bf 11}, 2579
 (1975).

\bibitem{sack} R. Roskies and P. Sackett, 
     J. Stat. Phys. {\bf 49}, 447 (1987).

\bibitem{mckest} S. McKenzie,  J. Phys. A  {\bf 16}, 2875 (1983). 
   
\bibitem{nick21}
 B. G. Nickel,
in {\it Phase Transitions: Cargese 1980}, 
edited by M. Levy , J. C. Le Guillou
and J. Zinn-Justin, (Plenum,  New York, 1982);
 S. Gartenhaus and W. Scott Mc Cullough, 
 Phys. Rev. B  {\bf 38}, 11688 (1988);
 W. P. Orrick, B. Nickel, A. J. Guttmann, and J. H. H. Perk, 
 J. Stat.\ Phys.\ {\bf 102}, 795-841 (2001). The expansion coefficients 
computed in this paper to order   323 for the $S=1/2$ susceptibility  
 can be found at the URL:  
http://www.ms.unimelb.edu.au/\~{ }tonyg/.


\bibitem{nr90} B. G.Nickel and J. J. Rehr,
J. Stat. Phys. {\bf 61}, 1 (1990).

\bibitem{lw} M. L\"uscher and P. Weisz, 
Nucl. Phys. B {\bf 300}, 325  (1988).


\bibitem{bakin}  G. A. Baker and J. M. Kincaid,
J. Stat. Phys. {\bf 24}, 469 (1981).

\bibitem{mck} S. McKenzie,  J. Phys. A  {\bf 16}, 3133 (1983).

\bibitem{bc2d} P. Butera, R. Cabassi, M. Comi, G. Marchesini,
Comp. Phys. Comm. {\bf 44}, 143 (1987); 
P. Butera, M. Comi and G. Marchesini, 
 Nucl. Phys. B {\bf 300}, 1 (1988).
\bibitem{bc2d1} P. Butera, M. Comi and G. Marchesini, 
  Phys. Rev. B{\bf 41}, 11494 (1990);
 P. Butera and M. Comi, Phys. Rev. B {\bf 46}, 11141 (1992); 
\bibitem{bc2d2} P. Butera and M. Comi,
Phys. Rev. B {\bf 47}, 11969  (1993);
Phys. Rev. B {\bf 50}, 3052 (1994); Phys. Rev. B {\bf 54}, 15828 (1996).

\bibitem{bc}P. Butera and M. Comi,  
Nucl. Phys.B (Proc. Suppl.) {\bf 63} A-C, 643 (1998); 
Phys. Rev. E {\bf 55}, 6391 (1997);
Phys. Rev. B  {\bf 52}  6185 (1995);  
Phys. Rev. B  {\bf 56}, 8212 (1997);
Phys. Rev. B {\bf 58}, 11552 (1998); 
Phys. Rev. B {\bf  60}, 6749 (1999);
Annals of Combinatorics {\bf 3}, 277 (1999).
   
\bibitem{bc23}P. Butera and M. Comi, 
Phys. Rev.   B {\bf 62}, 14837 (2000).

\bibitem{bcesse} P. Butera and M. Comi, 
 Phys. Rev. B. {\bf 65}, 144431 (2002); hep-lat/0112049.
 
\bibitem{zinn81}J. Zinn-Justin, J. Physique (France), {\bf42}, 783 (1981).

\bibitem{fishnick}J. H. Chen, M. E. Fisher and B. G. Nickel, 
 Phys. Rev. Lett. {\bf 48}, 630 (1982).

\bibitem{lcm} F. Englert,  Phys.Rev. {\bf 129}, 567 (1963);
M. Wortis, D. Jasnow and M. A. Moore, Phys.Rev. {\bf 185}, 805 (1969); 
M. Wortis, in {\it Phase Transitions and critical Phenomena}, 
edited by C. Domb and M.S. Green, (Academic,
London, 1974),Vol. 3;  S. McKenzie,
  in {\it Phase Transitions Cargese 1980}, edited by
M. Levy, J. C. Le Guillou and J. Zinn Justin, (Plenum, New York,
1982);  G. A. Baker, {\it Quantitative
Theory of Critical Phenomena}, (Academic, Boston, 1990).

\bibitem{cam} M. Campostrini, J. Stat. Phys. {\bf 103}, 369 (2001).

\bibitem{aha}V. Privman, P. C. Hohenberg and A. Aharony, 
 in {\it Phase Transitions and critical Phenomena}, edited by 
C. Domb and J. Lebowitz, (Academic, New York, 1989) Vol. 14.
\end{references}
\end{document}